\documentclass{aa}  

\usepackage{graphicx}
\usepackage{txfonts}
\begin{document}

   \title{Efficiency of non-thermal desorptions in cold-core conditions}

   \subtitle{Testing the sputtering of grain mantles induced by cosmic rays}

   \author{V. Wakelam
          \inst{1}
          \and
          E. Dartois\inst{2}  \and M. Chabot\inst{3}  \and S. Spezzano\inst{4}  \and D. Navarro-Almaida\inst{5} \and J.-C. Loison\inst{6}  \and A. Fuente\inst{5}   }

   \institute{Laboratoire d'astrophysique de Bordeaux, Univ. Bordeaux, CNRS, B18N, all\'ee Geoffroy Saint-Hilaire, 33615 Pessac, France\\
              \email{valentine.wakelam@u-bordeaux.fr}
         \and
             Institut des sciences Mol\'eculaires d'Orsay, CNRS, Universit\'e Paris-Saclay, B\^at 520, Rue Andr\'e Rivi\`ere, 91405 Orsay, France \\
             \and
             Universit\'e Paris-Saclay, CNRS/IN2P3, IJCLab, 91405 Orsay, France\\
             \and
             Max Planck Institute for Extraterrestrial Physics, Giessenbachstrasse 1, 85748 Garching, Germany\\
             \and
             Observatorio Astron\'omico Nacional (OAN), Alfonso XII, 3, 28014, Madrid, Spain\\
             \and
             Institut des Sciences Mol\'eculaires(ISM), CNRS, Univ. Bordeaux, 351 cours de la Lib\'eration, 33400, Talence, France \\
             }

   \date{Received xxx; accepted xxx}

% \abstract{}{}{}{}{} 
% 5 {} token are mandatory
 
  \abstract
   {Under cold conditions in dense cores, gas-phase molecules and atoms are depleted from the gas-phase to the surface of interstellar grains. Considering the time scales and physical conditions within these cores, a portion of these molecules has to be brought back into the gas-phase to explain their observation by milimeter telescopes. }
   {We tested the respective efficiencies of the different mechanisms commonly included in the models (photo-desorption, chemical desorption, and cosmic-ray-induced whole-grain heating). We also tested the addition of sputtering of ice grain mantles via a collision with cosmic rays in the electronic stopping power regime, leading to a localized thermal spike desorption that was measured in the laboratory.}
   {The ice sputtering induced by cosmic rays has been added to the Nautilus gas-grain model while the other processes were already present. Each of these processes were tested on a 1D physical structure determined by observations in TMC1 cold cores. We focused the discussion on the main ice components, simple molecules usually observed in cold cores (CO, CN, CS, SO, HCN, HC$_3$N, and HCO$^+$), and complex organic molecules (COMs such as CH$_3$OH, CH$_3$CHO, CH$_3$OCH$_3$, and HCOOCH$_3$). The resulting 1D chemical structure was also compared to methanol gas-phase abundances observed in these cores.}
   {We found that all species are not sensitive in the same way to the non-thermal desorption mechanisms, and the sensitivity also depends on the physical conditions. Thus, it is mandatory to include all of them. 
   %We were able to reproduce the observed methanol abundances within a factor of 10 at $3.6\times 10^5$~yr for all densities. 
   Chemical desorption seems to be essential in reproducing the observations for H densities smaller than $4\times 10^4$~cm$^{-3}$, whereas sputtering is essential above this density. The models are, however, systematically below the observed methanol abundances. 
%  A more efficient chemical desorption (as the one on bare grains for low density) and a more efficient sputtering (as the one on CO$_2$ ices for high densities) could reproduce better the observations.
      A more efficient chemical desorption and a more efficient sputtering could better reproduce the observations.}
   {In conclusion, the sputtering of ices by cosmic-rays collisions may be the most efficient desorption mechanism at high density (a few $10^4$~cm$^{-3}$ under the conditions studied here) in cold cores, whereas chemical desorption is still required at smaller densities. Additional works are needed on both mechanisms to assess their efficiency with respect to the main ice composition.}

   \keywords{Astrochemistry, ISM: molecules, ISM: abundances, ISM: evolution, ISM: cosmic rays
               }

   \maketitle
%
%-------------------------------------------------------------------

\section{Introduction}

In recent years, our knowledge of the composition of the interstellar medium (ISM) and the process of star formation has entered a new area thanks to the powerful capabilities (in term of sensitivity, spatial resolutions, and new wavelength windows) of recent ground-based and space observatories (Spitzer, Herschel, ALMA, NOEMA, and the new EMIR receiver on the IRAM 30m). These new instruments have allowed for the detection of new molecules (in the gas-phase or in the ices), namely, either simple ones such as HCl$^+$ \citep{2012ApJ...751L..37D} or complex ones such as (NH$_2$)$_2$CO \citep{2019A&A...628A..10B}. In addition to revealing the richness of the ISM chemistry, they have revealed the non-uniform distribution of their abundances from object to object or even within a certain type of object (with similar physical conditions).  One of the key questions astrochemists want to answer is how these molecules in the gas-phase or in the ices form and whether we can reproduce their variability in abundance.\\
The gas-phase composition of cold cores has been observed for a long time using ground-based telescopes in the mm and submm range. Spectral surveys have also revealed a diversity of relative abundances \citep{1992IAUS..150..171O,1997ApJ...486..862P,2000ApJ...542..870D,2016ApJS..225...25G}. Current chemical models do a good job at reproducing most of the observed abundances \citep{2006A&A...451..551W}, although there are still some molecules that have continue to present challenges, even though the chemistry has been  considered thoroughly \citep[e.g., CH$_3$CCH and S$_2$H, ][]{2016MolAs...3....1H,2017ApJ...851L..49F}. In recent years, however, the detection of complex organic molecules (such as HCOOCH$_3$, CH$_3$OCH$_3$, and CH$_3$CHO) in the cold environment of dark clouds has revived the interest for cold core chemistry \citep{2012A&A...541L..12B,2014ApJ...795L...2V}. Until this discovery, these molecules were believed to be formed only in the vicinity of hot protostars, where the high dust temperature would effectively promote their formation on the grains \citep{2009ARA&A..47..427H}. Now both their formation at low temperature and the possibility to observe them in the cold gas-phase has again raised questions around their origin 
\citep{2013ApJ...769...34V,2015MNRAS.449L..16B,2015MNRAS.447.4004R,2016ApJ...830L...6J,2017ApJ...842...33V} and determined their appearance at much earlier phases during the formation of stars. While there is a debate on the chemical path (gas-phase or surface) forming these molecules, it remains clear that some of them or their precursors need to desorb from the grains at low temperatures. \\
Even for simple molecules commonly observed in the cold gas of dense cores, the physical conditions and the time scales produce a depletion of molecules onto the grains \citep[see, e.g.,   ][]{2007A&A...467.1103G}. Astrochemical models simulating such environments need to invoke non-thermal desorption mechanisms to bring back into the gas-phase molecules that would have otherwise disappeared from the gas . Several mechanisms have been proposed and some of them have been studied in the laboratory, but their inclusion in astrochemical models is far from being perfect and their efficiency is far from being explicit. 
 We cite a number of them here, but this is not an exhaustive list. The effect of cosmic rays on the ice chemistry has recently gained a lot of attention. For instance, \citet{2018PCCP...20.5359S} and \citet{2018ApJ...861...20S} have proposed methods that include a cosmic-ray driven radiation surface chemistry into astrochemical models \citep[see also][]{2019ApJ...876..140S,2020ApJ...888...52S}. Once a cosmic-ray particle collides with a grain, it produces radiolysis of species on the grains. Some products of these radiolysis can be supra-thermal and then very reactive, increasing the production of complex organic molecules. Such chemistry would certainly benefit from more experimental measurements. \citet{2013A&A...554A.111K} have developed a model in which the bulk of the ices (separated from the surface) is made of cavities in which the diffusion of the species is more efficient. The sizes and locations of these cavities are changed by the cosmic-ray impacts. \citet{2015ApJ...805...59I} has theoretically revisited the effect of impulsive spot heating induced by cosmic-ray collisions. These two last papers discuss a non-thermal desorption mechanism called "explosive desorption." In this process, when energy is brought to an ice, in which a fair amount of radicals would be frozen, this energy allows the sudden diffusion of these radicals. A chain of exothermic reactions between these radicals would then take place, provoking an explosive desorption of the surfaces. The idea was studied experimentally by \citet{1982A&A...109L..12D} through the photolysis of ices. This process has been included in some models \citep[see, e.g.,][]{2004A&A...415..203S,2013MNRAS.430..264R}, however more experimental data on this process would be needed, as stated in these findings. Even the very exothermic formation of H$_2$ on the grains has been proposed to be responsible for non-thermal desorption localized around the location of the H$_2$ formation \citep{1993MNRAS.260...37D,1994MNRAS.267..949W,2007MNRAS.382..733R}. Such a process does not, however,  appear to be efficient in the lab \citep{2016A&A...585A..24M,2017MolAs...6...22W}. The effect of fast grain rotation on the species desorption rates have been investigated by \citet{2019ApJ...885..125H}. To explain some methanol gas-phase observations in a cold core, \citet{2020ApJ...895..101H} proposed the grain-grain collisions to be an efficient non-thermal desorption mechanism.\\
In this paper, we test the respective efficiencies of the different mechanisms commonly included in the models (photo-desorption, chemical desorption, and cosmic-ray-induced whole-grain heating) and tested the addition of sputtering of ice grain mantles via collision with cosmic rays in the electronic stopping power regime, leading to a localized thermal spike desorption (whose efficiency was recently measured). 
We describe in Section 2 the chemical and physical model used to study the efficiency of each of these processes. The model results are given in Section 3 and then compared to methanol observations in Section 4. In Section 5, we discuss some of the model assumptions and we present our conclusions in the last section.

\section{Chemical model}\label{chem_model}

\subsection{General presentation}
To study the non-thermal desorption processes, we used the Nautilus gas-grain code. Nautilus is widely detailed in \citet{2016MNRAS.459.3756R} and  we only briefly describe it here. It is a numerical model that simulates the chemistry under interstellar conditions. For simplicity, only one grain size of 0.1$\mu$m is considered.
The chemical processes are described as chemical reactions and the model computes the efficiency of each process based on a number of parameters. For bimolecular gas-phase reactions, for instance, the efficiency of each reaction is computed with a modified Arhenius temperature dependent law with three parameters. In some cases, these parameters are determined by experiments or theoretical calculations. In many other cases, the rate coefficients are guessed based on similarities with other systems. The considered gas-phase processes are bimolecular reactions (involving neutral-neutral and ion-neutral reactions), direct cosmic-ray ionization or dissociation, ionization or dissociation by UV photons, ionization or dissociation produced by photons induced by cosmic-ray interactions with the medium \citep{1983ApJ...267..603P}, and electronic recombinations. Details on how the efficiency of each gas-phase process is computed are given in \citet{2012ApJS..199...21W}. \\
The chemical species in the gas-phase can be physisorbed on dust surfaces while colliding with the grains based on equations of \citet{1992ApJS...82..167H}. We used a sticking probability of 1, except for H and H$_2$ for which we use temperature dependent values \citep[based on laboratory experiments from][]{2010JChPh.133j4507M,2012A&A...538A.128C}. For species on the surfaces, we make a distinction between the species in the most external layers (2 monolayers here) that we call surface species, and the species below these layers called mantle species. The species are adsorbed on the surface and become the mantle during the construction of the ices. Similarly, when the species desorb, only the species from the surface can desorb but surface species are gradually replaced  by the mantle species. 
Once they are on the grains, these species can diffuse through thermal hopping or tunneling through diffusion barriers. For this particular model, we assumed that all species can diffuse through tunneling with an efficiency that depends on the mass of the species. The equations for diffusion through tunneling are based on Eq. 10 from \citet{1992ApJS...82..167H}. The thickness of the diffusion barrier is taken to be 2\AA   \citep{Asgeirsson2017}. Both surface and mantle species are capable of diffusion but the process in the mantle is much less efficient. We assume a ratio of 0.4 for the surface species and 0.8 for the mantle species between the diffusion and binding energies. In addition, following \citet{2015PCCP...1711455G}, we assume that water drives the diffusion in the mantle, so that all mantle species with binding energy smaller than that of water are set to the water value. \\
The surface reactions are assumed to proceed through the Langmuir-Hinshelwood mechanism. The probability of reaction is assumed to be one for exothermic and barrier less reactions. For reactions with barriers, the probability of reaction is computed taking into account the competition between diffusion and reaction as explained by \citet{2016MNRAS.459.3756R}. This probability also depends on the efficiency of the reaction to tunnel through the chemical barrier and the reduced mass of the system. The width of the chemical barrier is taken to be 1\AA. 
%For the reactions with barriers that represent an exchange in hydrogen, the probability is computed using only the mass of the hydrogen. 
In addition to surface reactions between two adsorbed species, the model includes the ionization or dissociation of the surface and mantle species by UV photons and photons induced by cosmic-ray particles interactions with the gas. The values of the rate coefficient for these processes are the same as in the gas-phase because of a lack of laboratory or theoretical data. \citet{2018MNRAS.478.2753K} proposed applying a scaling factor to the photodissociation rates to use them for surface species. By comparing his model results with ice mantle observations, he determined that the best agreement was obtained for a scaling factor of 0.3. Such a result would, however, be strongly model-dependent.   \\
Thermal desorption is included in the model as well as a number of non-thermal desorption mechanisms that will be discussed below. In essence, the model described above is the same as in \citet{2016MNRAS.459.3756R}, except for the diffusion through tunneling for all species. The gas and ice chemical networks are the same as in \citet{2019MNRAS.486.4198W}.
%The chemical network is however very different. As already fully described in Manigrand et al. (submitted), the gas and grain networks have been completely rebuilt, limiting to skeleton up to five atoms of carbons for linear chains, including oxygen nitrogen and sulfur-bearing, neutral and ionic compounds. DIRE PLUS. 

\subsection{Non-thermal desorption processes}

The Nautilus model from \citet{2016MNRAS.459.3756R} includes three non-thermal desorption processes: photodesorption, chemical desorption, and cosmic-ray (CR) heating. We included in Nautilus a new process, which is the sputtering of the grains by the cosmic-ray particles (CR sputtering). Photo-desorption, CR heating, and chemical desorption are only allowed for surface species, while CR sputtering can occur for both surface and mantle species. 

\subsubsection{Photo-desorption}

%Photo-desorption is a process in which a UV photon hits a grain and transfers its energy to the molecules present close to the surface of the grains. The molecules can directly desorb if their binding energy is smaller than the transferred energy. {\bf Emmanuel: COULD YOU PLEASE CHECK THIS SENTENCE, I DO NOT KNOW WHAT I MEANT They can alternatively transfer this energy to molecules below the surface and these molecules can interact with the upper layers of molecules or desorb after diffusing through the surface.} 

Photodesorption occurs when a single energetic UV photon absorbed close to the grain surface induces the desorption of molecules or radicals. Molecules at the surface can desorb immediately if the energy transferred by the UV photon overcome their binding energy. The energy can also be transferred to a neighboring atom or molecule, inducing an indirect desorption process. Molecules absorbing photons below the surface can also interact with upper layers of molecules or diffuse through the surface and desorb.
Another possibility is called co-desorption, whereby the species below the surface takes with it the species above it. The process can in fact be even more complicated when including
the photodissociation. For instance, \citet{2016ApJ...817L..12B} and \citet{2016A&A...592A..68C} showed experimentally that methanol would break and its fragments would desorb. Depending on the energy brought to the system, the photo-fragments could also stay close to the surface and recombine.  For simplicity, we considered independently the dissociation (see Section \ref{chem_model}) and the desorption. In other words, photo-desorption leaves the product intact but photo-dissociations on the surface can occur and then the products can photo-desorb. \\
For the photo-desorption, we have used the simplified formalism for all species described in \citet{2016MNRAS.459.3756R}:
\begin{equation}\label{eq1}
\rm k_{des,UV} = F_{UV}S_{UV}exp(-2A_V)Y_{pd}\frac{\pi r_{dust}^2}{N_{site}}
,\end{equation}
\begin{equation}
\rm k_{des,UV-CR} = F_{UV-CR} \frac{\zeta}{10^{-17}} Y_{pd}\frac{\pi r_{dust}^2}{N_{site}}  \\
,\end{equation}
for  the desorption rates induced by direct $\rm k_{des,UV}$ and indirect $\rm k_{des,UV-CR}$ UV photons (in s$^{-1}$). The strength of the direct UV field is $\rm F_{UV} =1.0\times 10^8$ photons cm$^{-2}$s$^{-1}$ \citep{2007ApJ...662L..23O}.
We adopt a secondary (cosmic ray induced) UV field of $\rm F_{UV} =10^3$ photons cm$^{-2}$s$^{-1}$ scaled to an ionisation rate of $10^{-17}$s$^{-1}$. This value is an adapted value considering the $\sim$3100 photons cm$^{-2}$s$^{-1}$ for an ionisation rate of $3\times 10^{-17}$s$^{-1}$ from \citet{2004A&A...415..203S}, whereas \citet{1983ApJ...267..603P} gave about 2380 photons cm$^{-2}$s$^{-1}$ for an ionisation rate of $3\times 10^{-17}$s$^{-1}$. In addition,
$\rm S_{UV}$ is the scaling factor for the UV radiation field and $\zeta$ the cosmic ray ionisation rate in s$^{-1}$.
The yield ($\rm Y_{pd}$) is assumed to be $10^{-4}$ molecules per photons for all species \citep{2008A&A...491..907A}. Furthermore, $\rm N_{site}$ is the total number of surface sites on a grain while $\rm r_{dust}$ is the radius of the grains. In our case, $\rm N_{site} \sim 1.2\times 10^6$ and $\rm r_{dust}$ = 0.1$\mu$m.
The strength of the UV radiation field induced by the cosmic-rays is scaled with the local value of the CR ionization rate. The factor of 2 in the exponential of the photo-desorption rate by direct UV photons (Eq.~\ref{eq1}) is taken from \citet{1991ApJS...77..287R} and it takes into account the higher extinction of the grains in the UV wavelength range as compared to the visible (used to compute the Av). This is a mean value over the distribution of photons. A more robust approach would be to compute this scaling factor for each molecule of the ice by convolving the spectrum of photo-desorption for each molecule by the extinction curve of the grains. Such a calculation is, however, outside the scope of this paper, but should be considered in the future. 
%To be consistent with values used for the cosmic- ray sputtering, we have also changed the computation of the secondary UV photons (used to compute the photodesorption induced by secondary photons) produced by cosmic-rays: 1000 photons/cm$^{-2}$ for $10^{-17}$~s$^{-1}$ scaled with $\zeta$. 

\subsubsection{Grain heating by cosmic rays}

The energy released by the collision between high velocity cosmic-ray particles and grains can produce a global heating of the grain or a localized one \citep{1985A&A...144..147L,2004A&A...415..203S}, which then cools down to its initial temperature. To include this process in Nautilus, we followed the simple formalism proposed by \citet{1993MNRAS.261...83H} for the whole grain heating: $\rm k_{des,cosmic} = f_{CR} \times k_{des,therm}(T_{peak}),$ with $\rm f_{CR}$ being the ratio between the cooling time of the grain and the time between two collisions, while T$_{\rm peak}$ is the peak temperature reached by the grain. The $\rm f_{CR}$ and T$_{\rm peak}$ parameters critically depend  on the size of the grains as well as their nature and their coverage, as the cooling of the grains occurs mostly by molecular evaporation \citep{2006PNAS..10312257H}. In the current model, we kept the prescription by \citet{1993MNRAS.261...83H}: $\rm f_{CR}  = 3.16\times 10^{-19}$ and T$_{\rm peak}$ = 70~K. 
  This process is stochastic and our simple approach does not fully reproduce  its complex nature. \citet{2019MNRAS.486.2050K,2020A&A...633A..97K,2020A&A...641A..49K}, for instance, studied this process in detail  and proposed other ways to include it into astrochemical models. One  limitation of our approach is that we used only one grain size, thus restricted to a representative grain size, in order to limit the number of free parameters. \citet{2006PNAS..10312257H} did a theoretical study of the efficiency of whole grain heating with the size and composition of the grains. They showed that for silicate grains, the peak temperature reached by small grains upon cosmic ray collision increases as the grain radius decreases. \citet{2018A&A...615A..20I}  showed that considering a grain size distribution in these complex astrochemical models increases strongly the importance of this process.

\subsubsection{Chemical desorption}

The energy released by exothermic surface reactions can produce partial evaporation of the products. We call this process chemical desorption. The exact description of the process is yet to be understood. As a result, several models have been proposed to include this process into kinetic models. \citet{2019MNRAS.490..709Y} is the latest published model and compares its results with the two previously published ones: \citet{2007A&A...467.1103G} and \citet{2016A&A...585A..24M}. The model proposed by \citet{2016A&A...585A..24M} is the only one based on some experimental measurements but only for a few small systems (O + H, OH + H, and N + N). The three proposed models are all based on a number of unknown parameters. The model from \citet{2007A&A...467.1103G} depends on the fraction of the energy released by the reaction that is transfered to the produced species. The model from \citet{2016A&A...585A..24M} depends on the effective mass of the surface (which also defines the fraction of the energy retained by the product). The recent model from \citet{2019MNRAS.490..709Y} is based on a different theory. It does not assume that the products retain some energy, which is then transformed to kinetic energy, but it assumes that the released energy heats the surface. So their model depends on the thermal diffusivity and specific heat of the surface. \citet{2019MNRAS.490..709Y} compared the efficiency of the three models for bare grains and showed that this gave similar results.  \citet{2016A&A...585A..24M} showed experimentally that the chemical desorption would be very much less efficient on water ice surfaces than on bare grains. In the simulations presented here, we anticipate such concerns by stating that the grains are covered by water ices (with a significant fraction of CO$_2$). Since 
the model of Garrod does not explicitly depends on the nature of the surface (except by decreasing the $a$ parameter to an unknown value) and that the model of Yamamoto does not give the parameters to be used for water ices, we included only the model from \citet{2016A&A...585A..24M}. \\
Similarly to what we did in \citet{2017MolAs...6...22W}, the fraction of evaporation for singly produced species is computed following:
\begin{equation}
\rm f = e^{-\frac{E_D}{\epsilon E_{reac}/N}}
,\end{equation}
with N the number of degree of freedom of the produced species (N = 3n with n the number of atoms in the produced species) and $\epsilon = \frac{(M-m)^2}{(M+m)^2}$ the fraction of energy kept by the product with a mass m. M is the effective mass of the surface. We first compute the effective mass for bare grains of 120 amu and divide it by 10 to obtain the values for water ices. For the three measured systems, we used the measured values ($\rm f_{O+H} = 30$\%, $\rm f_{OH+H}$ = 25\%, and $\rm f_{N+N}$ = 50\%). For channels producing more than one product, we set the chemical desorption efficiency to zero.

\subsubsection{Sputtering by cosmic-rays}

When  CRs impact solids, many  excited electron states are created. In less than a picosecond, these excited states relax in atomic motions and lead to a thermal spike \citep{2000NIMPB.166..903T}. As the result of the hot spot, some of the material is sputtered. The sputtering is scaling as the square of the stopping power, or the energy loss by thickness unit dE/dx \citep{2015NIMPB.365..472D,2015NIMPB.365..477M}. It sets heavy and low-energy CRs as the main contributor to this process.

We added this new type of process into Nautilus assuming that both surface and mantle species can desorb simultaneously. For simplicity, we assume that species desorb without breaking but this may not be the case all the time (as for photo-desorption).
The rate coefficient of these reactions are computed as follows:
\begin{equation}
%\rm k_{des,CR}(i) = (\zeta/3\times10^{-17}) \times Y_{eff} \times erf(n_{layers}/\alpha) \times \pi \times r_{dust}^2/N_{site}
\rm k_{des,CR}(i) = (\zeta/3\times10^{-17}) \times Y_{eff} \times \pi \times r_{dust}^2/N_{site}
,\end{equation}
with
\begin{equation}
\rm Y_{eff} = Y^{\infty} \times (1-e^{-n_{layers}/\beta)^\gamma})
,\end{equation}
where $\zeta$ is the cosmic-ray ionisation rate of H$_2$ (in s$^{-1}$),  $\rm r_{dust}$ the radius of a grain, and $\rm N_{site}$  the number of sites on one grain. $\rm Y_{eff}$ is the efficiency of desorption integrated over a cosmic-ray spectral distribution, which is a function of the number of layers of ices ($\rm n_{layers}$) \citep{2018A&A...618A.173D}. 
$Y^{\infty}$ is the sputtering yield for thick ices and $\beta$,  $\gamma$ are two parameters associated to nature of the ice. For the physical structure adopted here, water ices remain the dominant ice component although CO$_2$ is also very abundant in the ice. This is due to the high dust temperature assumed (see discussion in Section 3.1). Sputtering on CO$_2$ ice is much more efficient than on water. We used the sputtering parameters for water ices: $Y^{\infty} = 3.63$ , $\beta = 3.25$, and $\gamma = 0.57$ \citep{2018A&A...618A.173D} and we test the values for CO$_2$ ices in Section 4. 
 For the grain sputtering, in the experiments, the diameter of the craters made by the heavy (above C) cosmic-ray particles in the ices is on the order of a few nanometers. Thus, for grains with sizes above 10 nm, the measured yield should apply. The ice mantle thickness influence on the yield is already introduced in the formalism.  For smaller sizes, the sputtering yield should be higher and capable of being summed up with the global heating process. 

%Note that assuming the water sputtering efficiency for any ice mantle is conservative, and provides a minimum to the effective cosmic ray sputtering. When ices start to be composed of significant amounts of species such as CO$_2$ and CO ices, the effective sputtering rate is expected to increase. 

\subsection{1D physical structure}

\begin{figure*}
\centering
\includegraphics[width=0.46\linewidth]{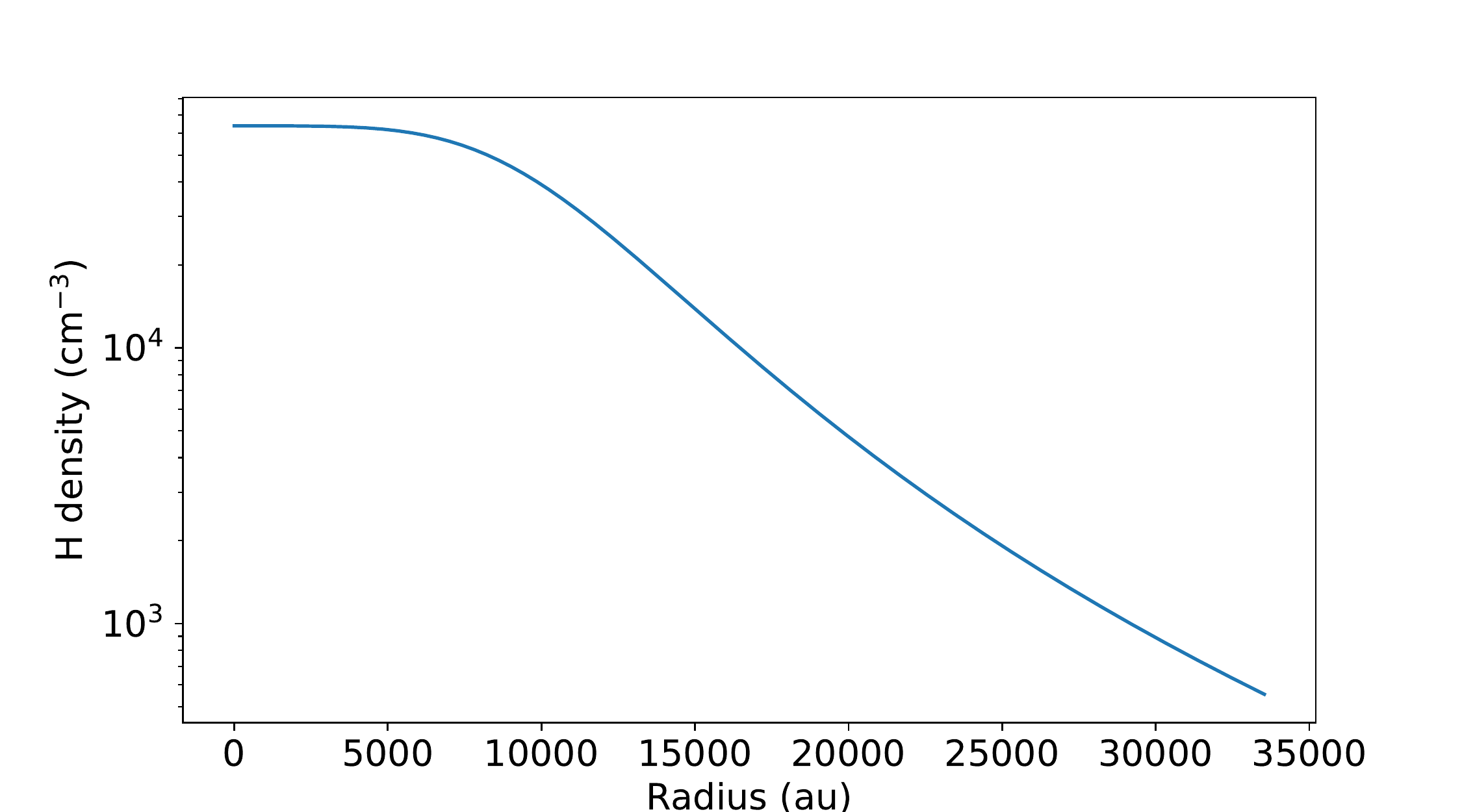}
\includegraphics[width=0.46\linewidth]{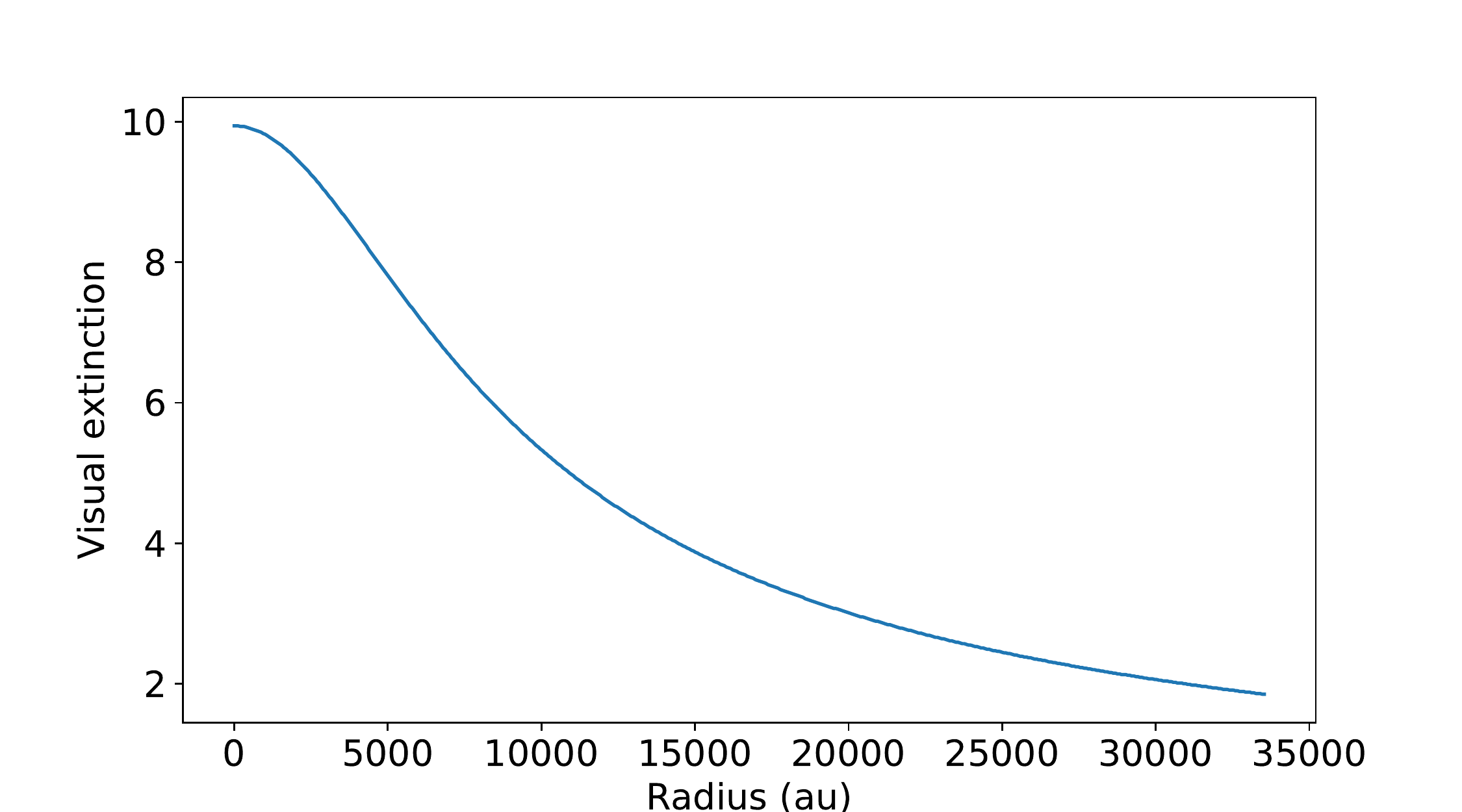}
\includegraphics[width=0.46\linewidth]{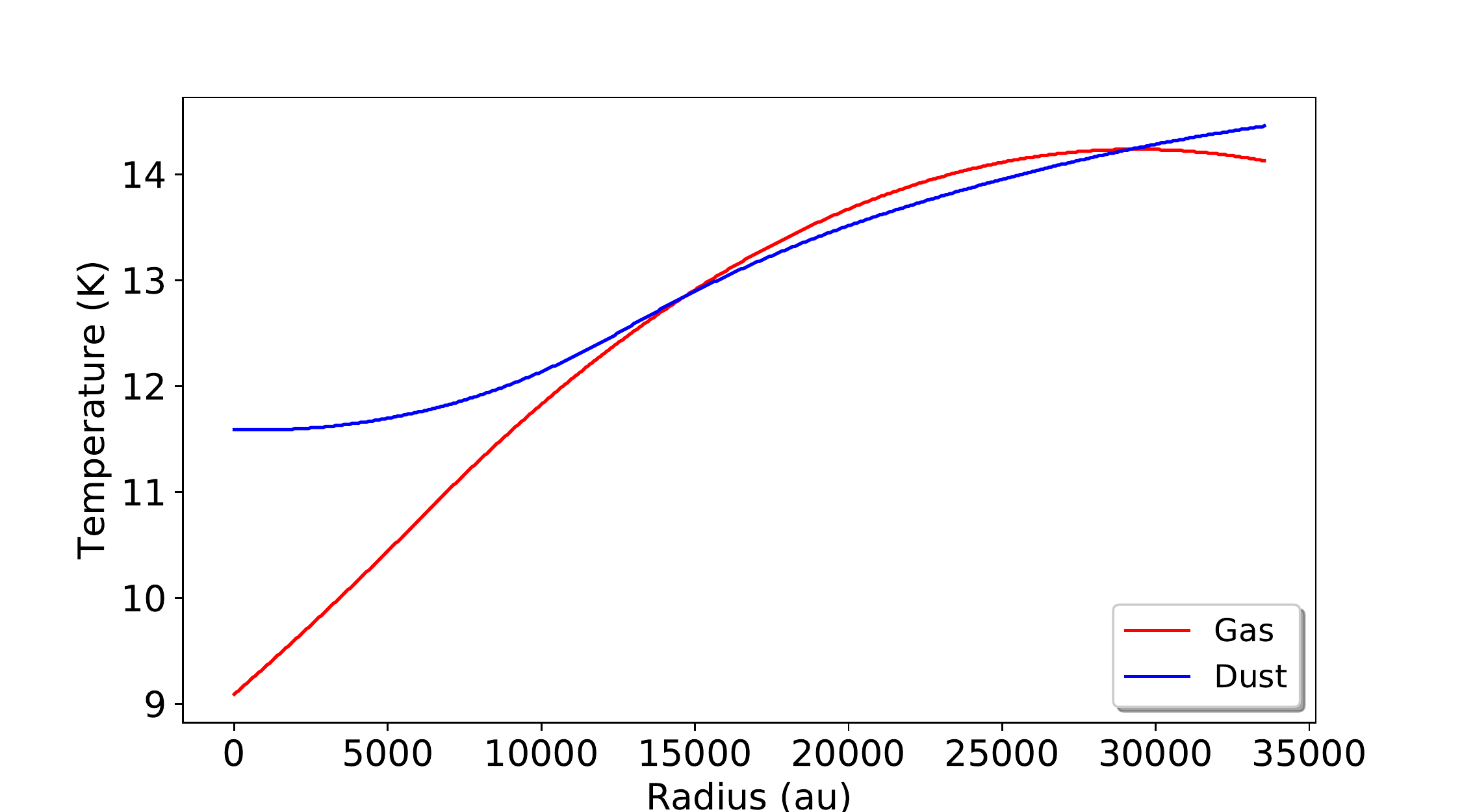}
\includegraphics[width=0.46\linewidth]{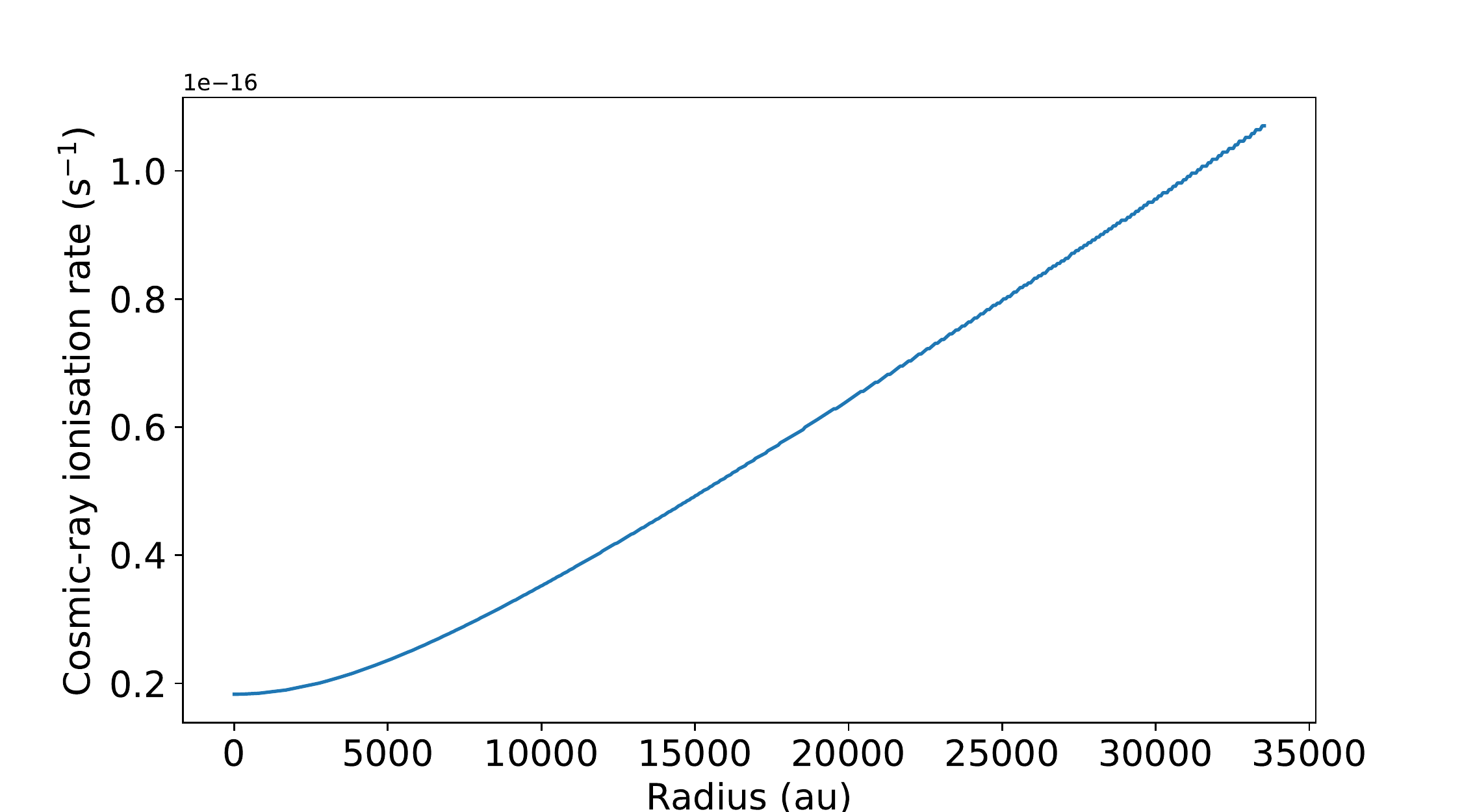}
\caption{1D physical structure used for TMC1.\label{1D_structure}}
\end{figure*}

% David here: They derived the density and gas temperature at several positions in the TMC1-C and TMC1-CP cores using CS and CS isotopologues and the RADEX radiative transfer code \citep{2007A&A...468..627V}, together with the Markov Chain Monte Carlo (MCMC) approach (see, e.g., \citet{2019A&A...628A..16R}). The density structure of TMC1 was then obtained fitting these densities to a Plummer-like analytical density profile, a widely used density profile for prestellar cores \citep{2002ApJ...569..815T, 2018AJ....156...51P}. The visual extinction profile of several positions in the cloud were obtained from the Herschel extinction maps in \citet{2020A&A...637A..39N}. Assuming spherical symmetry and isotropic UV illumination, the UV shielding at every point inside the cloud is due to an extinction that were taken as half of that measured in the extinction maps. These values of extinction at each position were then interpolated using cubic spline interpolation. Finally, the dust and gas temperature profiles were determined by a cubic spline interpolation of the dust temperatures from the Herschel maps and the gas temperatures from the MCMC approach.

To study the effect of the non-thermal desorption processes, we used the 1D cold core physical structure determined by \citet{2020A&A...637A..39N} from observations. These authors derived the density and gas temperature at several positions in the TMC1-C and TMC1-CP cores using CS and CS isotopologs and the RADEX radiative transfer code \citep{2007A&A...468..627V}, together with the Markov Chain Monte Carlo (MCMC) approach \citep[see, e.g., ][]{2019A&A...628A..16R}. The density structure of TMC1 was then obtained by fitting these densities to a Plummer-like analytical density profile, a widely used density profile for prestellar cores \citep{2002ApJ...569..815T, 2018AJ....156...51P}. The visual extinction profiles of several positions in the cloud were obtained from the Herschel extinction maps in \citet{2020A&A...637A..39N}. Assuming spherical symmetry and isotropic UV illumination, the UV shielding at every point inside the cloud is due to an extinction that was taken as half of that measured in the extinction maps. These values of extinction at each position were then interpolated using cubic spline interpolation. Finally, the dust and gas temperature profiles were determined by a cubic spline interpolation of the dust temperatures from the Herschel maps and the gas temperatures from the MCMC approach. The density, gas and dust temperatures, and visual extinctions are shown in Fig.~\ref{1D_structure} as a function of radius from the center of the core. The density starts around $5\times 10^3$~cm$^{-3}$ at $3.4\times 10^4$ au, increases up to $6\times 10^4$~cm$^{-3}$ at 5000 au and then remains flat. The visual extinction is small outside (around 2) and increases up to 10 inside. The gas and dust temperatures decreases toward the inside. They are between 9 and 14.5~K, but the dust temperature is always slightly larger than the gas temperature. 

We added to this structure a radius dependent cosmic-ray ionization rate. Indeed, the CRs coming from the outside of the cloud are slowing down by the gas and following cosmic-ray ionization rate is decreasing with penetration \citep{2009A&A...501..619P,2016A&A...585A..15C}. Observations of  cosmic-ray ionization rates using molecular ions, show a correlation with the column density \citep{2017ApJ...845..163N,2012ApJ...745...91I}.  The H$_2$ cosmic-ray ionization rate (in s$^{-1}$) was computed following the dashed red line fit presented in the lower panel of Fig 6 of \cite{2017ApJ...845..163N}:
\begin{equation}
 \zeta (Av) = 10^{-1.05\times \log_{10}(Av) - 15.69} \rm \;for\;Av > 0.5.
 \end{equation}
For low Av ($\leq$ 0.5), we adopted a constant CR rate corresponding to a non attenuated CR flux of:
 \begin{equation}
  \zeta (Av) = 10^{-1.05 \times \log_{10}(0.5)-15.69} \sim 2.9\times 10^{-16}.
\end{equation}

 We note that we are using a static 1D physical structure that does not evolve with time. This is clearly a simplification of the model, but we do not yet have  a good constraint on the dynamical evolution of the studied cores. In addition, we are interested in comparing the efficiency of several chemical processes. Such a comparison would be more difficult to make if we added a time dependency on the physical conditions and moving structures.

\subsection{Other model parameters}

Starting from a mix of atoms (with abundances listed in Table \ref{ab_ini} and apart from  H, which is assumed to be in its molecular form), we ran the chemical model using the 1D physical structure and for a time span of $10^7$~yr. The impact of the initial atomic hydrogen abundance is discussed in Section \ref{initial_H}. The external incident UV flux is assumed to be five times the Draine's field, as suggested by the observations \citep{2019A&A...624A.105F}. To independently study  the effect of the non-thermal desorptions, we switched off all of them (no desorption model) and then turn them on one at a time. In the "no desorption model," the thermal desorption is still active, but we checked that in the conditions used here, it does not change the results if no desorption at all is assumed. The exceptions are H$_2$ and He, which have to be allowed to thermally desorb, otherwise most of the gas would end up depleted.
\begin{table}
\caption{Initial abundances (with respect to the total proton density).}
\begin{center}
\begin{tabular}{c|c}
\hline
\hline
Species & Abundance\\
\hline
He  &  $9\times 10^{-2}$ \\
N  & $6.2\times 10^{-5}$ \\ 
O  & $2.4\times 10^{-4}$ \\ 
H$_2$ & 0.5 \\
C$^+$  & $1.7\times 10^{-4}$\\ 
S$^+$ & $1.5\times 10^{-5}$\\ 
Si$^+$ & $1.8\times 10^{-6}$ \\
Na$^+$ &  $2.3\times 10^{-7}$\\
Mg$^+$ & $2.3\times 10^{-6}$ \\
P$^+$   & $7.8\times 10^{-8}$ \\
Fe$^+$  & $1\times 10^{-8}$\\ 
Cl$^+$ & $1\times 10^{-9}$\\ 
F & $6.68\times 10^{-9}$\\
\hline
\end{tabular}
\end{center}
\label{ab_ini}
\end{table}%

\section{Model results}\label{model_results}

The four non-thermal processes presented in the previous section do not depend on the same quantities. The chemical desorption will depend on the abundance of the reactants and the efficiency of diffusion (which depends on the dust temperature). The photo-desorption depends on both the visual extinction and the cosmic-ray ionisation rate, while the two last processes depend on the cosmic-ray ionization rate. In addition, all species will not be affected the same way. The effect for species essentially formed on the surface should be direct while the effect for species formed in the gas-phase is more complex as it can impact their gas-phase precursors. We separated several groups of molecules and we present the impact of the different non-thermal desorption processes.

\subsection{Main ice constituents}\label{main_ice}

\begin{figure*}
\centering
\includegraphics[width=0.46\linewidth]{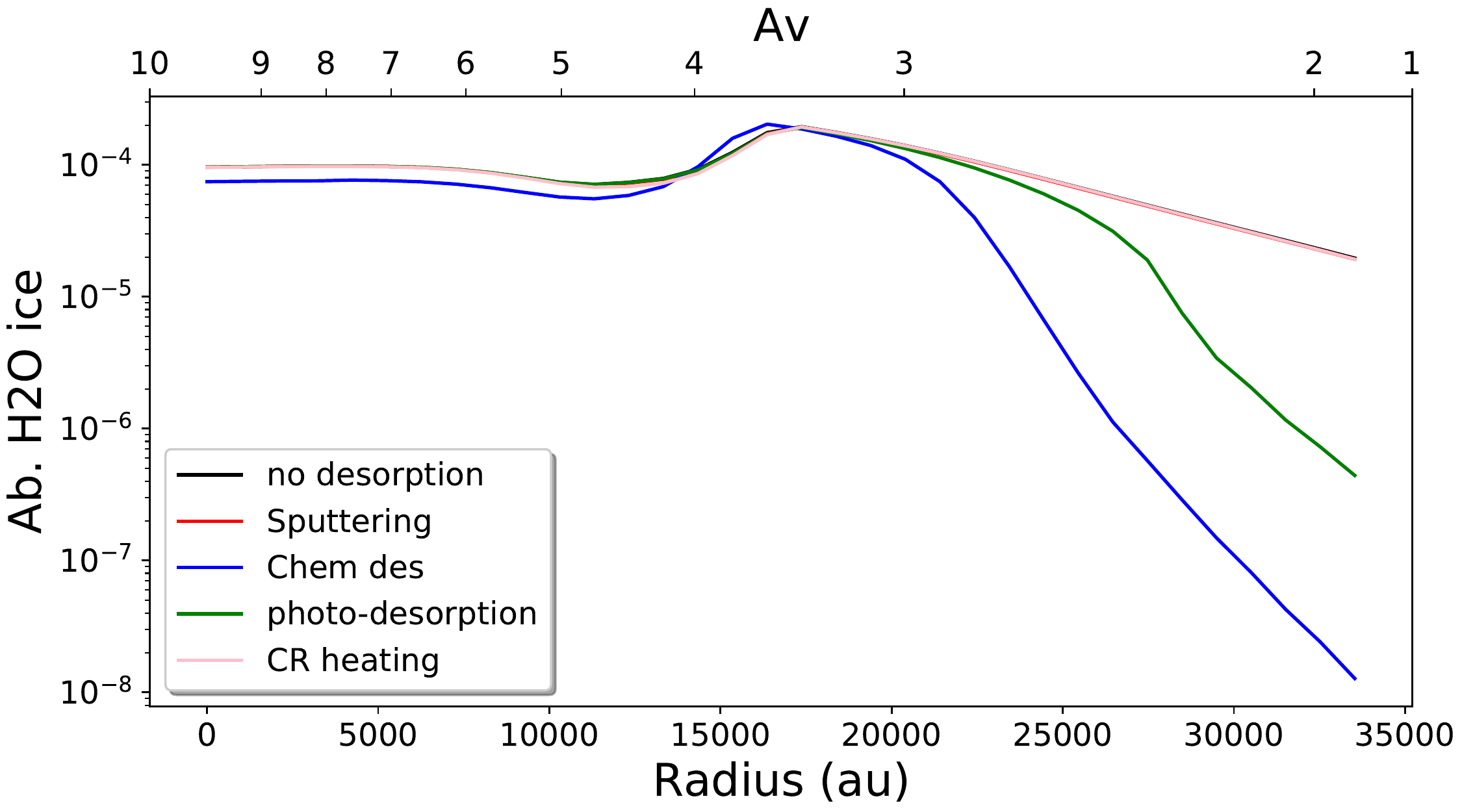}
\includegraphics[width=0.46\linewidth]{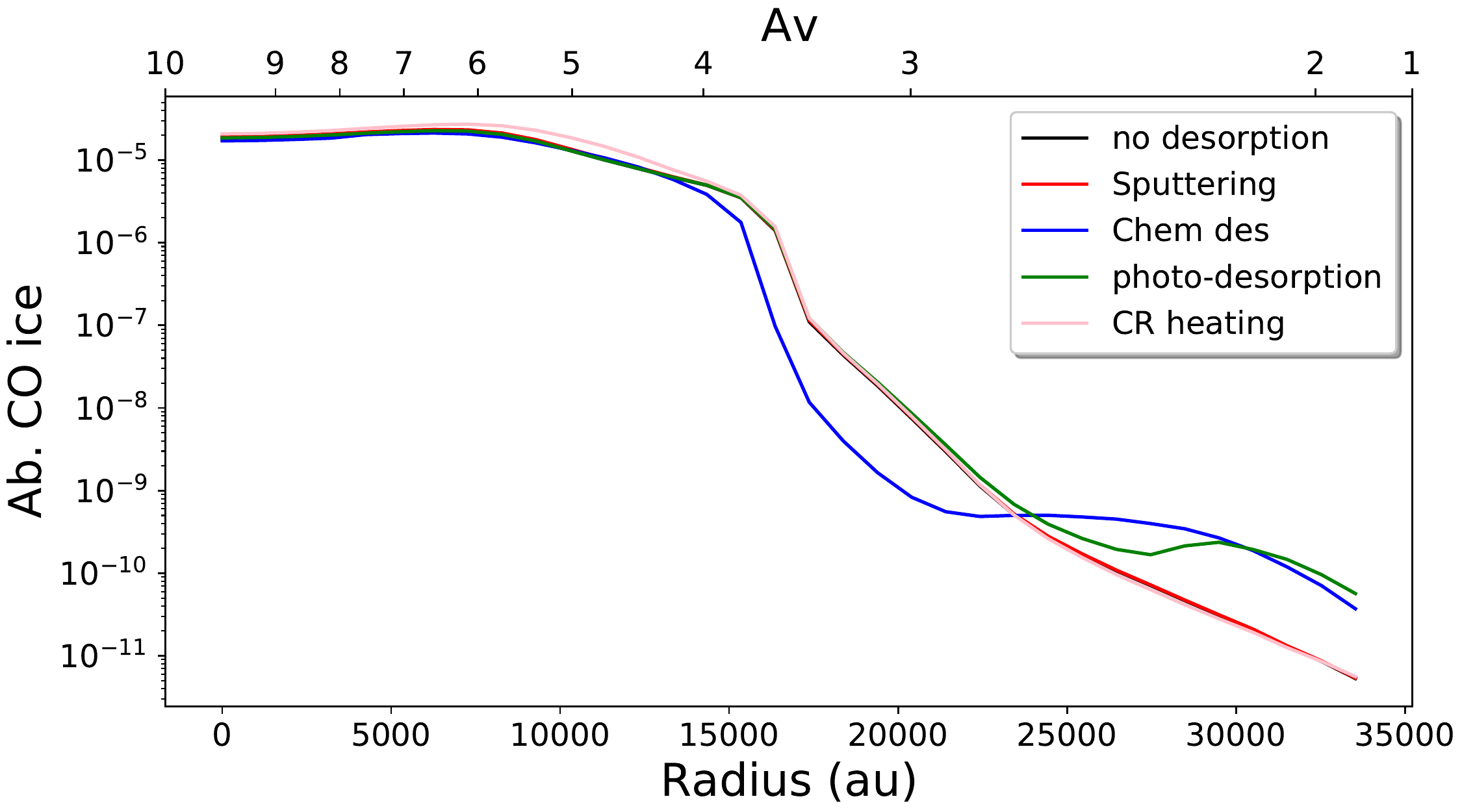}
\includegraphics[width=0.46\linewidth]{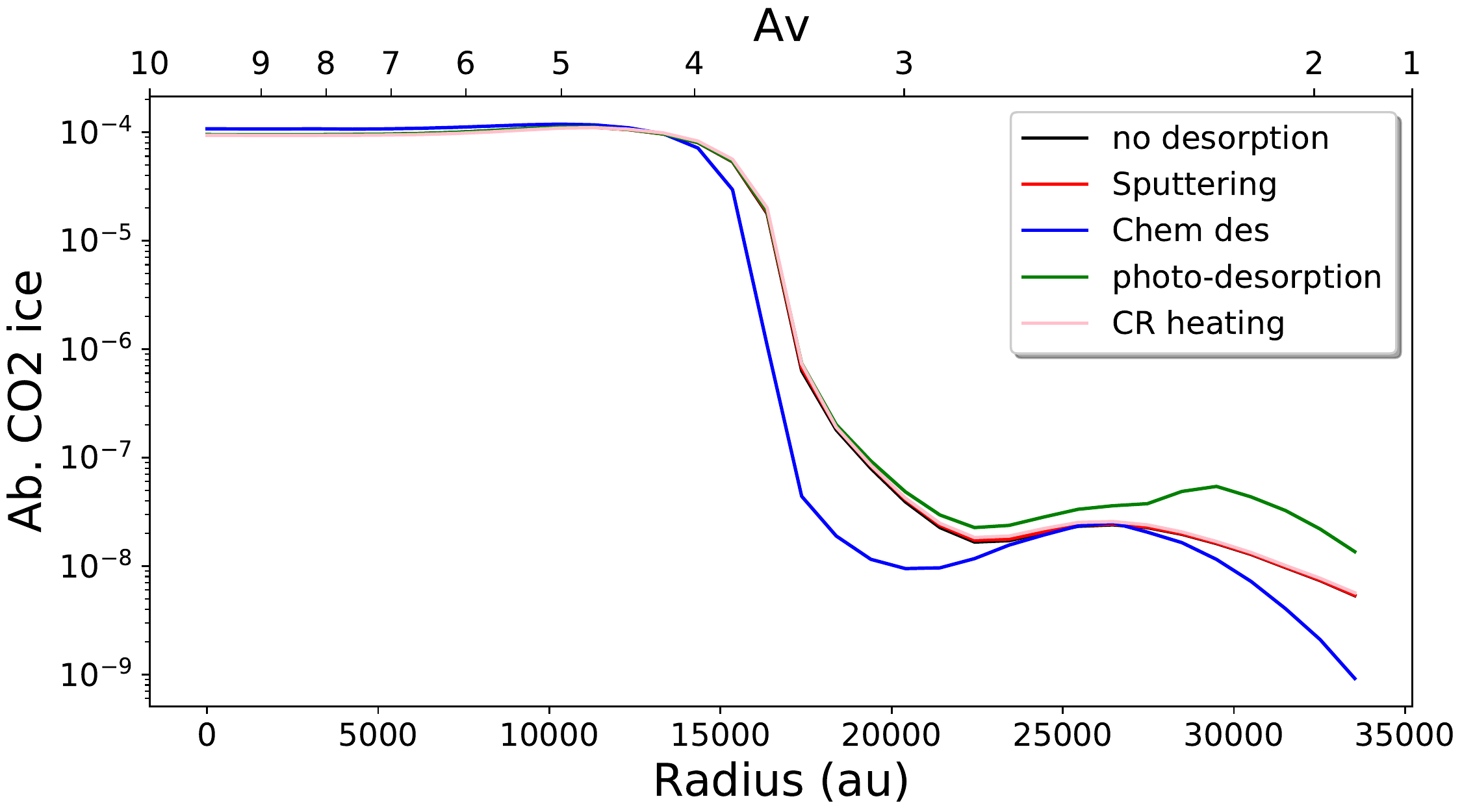}
\includegraphics[width=0.46\linewidth]{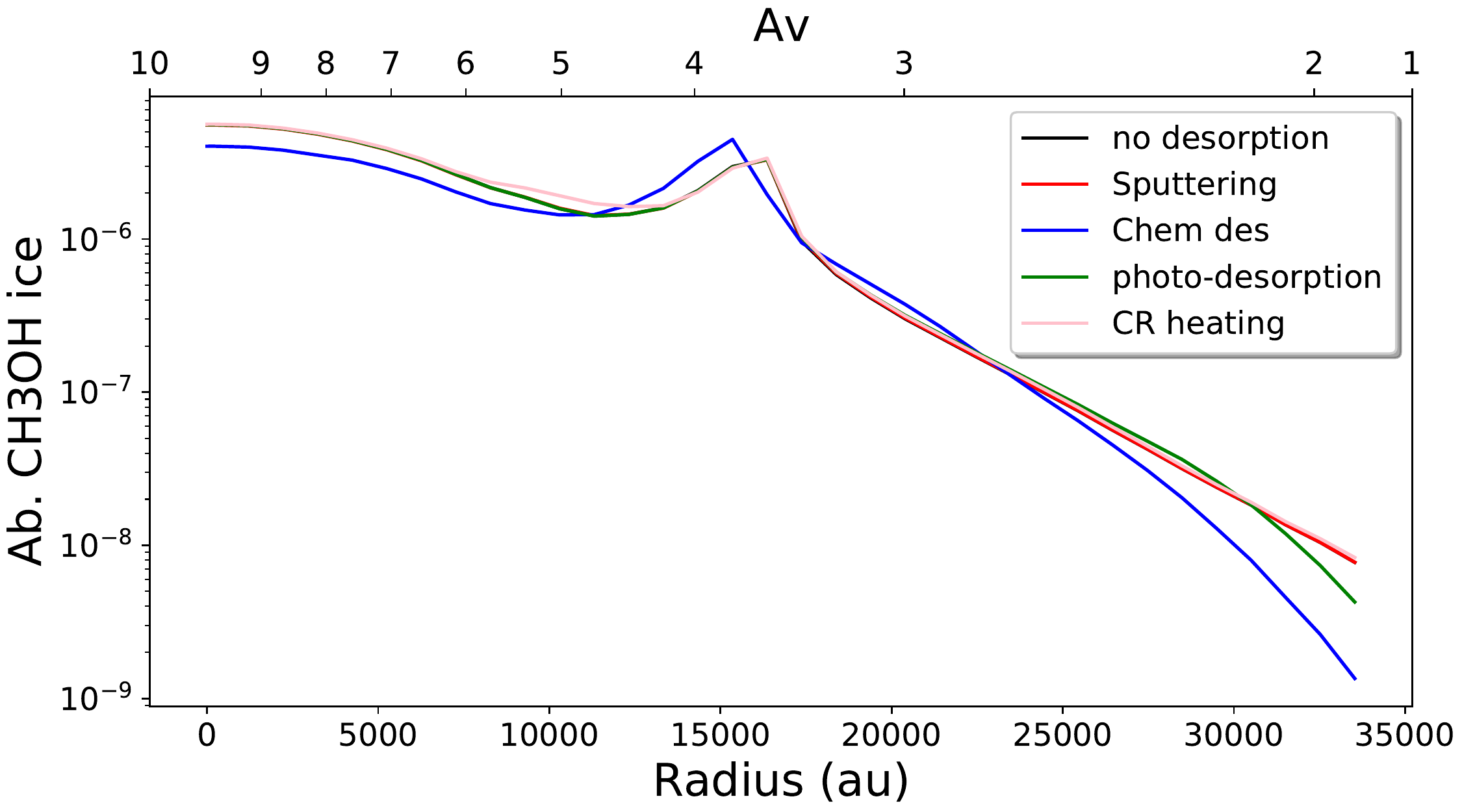}
\includegraphics[width=0.46\linewidth]{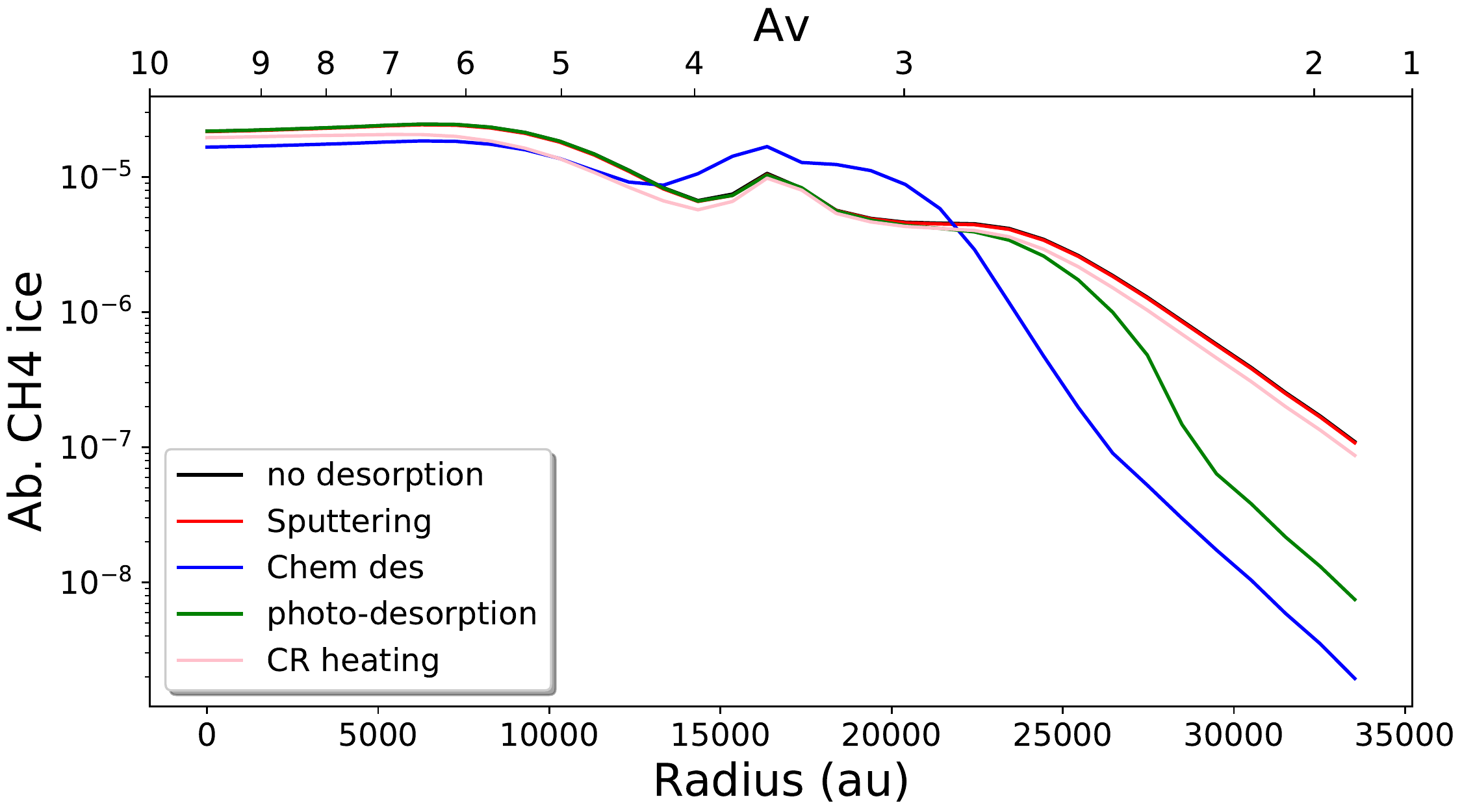}
\includegraphics[width=0.46\linewidth]{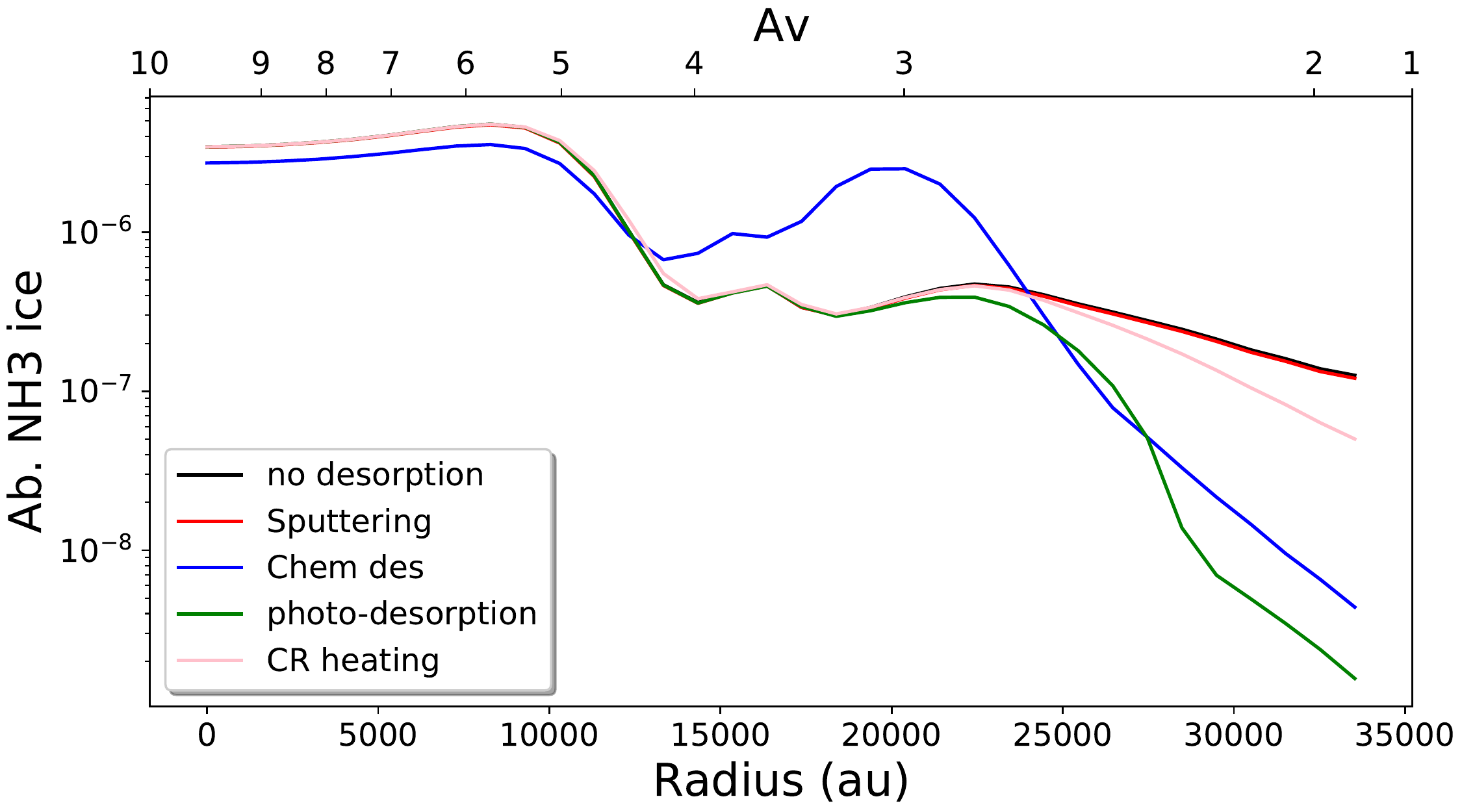}
\caption{Abundance of the main ice components as a function of radius (and visual extinction) for a time of $6\times 10^5$~yr. The "no desorption" curve is almost the same as the "sputtering" one.}
\label{ice_fig}
\end{figure*}

Figure \ref{ice_fig} shows the abundance of the main constituents of the ices (H$_2$O, CO, CO$_2$, CH$_3$OH, CH$_4$, and NH$_3$) as a function of radius for an integration time of $6\times 10^5$~yr, which is the typical age of an evolved pre-stellar core. We chose this time to emphasize the differences between the models. Indeed, at later times, the differences between the model results are much smaller and are negligible at $2\times 10^5$ yr and earlier because the interactions with the grains are less efficient. 
The figures show that the abundance of the ice species increases toward the higher densities (smaller radii). All desorption mechanisms produce similar ice abundances at high density and similar to the model without desorption. Here, H$_2$O and CO$_2$ dominate the ices for H densities larger than $10^4$~cm$^{-3}$ (Av = 4, inside 1500 au). Going outward, the CO$_2$ ice abundance drops and water clearly dominates until an Av of 4 (H density of about $(5-6)\times 10^3$~cm$^{-3}$). The amount of water then depends on the non-thermal mechanism considered, chemical desorption being the most efficient to decrease it while sputtering is the least efficient. 
%The ices composition is dominated by H$_2$O and CO$_2$ over all radii in all models. Going from high to low density, CO$_2$ and H$_2$O are equally abundant up to 6000 au, then CO$_2$ is more abundant between 6000 and about 20000 au while H$_2$O is largely the most abundant ice species in the external radii. 
The large CO$_2$ abundance over CO reflects the grain temperature that is slightly above 10~K in the entire structure. The dust temperature used in these simulations comes from Herschel data \citep{2012A&A...544A..50M} and was derived by fitting the SED with gray-body emission \citep{2020A&A...637A..39N}. This procedure is known to overestimate (by about 1-2 K) the dust temperature in the center of the cores because of the contribution of the warmer grains in the external layers of the core surface that are located along the line of sight. The effect of dust temperature is discussed in Section~\ref{sect_dust_t}.
%Decreasing the dust temperature to values as low as 7~K at high density, will only impact the CO/CO$_2$ conversion and will not affect the other species abundances. 
At an Av lower than 3, hydrogenated species such as CH$_4$, NH$_3$, and CH$_3$OH ices can be more abundant than CO$_2$ because there is more free hydrogen thanks to H$_2$ photo-dissociation. 
 
%The chemical desorption is the only process that produces a difference and only at radii larger than 15000 au. H$_2$O ice is less abundant in this model mostly because of the high desorption efficiency of the reactions O + H and OH + H. %EXPLIQUER LE RESTE?

\subsection{Simple abundant species}\label{simple_species}

\begin{figure*}
\centering
\includegraphics[width=0.46\linewidth]{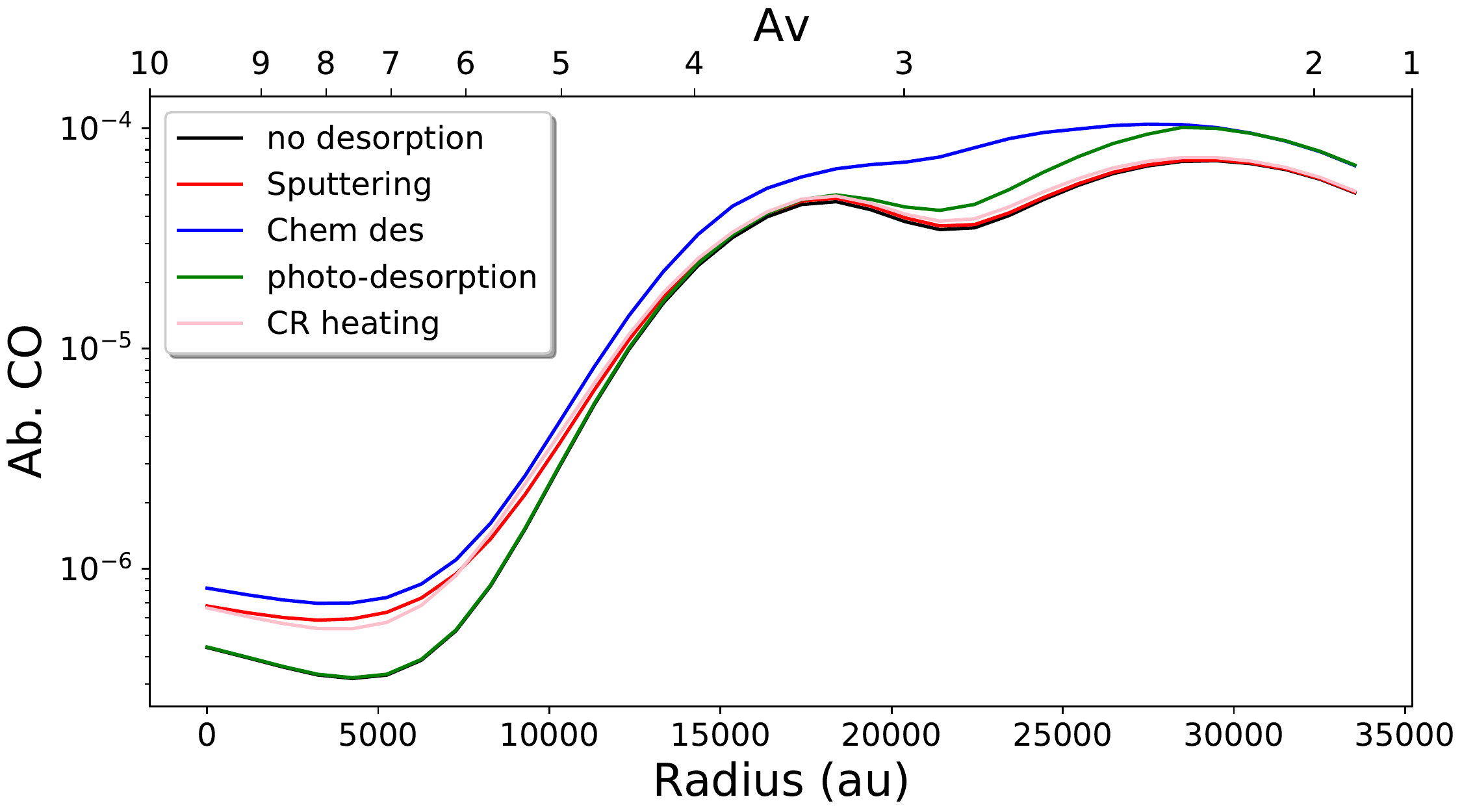}
\includegraphics[width=0.46\linewidth]{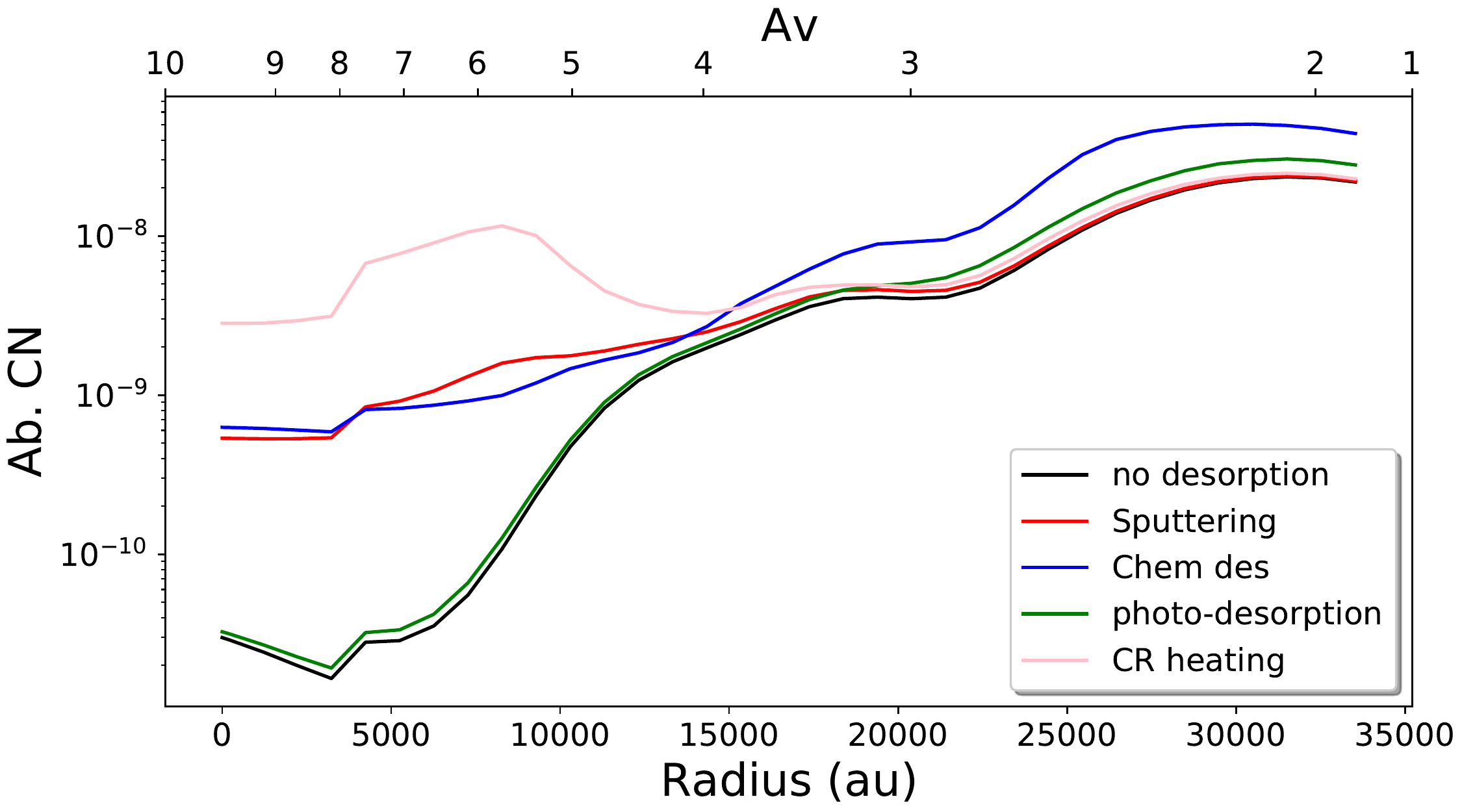}
\includegraphics[width=0.46\linewidth]{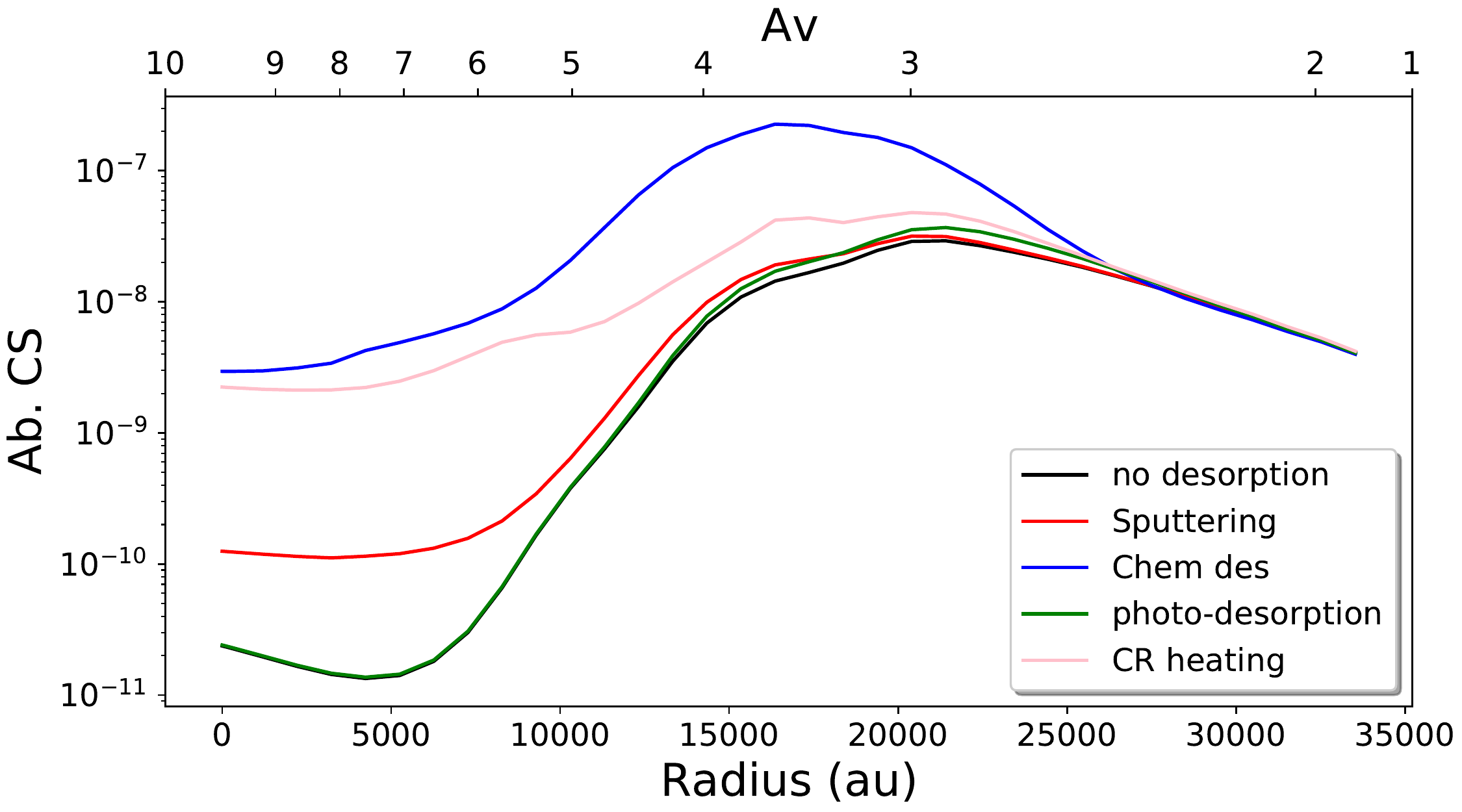}
\includegraphics[width=0.46\linewidth]{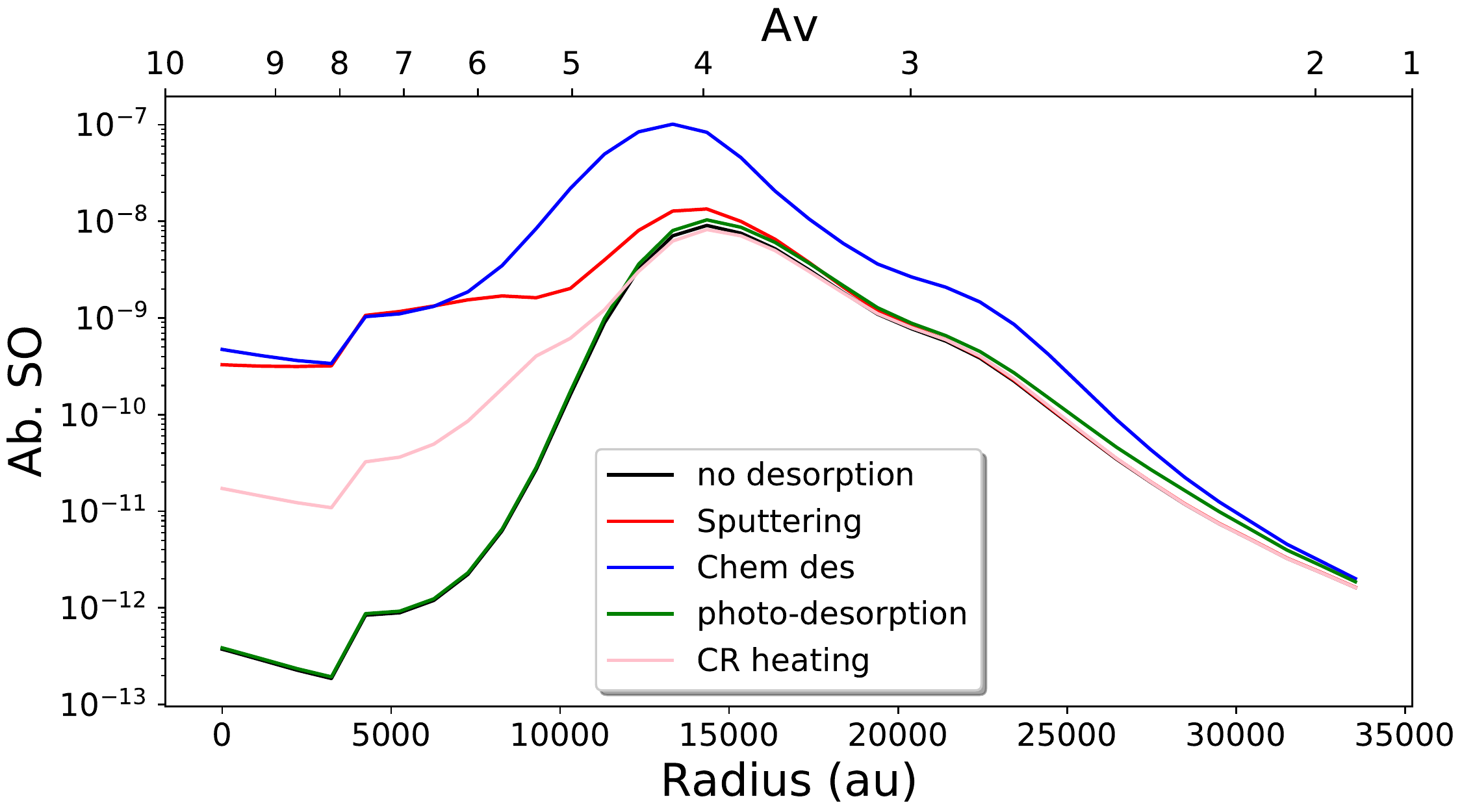}
\includegraphics[width=0.46\linewidth]{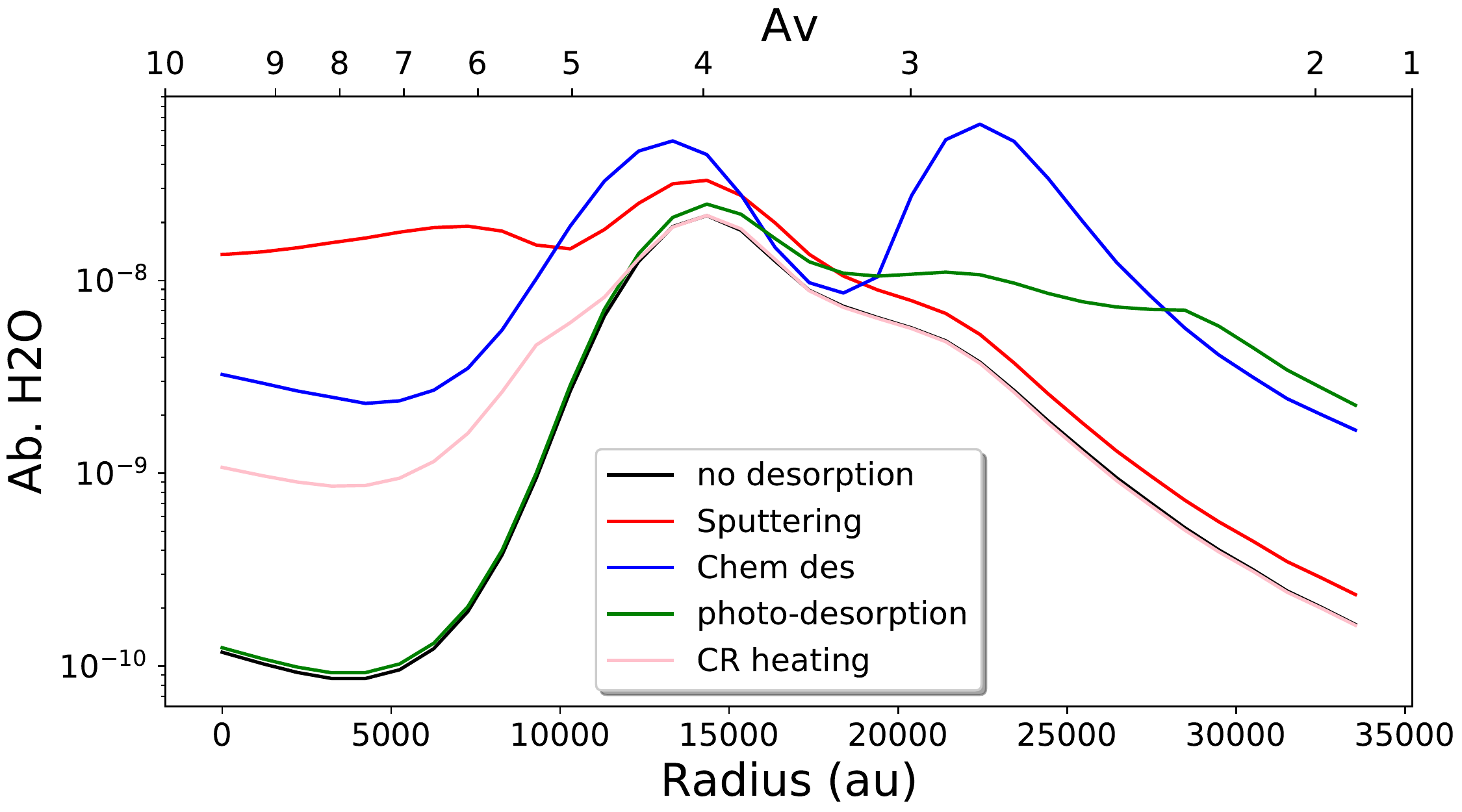}
\includegraphics[width=0.46\linewidth]{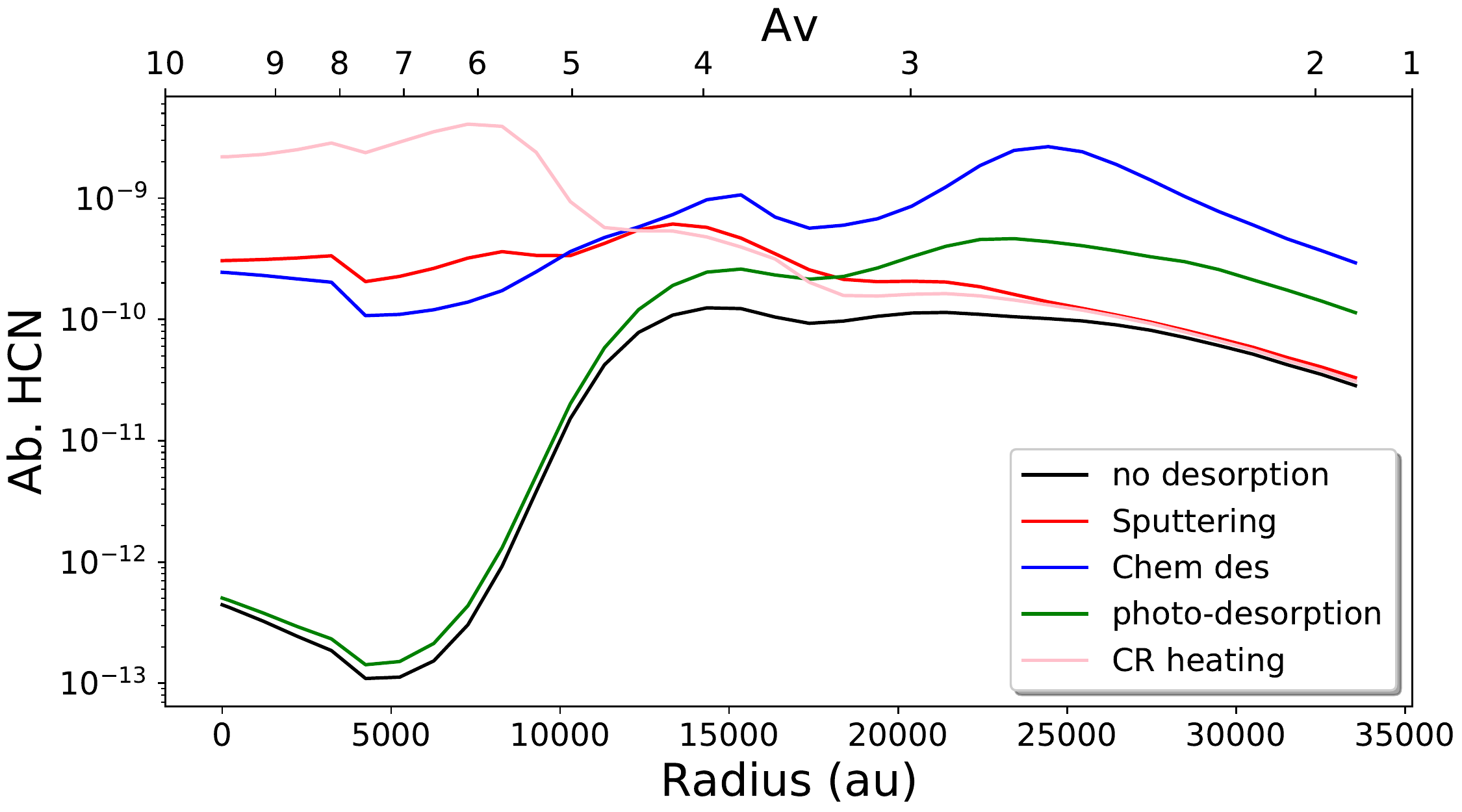}
\includegraphics[width=0.46\linewidth]{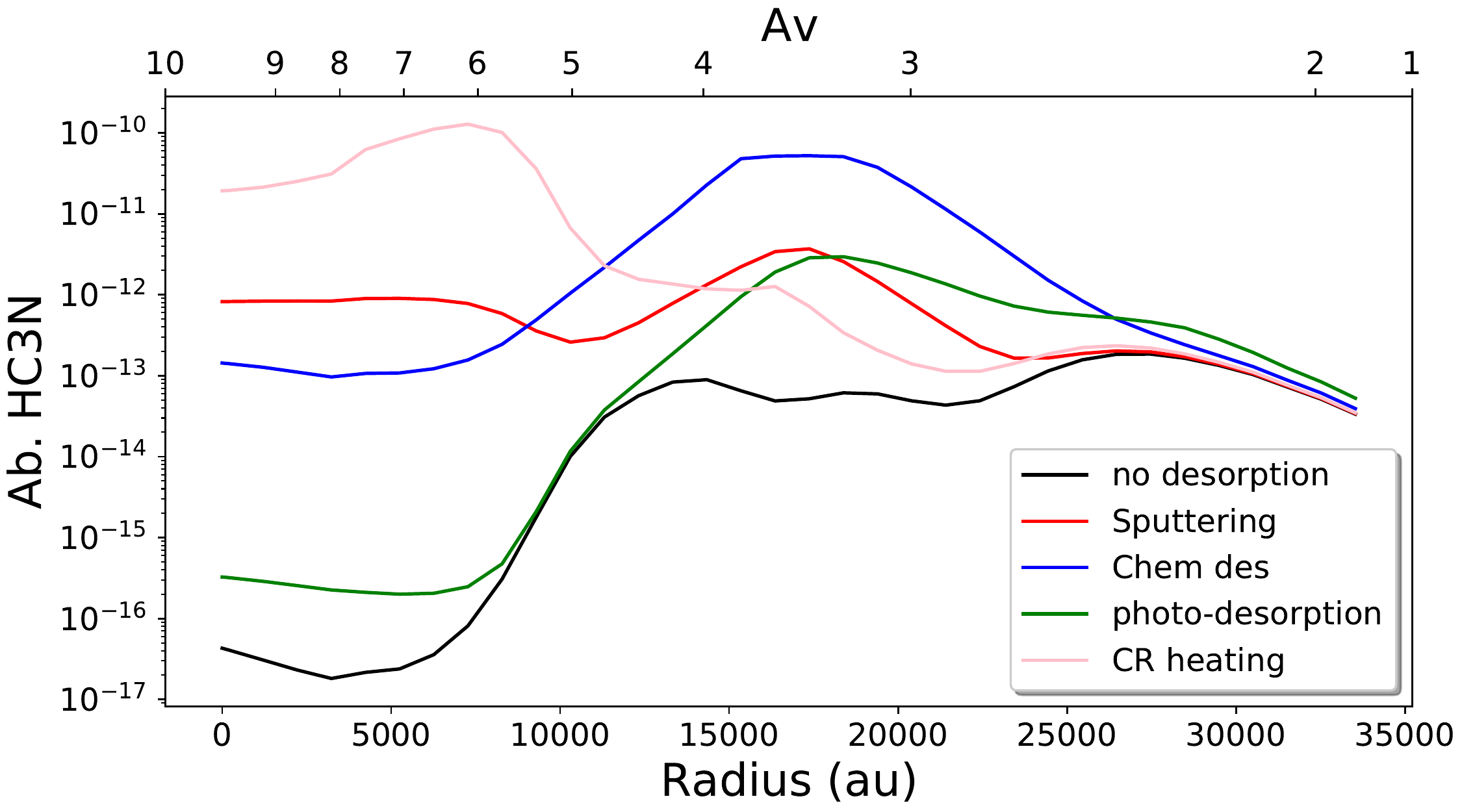}
\includegraphics[width=0.46\linewidth]{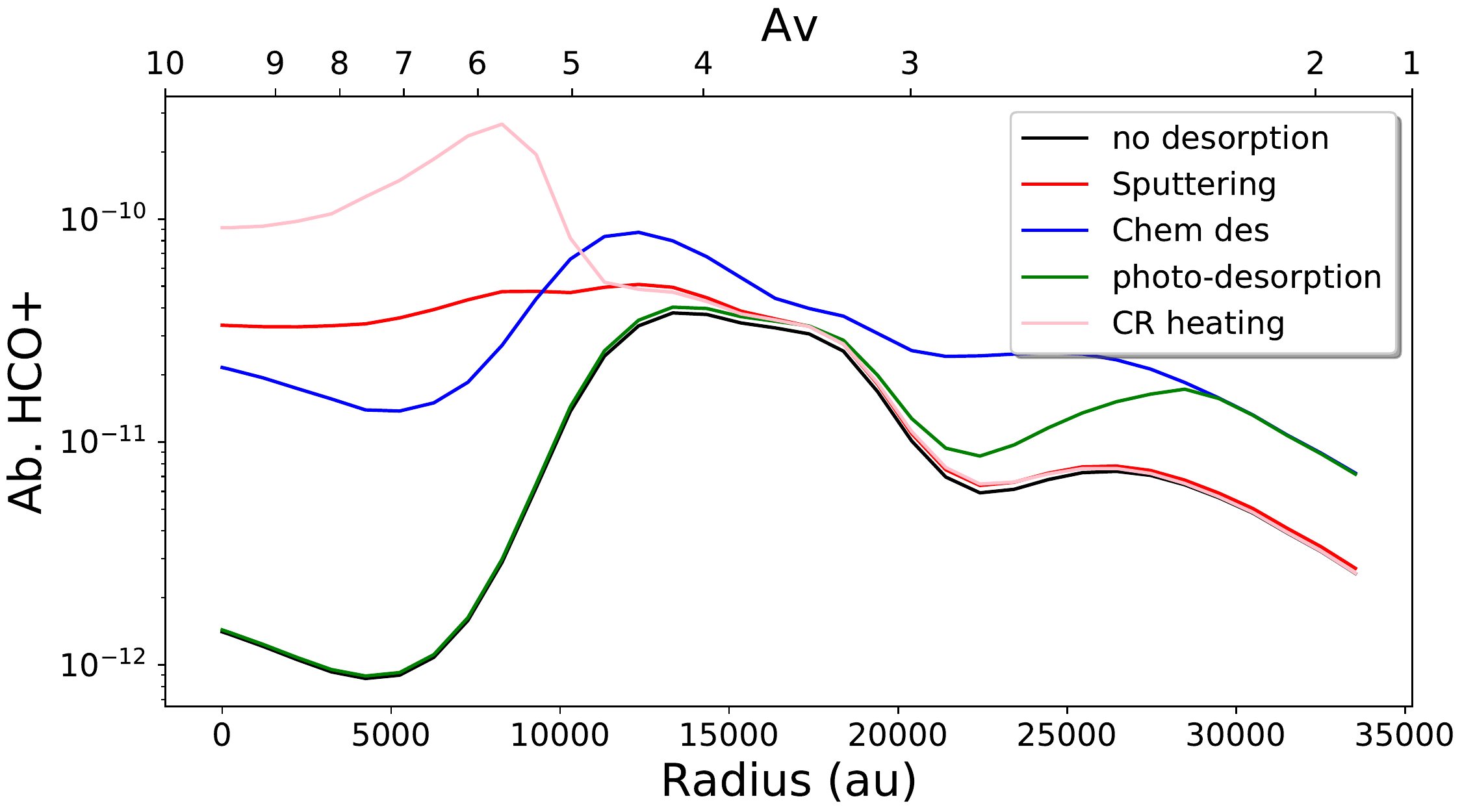}
\caption{Abundance of simple gas-phase molecules as a function of radius (and visual extinction) for a time of $6\times 10^5$~yr.}
\label{simplemol_fig}
\end{figure*}

In Fig.~\ref{simplemol_fig}, we show the model results for simple gas-phase molecules often observed in cold cores. Except for water, these molecules are not directly formed on the grains but we can see that their gas-phase abundances are strongly sensitive to the non-thermal desorption processes because: 1) some precursors can be desorbed from the grains; and 2) after being formed in the gas-phase, they are depleted on the grains and non thermal desorption brings them back into the gas-phase. \\
The CO gas-phase abundance is not strongly sensitive to the non-thermal desorptions. 
The other molecules (except for water) present a lower level of sensitivity in the outer parts, where the densities are smaller and, thus, the depletion is smaller. In the inner, denser regions, the larger gas-phase abundances are not produced by the same processes for all species. For CN, HCN, HC$_3$N, and HCO$^+$, cosmic-ray heating produces the largest abundances, then sputtering followed by chemical desorption and photo-desorption (the two last ones being equally efficient for CN). Here, 
%If is it difficult to identify the gas-phase pathways in these complex networks, 
CN, HCN (and HNC), and HC$_3$N molecules are chemically linked. The CN molecule is mostly formed by N + CH, then can react with H$_3^+$ to form HCNH$^+$. While recombining with electrons, HCNH$^+$ produces HCN (and HNC); CN can also react with C$_2$H$_2$ to form HC$_3$N. The main effect of CR heating is to desorb CH$_4$ from the ices. Removing this process leads to results similar to the photo-desorption. Once in the gas-phase, CH$_4$ participates to the ion-neutral chemistry by providing CH radicals. CH$_4$ for instance reacts with H$_3^+$ to form CH$_5^+$, which recombines with electrons to produce CH$_3$. CH$_3$ then reacts with atomic carbon to produce C$_2$H$_2$ involved in the formation of HC$_3$N. The increase of HCO$^+$ abundance is also due to the CH$_4$ CR heating desorption. One of the paths to form HCO$^+$ is through the reaction CO + CH$_5^+$. 

The gas-phase SO abundance is enhanced by both the chemical desorption and the sputtering in the inside. This molecule is formed by the neutral-neutral reaction S + OH and O + HS. In the case of chemical desorption, both the chemical desorption of HS and OH (during hydrogenation of S and O on the surfaces) are at the origin of the SO increase. We need to remove both of them to decrease the SO gas-phase abundance. In the case of sputtering, OH gas-phase abundance is highly increased. The CS gas-phase abundance is mostly enhanced by the chemical desorption and the CR heating. For the CR heating, it is again the desorption of CH$_4$ ice that is responsible for this increase. In the CR heating model, gas-phase CS is mostly produced by three reactions: HCS$^+$ + e$^-$, H + HCS, and H$_3$CS$^+$ + e$^-$. All three precursors, HCS, HCS$^+$, and H$_3$CS$^+$, are formed by reactions between neutral or ionized atomic sulfur with CH$_2$ or directly CH$_4$. Similarly to SO, it is the chemical desorption that increases most the gas-phase CS abundance and this is due to the larger HS abundance as it is formed by C + HS in the chemical desorption model.

\subsection{Complex organic molecules observed in cold cores}\label{COMs}

\begin{figure*}
\centering
\includegraphics[width=0.46\linewidth]{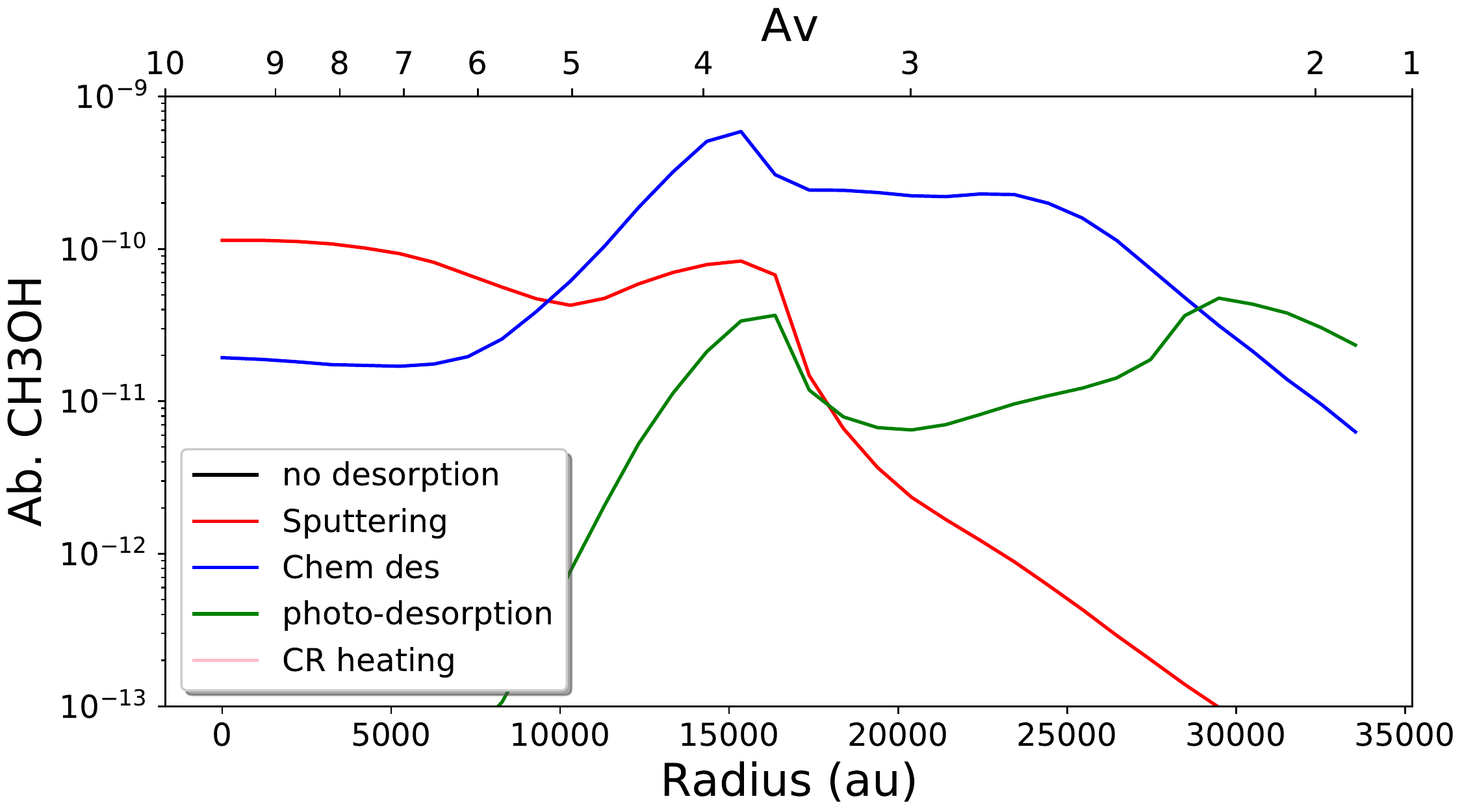}
\includegraphics[width=0.46\linewidth]{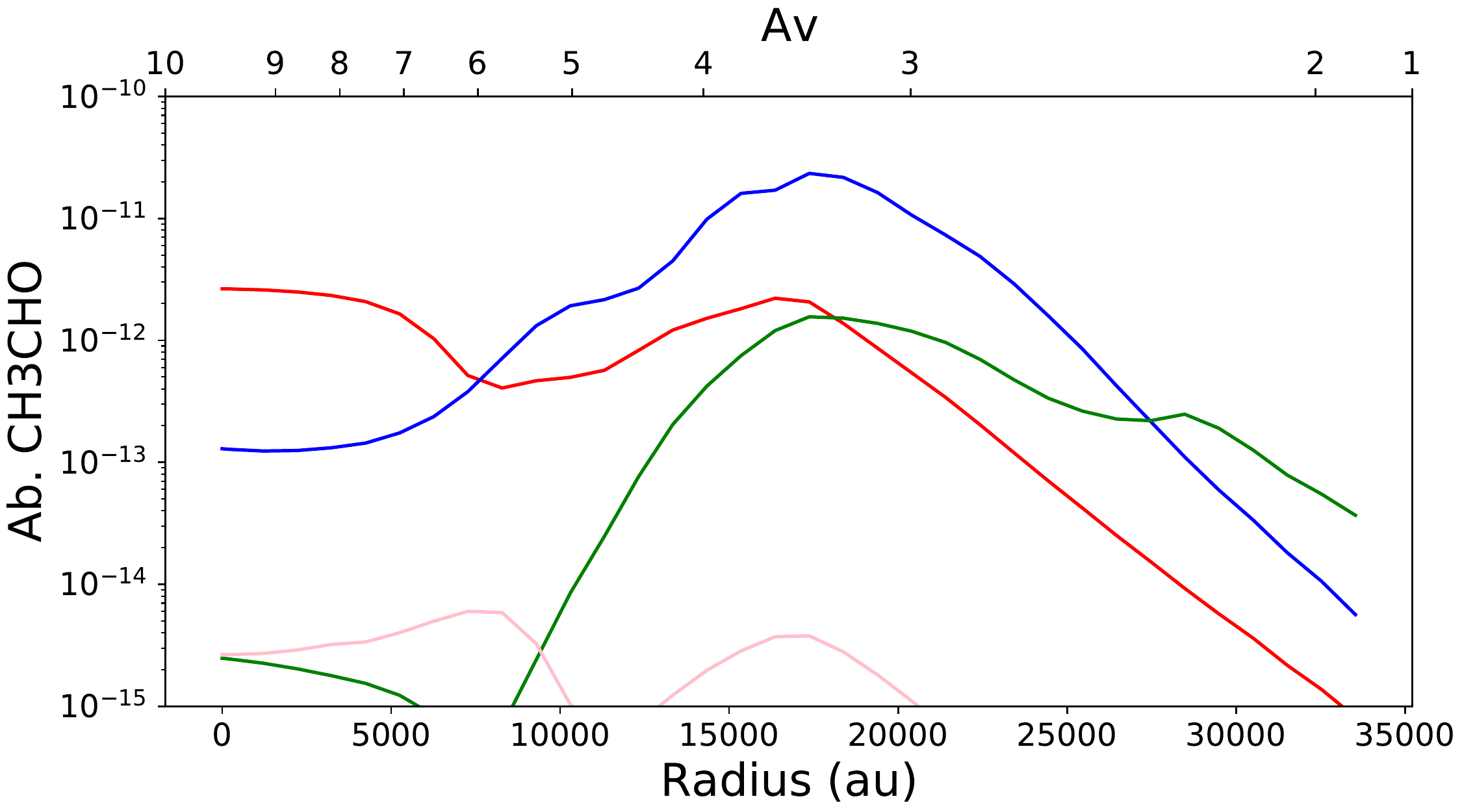}
\includegraphics[width=0.46\linewidth]{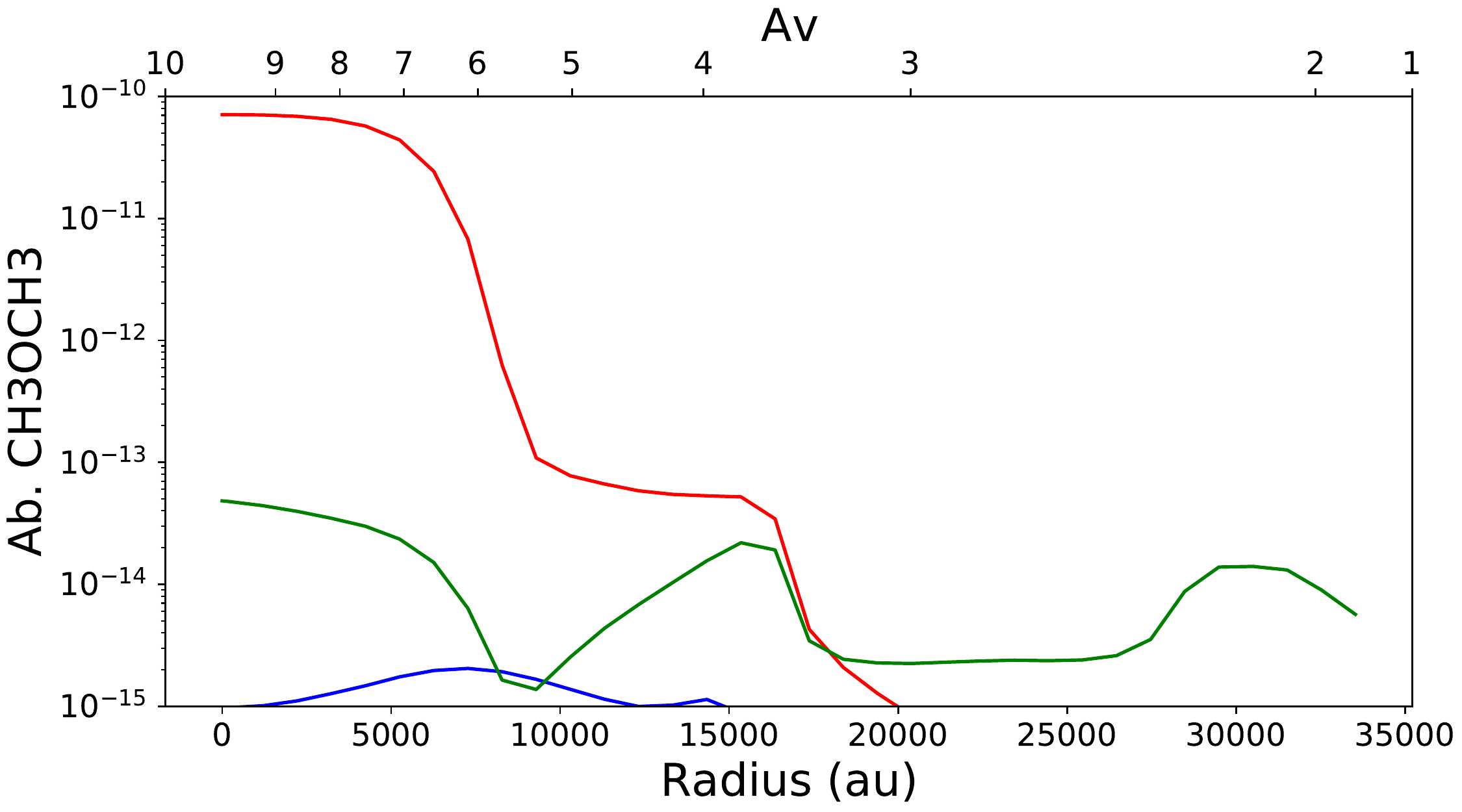}
\includegraphics[width=0.46\linewidth]{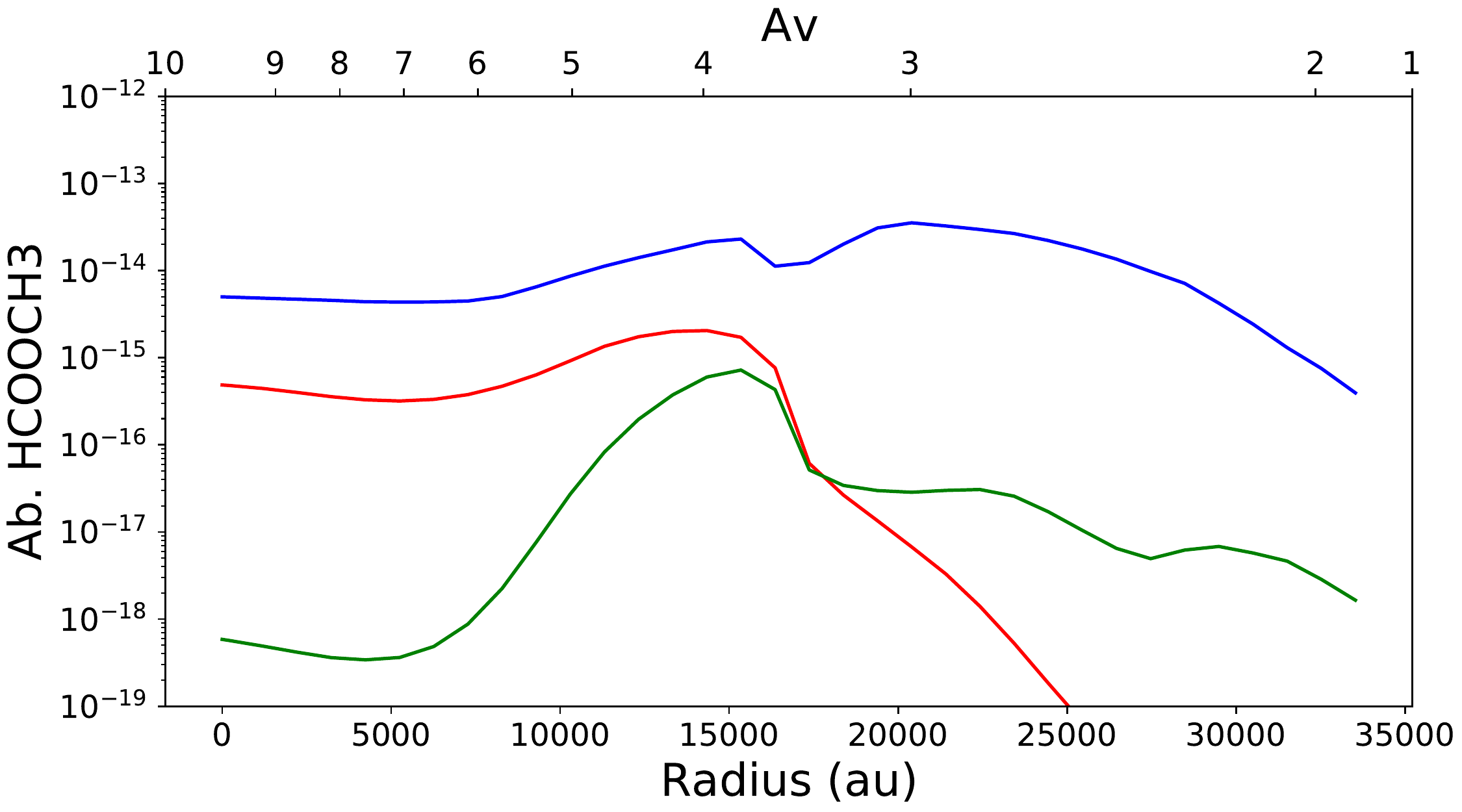}
\caption{Abundance of the complex organic molecules as a function of radius (and visual extinction) for a time of $6\times 10^5$~yr.\label{COM_fig}}
\end{figure*}

Figure~\ref{COM_fig} shows the abundance of CH$_3$OH, CH$_3$CHO, HCOOCH$_3$, and CH$_3$OCH$_3$ as a function of radius and for the same time as previously. Only three models are seen on the figure (except for CH$_3$CHO) because the two other models (i.e., no desorption and CR heating) produce abundances in the gas-phase that are very small. These molecules are produced on the grains, or from precursors formed on the grains, and only efficiently evaporated by sputtering, chemical or photo-desorption. Models with no desorption or only heating by CRs produces negligible gas-phase abundances of these species. \\
The model with chemical desorption is the one producing the largest gas-phase abundance of HCOOCH$_3$  at all radii (although smaller than $10^{-13}$).  
In our model, HCOOCH$_3$ is formed in the gas-phase by the reaction  CH$_3$OCH$_2$ + O $\rightarrow$ HCOOCH$_3$ + H. CH$_3$OCH$_2$ is formed on the grain surfaces by the reaction s-C...CH$_3$OH + s-H followed by chemical desorption of the produced species CH$_3$OCH$_2$. s-C...CH$_3$OH is a Van Der Waals complex included in Nautilus by \citet{2015MNRAS.447.4004R}. \\ 

 The gas-phase abundances of CH$_3$OH and CH$_3$CHO are larger with the model including sputtering at high density while chemical desorption produces the largest abundances at radii larger than 5000 au. Finally, photo-desorption is the least efficient of the three models at high density but becomes much more efficient than sputtering at radii larger than 15000 au and more efficient than chemical desorption at radii larger than 27000 au. Methanol, CH$_3$OH, is formed on the surfaces by successive hydrogenation of CO. The efficiency of the sputtering and photo-desorption are proportional to the abundance of CH$_3$OH on the grains whereas chemical desorption depends on the abundance of the reactants. We note that in our model, the photo-desorption of methanol is not destructive while experiments by \citet{2016ApJ...817L..12B} showed that is should be partly destructive. In addition, sputtering of the entire mantle along with the surface is allowed, whereas chemical desorption and photo-desorption are only possible for species on the surfaces. Although the cosmic-ray ionization rate decreases inside the cloud, the flux of production of CH$_3$OH is about ten times larger in the sputtering model at the inner point than with the chemical desorption model. Photo-desorption is more efficient at low visual extinction (i.e., in the outer part of the cloud). 
%{\bf Note that our model includes the back  \\
Although CH$_3$CHO shows a similar sensitivity to the different desorptions, its formation path is different. In the chemical desorption model (at all radii), it is the gas-phase reaction O + C$_2$H$_5$ that produces gas-phase CH$_3$CHO and C$_2$H$_5$ is produced mostly by H + C$_3$H$_7$, while C$_3$H$_7$ is desorbed at the end of a long hydrogenation chain of surface reaction of C$_3$. In the sputtering model, the O + C$_2$H$_5$ reaction plays also a role (C$_2$H$_5$ being evaporated by sputtering from the surfaces), especially in the outer parts, but the large increase of gas-phase CH$_3$CHO inside is due to the dissociative recombination of CH$_3$CHOH$^+$, itself produced by the reaction CH$_3$OCH$_3$ + H$^+$. The CH$_3$OCH$_3$ abundance at the highest density is indeed quite large, as seen in Fig.~\ref{COM_fig} for this model. \\

For CH$_3$OCH$_3$, it is the chemical desorption that is the least efficient whatever the radius while sputtering is the most efficient for radii inside 22000 au and photo-desorption outer this radius. CH$_3$OCH$_3$ is formed on the surfaces via reactions such as s-H + s-CH$_3$OCH$_2$ and s-CH$_3$ + s-CH$_3$O. The surface abundance of this species is similar in all the models. The binding energy of the species results in a very small fraction of chemical desorption while the other mechanisms are more efficient. \\

 The main production reactions discussed here were found by looking at the fluxes of the reactions producing these species. While doing various tests, we found that these COMs were particularly sensitive to the chemistry of Van Der Waals complexes from  \citet{2015MNRAS.447.4004R} that we introduced into the network. By removing these processes, we obtain much fewer of these species both in the gas-phase and on the grains. This is particularly true for CH$_3$OCH$_3$, whose surface precursor CH$_3$OCH$_2$ is formed by s-C...CH$_3$OH + s-H $\rightarrow$ s-CH$_3$OCH$_2$. In addition, s-CH$_3$CHO is also impacted because one of the channels producing this species on the surface involves C + s-CO $\rightarrow$ s-CCO and C + s-H$_2$CO $\rightarrow$ s-H$_2$CCO. The species presented in the two previous (Sections~\ref{main_ice} and \ref{simple_species}) are not sensitive to this chemistry. We note that our model predictions for COMs cannot be compared to \citet{2020ApJS..249...26J} because we have used a different physical model, which is much less dense than theirs. Their chemical model includes new non-diffusive surface mechanisms to enhance the production of COMs and does not include the chemistry of Van Der Waals complexes from  \citet{2015MNRAS.447.4004R}. They found that the chemical desorption was the main efficient process to desorb COMs from the grains, mostly because of the H-abstraction from grain-surface COMs, followed by recombination, amplifying the chemical desorption. We include these reactions as well and we find that sputtering could play a major role at the highest densities of our model.

\section{Observational constraints: Case of methanol}\label{obs_methanol}

\subsection{Considering abundances}

\begin{figure*}
\centering
\includegraphics[width=0.46\linewidth]{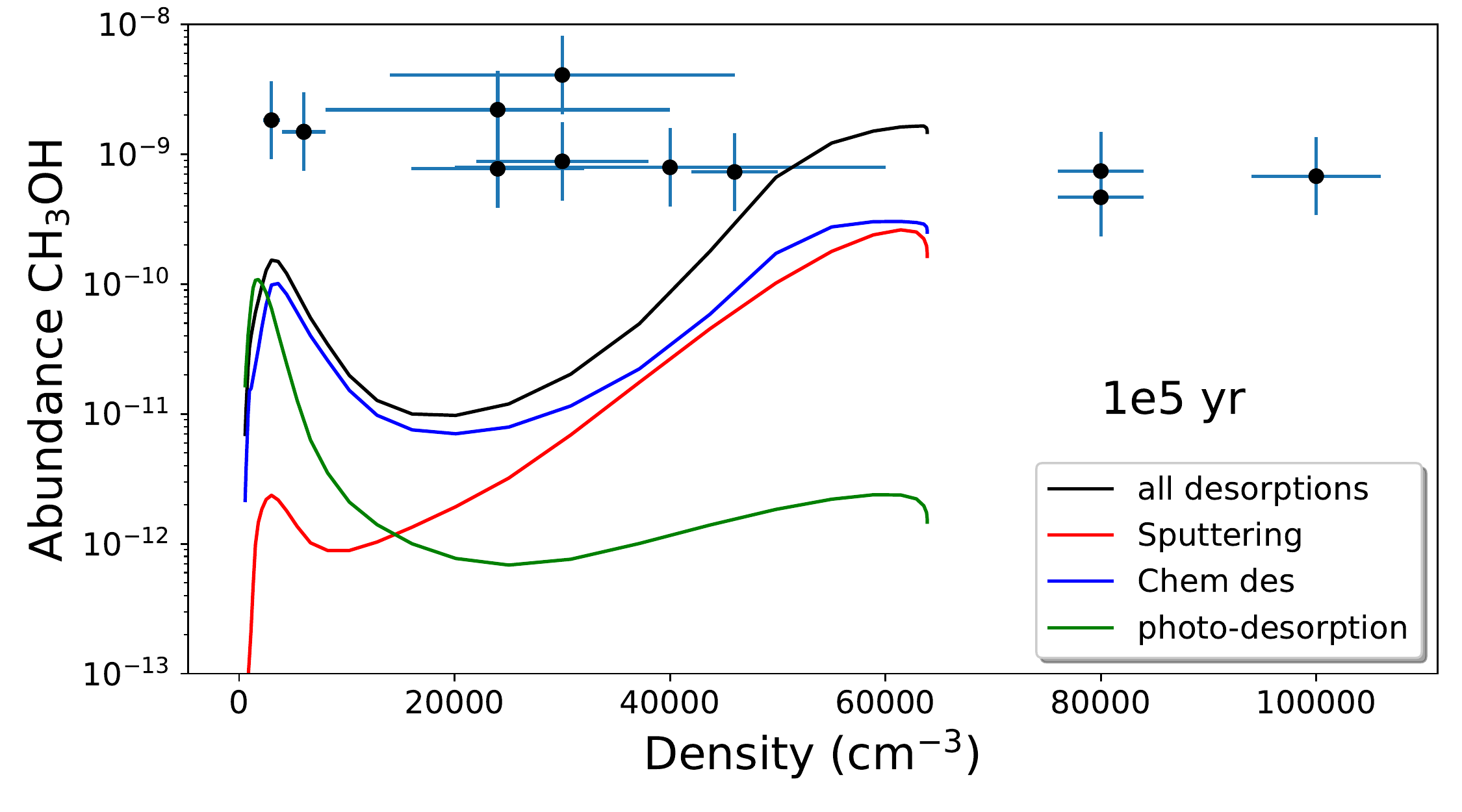}
\includegraphics[width=0.46\linewidth]{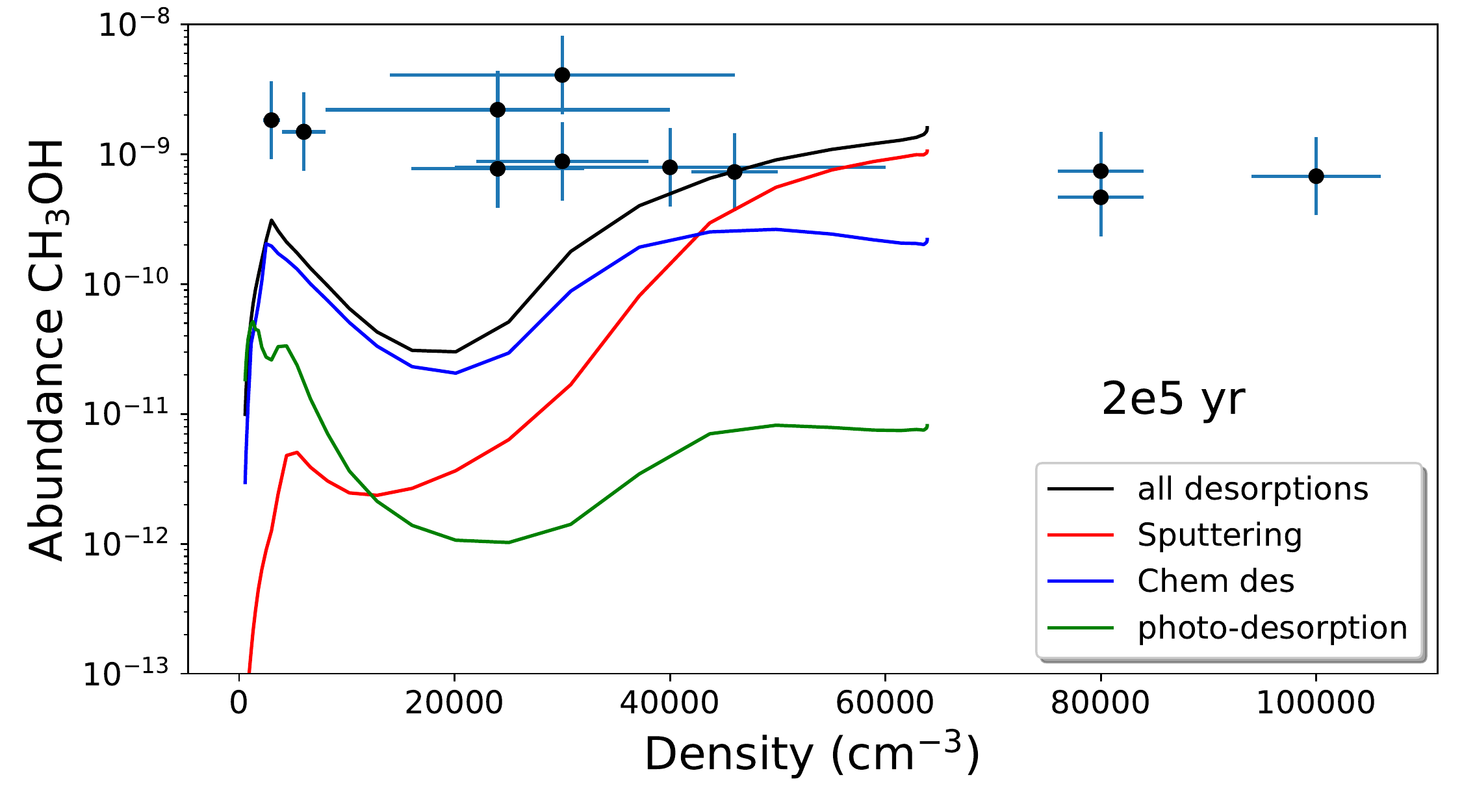}
\includegraphics[width=0.46\linewidth]{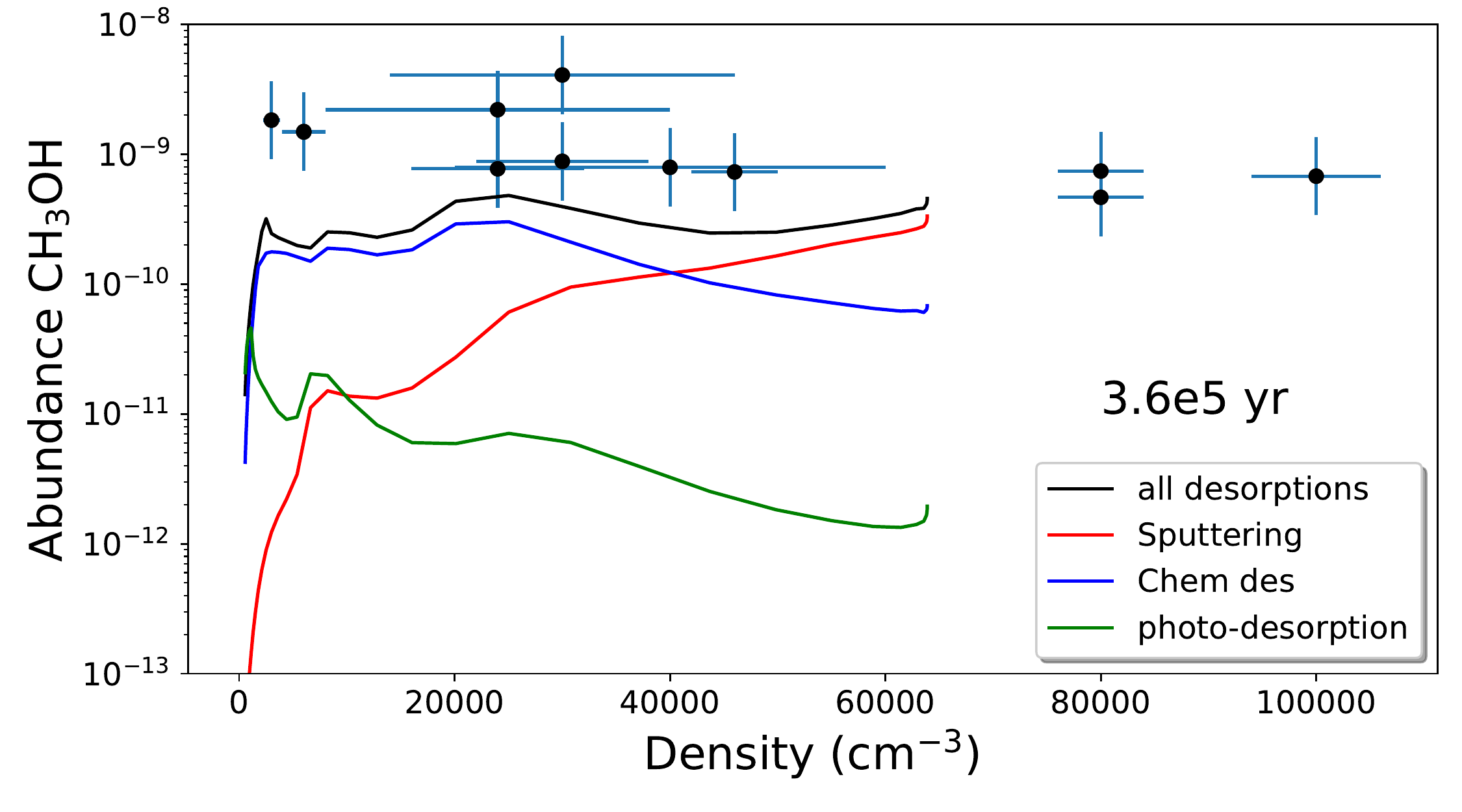}
\includegraphics[width=0.46\linewidth]{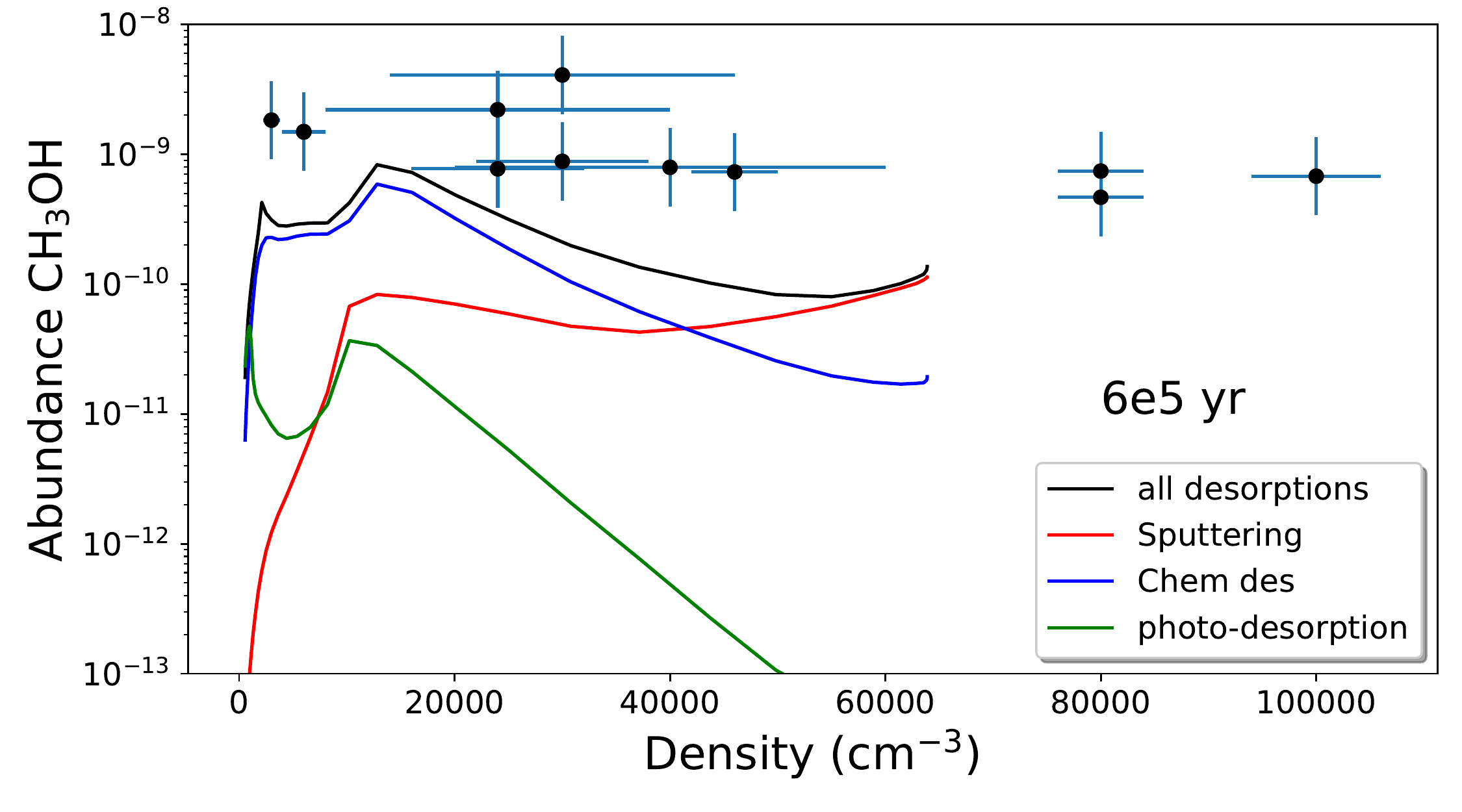}
\includegraphics[width=0.46\linewidth]{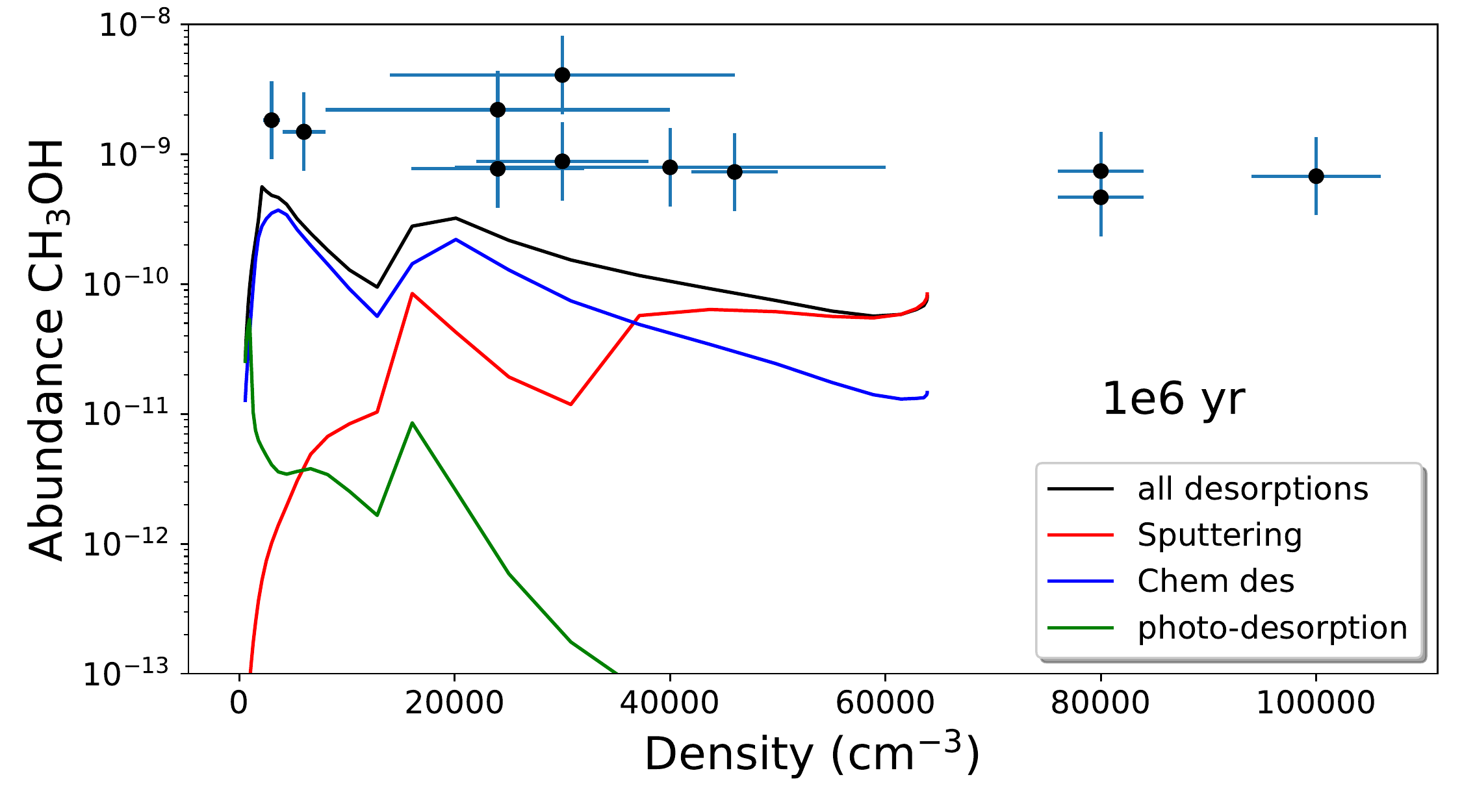}
\caption{Abundance of gas-phase methanol as a function of H density (in cm$^{-3}$). The different graphs represent different times. The lines in each box represent the models in which one of the non-thermal desorption process has been added or all of them. The points are the observed abundances, as described in the text.}
\label{CH3OH_allprocess}
\end{figure*}

Methanol is an interesting species as its formation is almost entirely on the grains as can be seen by the very small gas-phase abundance obtained with the model without any non-thermal desorption (Fig.~\ref{COM_fig}). The different non-thermal desorption processes studied in the previous section are potentially efficient at the same time with respective efficiencies although for methanol sputtering, chemical desorption, and maybe photo-desorption dominate. We have run an additional model in which we included all non-thermal desorption mechanisms with the same 1D physical structure. In Fig.~\ref{CH3OH_allprocess}, we show the gas-phase abundance of methanol computed by the model for different times (between $10^5$ and $10^6$~yr) as a function of density (rather than radius as shown in the previous section). We show the model results when each of the important non-thermal desorption mechanism is added or all of them. In the same figure, we also plot observed methanol abundances. After $10^6$~yr, the gas-phase CH$_3$OH abundance evolves less quickly.

The observed methanol abundances have been derived from observations of four rotational transitions of methanol toward TMC-1C, TMC-1(CP), TMC-1(NH3) at 3mm as a part of the "Gas phase Elemental abundances in Molecular Clouds" (GEMS) IRAM 30m large program \citep{2019A&A...624A.105F}. The observed lines have been modeled within the software CASSIS\footnote{http://cassis.irap.omp.eu}  \citep[developed by IRAP-UPS/CNRS,][]{2015sf2a.conf..313V} with the RADEX\footnote{http://www.strw.leidenuniv.nl/$\sim$moldata/radex.html} non-LTE radiative transfer code \citep{2007A&A...468..627V}, using the Markov Chain Monte Carlo method (MCMC), more details will be presented in an upcoming paper (Spezzano et al. in prep). Observed column densities at each cloud position, with the H$_2$ column densities, are given in Table~\ref{obs_CH3OH}.

\begin{figure*}
\centering
\includegraphics[width=0.46\linewidth]{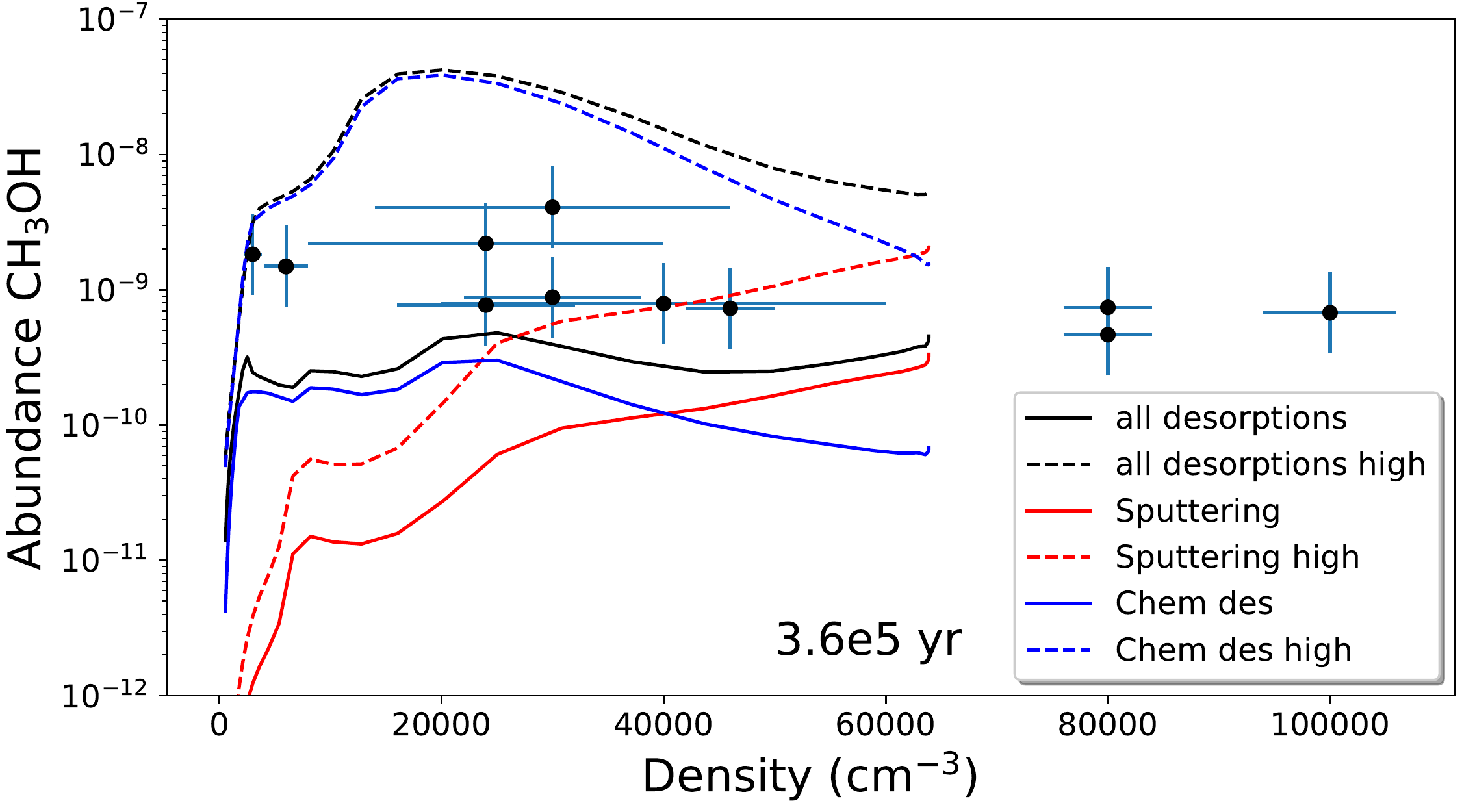}
\includegraphics[width=0.46\linewidth]{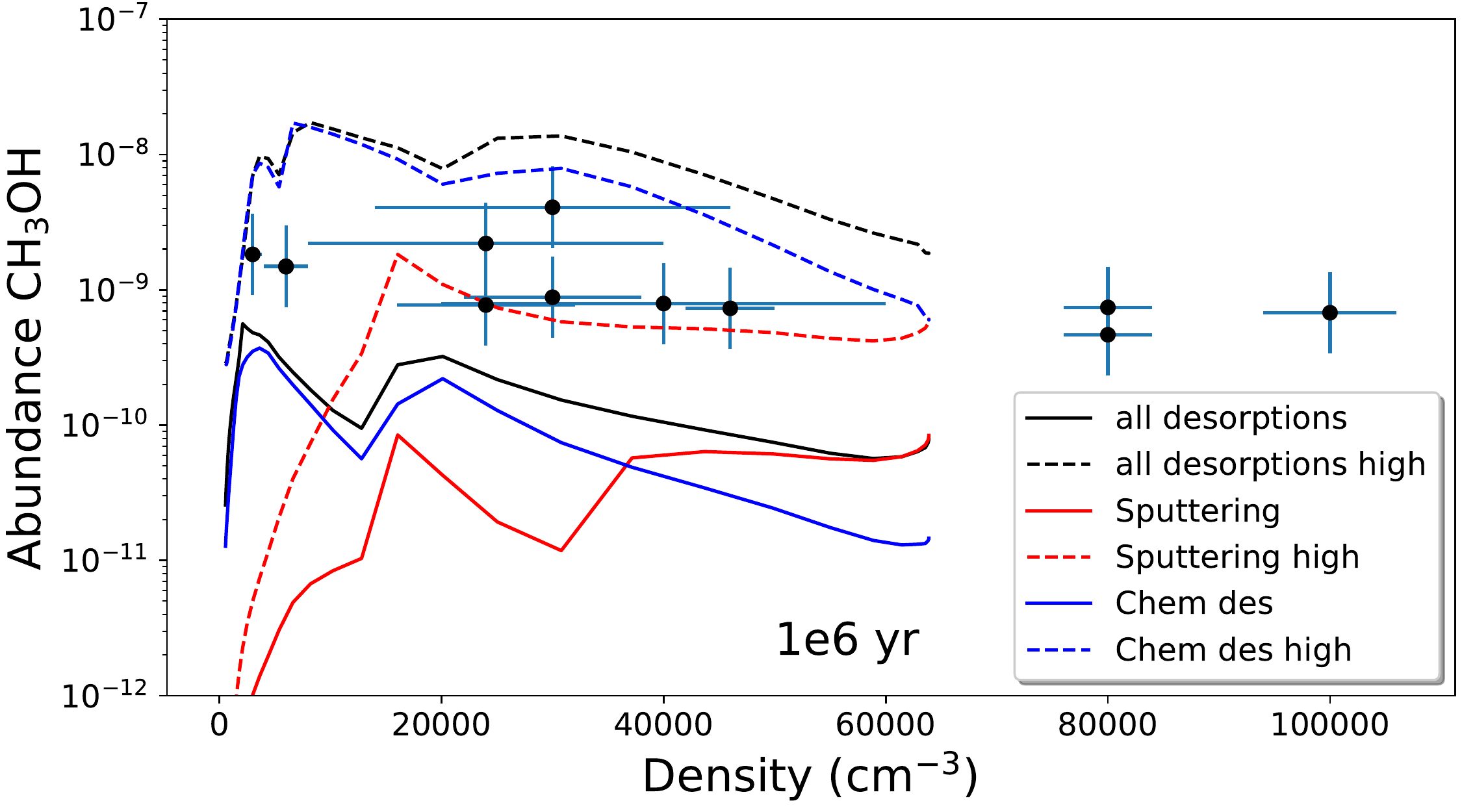}
\caption{Abundance of gas-phase methanol as a function of H density (in cm$^{-3}$), for two different times (left $3.6\times 10^5$~yr and right $1\times 10^6$~yr). The points are the observed abundances with error bars.}
\label{CH3OH_allprocess_alternate}
\end{figure*}

The modeled gas-phase methanol is high at high density at $10^5$~yr and decreases with time. At low density, on the contrary, the abundance is small and increases with time. The observations show a rather flat abundance of methanol with density, with a small increase toward the small densities. With respect to the observations, the model with all desorption mechanisms and times between $3.6\times 10^5$ and $10^6$~yr best reproduces  the observations. Around $3.6\times 10^5$~yr, the observations are best reproduced at high density thanks to the cosmic-ray sputtering while at smaller densities, the observations are best reproduced for later times thanks to the chemical desorption. At $3.6\times 10^5$~yr, the modeled and observed abundances are in agreement within a factor of 10. \\
An underproduction of gas-phase methanol at small densities ( $< 2\times 10^4$~cm$^{-3}$)  could indicate a more efficient chemical desorption under these conditions as it would be if the water ice coverage of the grains were small. Experiments from \citet{2016A&A...585A..24M} showed a much more efficient chemical desorption for bare grains and the efficiency would decrease as the grains are covered with water. \citet{2020A&A...637A..39N} found that the observed H$_2$S gas-phase abundance in the same regions with densities smaller than $2\times 10^4$~cm$^{-3}$ could only be reproduced by the chemical model assuming the high values of chemical desorption for bare grains. They suggested that this density would represent a change in the chemical composition of the surface of the grains. Our results on methanol goes in the same direction. 
%Using the current prescription of \citet{2016A&A...585A..24M} for water ice surfaces, the fraction of produced CH$_3$OH that desorb in the gas-phase while CH$_3$OH is formed in the surface (adding the two production channels sH + sCH$_2$OH and sH + sCH$_3$O) is 0.06\%. 
Using the current prescription of \citet{2016A&A...585A..24M} for water ice surfaces, the fraction of produced CH$_3$OH that desorb in the gas-phase is 0.06\%. We changed this fraction in the model and found that the observations at low density can be reproduced with an efficiency ten times larger (0.6\%). The experiments of
\citet{2016A&A...585A..24M}  could provide upper limits on the CH$_3$OH chemical desorption on water ice or bare surfaces of 8\%, which is much greater than what we would need. %Using their prescription for bare grains (with an effective mass of 120 amu), we obtain an efficiency of desorption of 0.7. This value depends on the binding energy of the product CH$_3$OH. \citet{2016A&A...585A..24M} assumed a smaller binding energy than us so their theoretical desorption efficiency is 2.3\%, which would be too high to reproduce our observations. \\
Alternatively, laboratory experiments on CR sputtering have shown to be more efficient with CO or CO$_2$ ices, rather than pure water ices, and our model indeed predict an ice CO$_2$ abundance as high as water and even larger under some conditions. \\
In summary, to reproduce the flat CH$_3$OH abundance as a function of density for a single time, we need to change the efficiency of the chemical desorption or the CR sputtering with the radius, which would be consistent with a change in the ice composition. There are too many uncertain parameters to be able to constrain quantitatively the efficiency of these processes this way but to illustrate the model sensitivity, we have run three additional models here representing extreme cases. The first one uses Minissale's bare grain prescription for the chemical desorption and the second model uses the sputtering parameters for CO$_2$ ices \citep[$Y^{\infty} = 21.9$, $\beta = 56.3$, and $\gamma = 0.6$,][]{2021A&A...647A.177D}. For the third model, we have considered all processes and the prescriptions for bare grains and CO$_2$ ices. These three models are shown in Fig.~\ref{CH3OH_allprocess_alternate} for two different times ($3.6\times 10^5$~yr and $1\times 10^6$~yr) and noted as "chem des high", "sputtering high", and "all desorptions high". In the same figure, we also report the previous models. The sputtering from an ice mostly composed of CO$_2$ is much more efficient than from water ice producing at high density a CH$_3$OH gas-phase abundance almost ten times larger. The enhanced chemical desorption produces larger CH$_3$OH gas-phase abundance than the enhanced sputtering at all densities. At a small density ($<2\times 10^4$~cm$^{-3}$), a high efficient chemical desorption, as would be expected to occur on bare grains, seems necessary to reproduce the high observed gas-phase methanol abundance. At larger densities ($>2\times 10^4$~cm$^{-3}$), however, the enhanced chemical desorption seems overly efficient, while the enhanced sputtering appears to be a major asset. \\
The modeling results are strongly dependent on the physical conditions. The external UV field for instance was set to 5 in Draine units based on observational constraints. 
Decreasing this value increases the methanol abundance. The 1D physical model used here does not cover the points observed at the highest densities (n$_{\rm H}$ = $8\times 10^4$~cm$^{-3}$ and higher). The fact that sputtering produces larger methanol abundance than chemical desorption occurs at higher density with these models than the less efficients ones (above n$_{\rm H}$ = $6\times 10^4$~cm$^{-3}$ instead of $4\times 10^4$~cm$^{-3}$). We checked that this tendency seen on the curves is confirmed by running higher density models.

\subsection{Considering column densities}

\begin{figure*}
\centering
\includegraphics[width=0.46\linewidth]{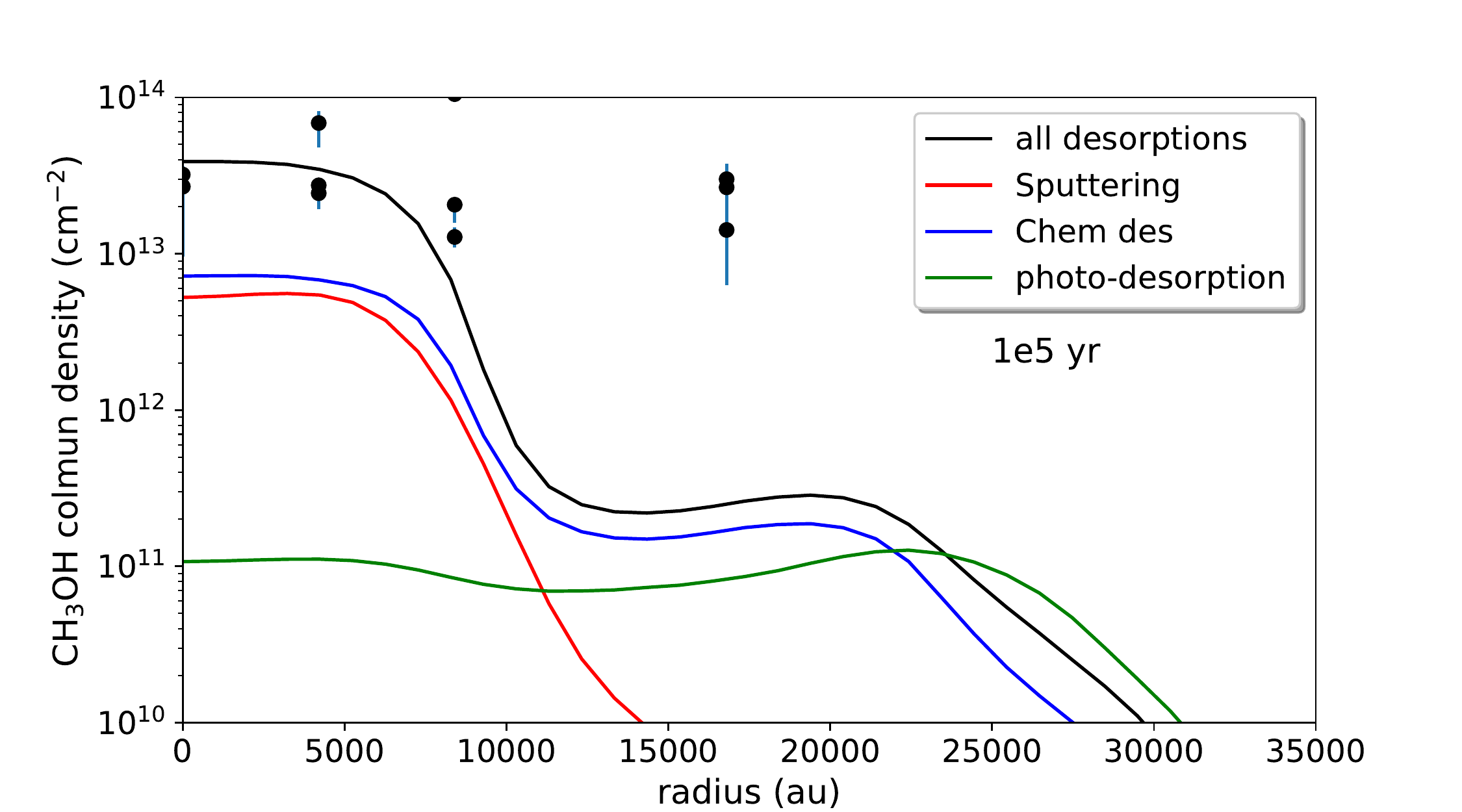}
\includegraphics[width=0.46\linewidth]{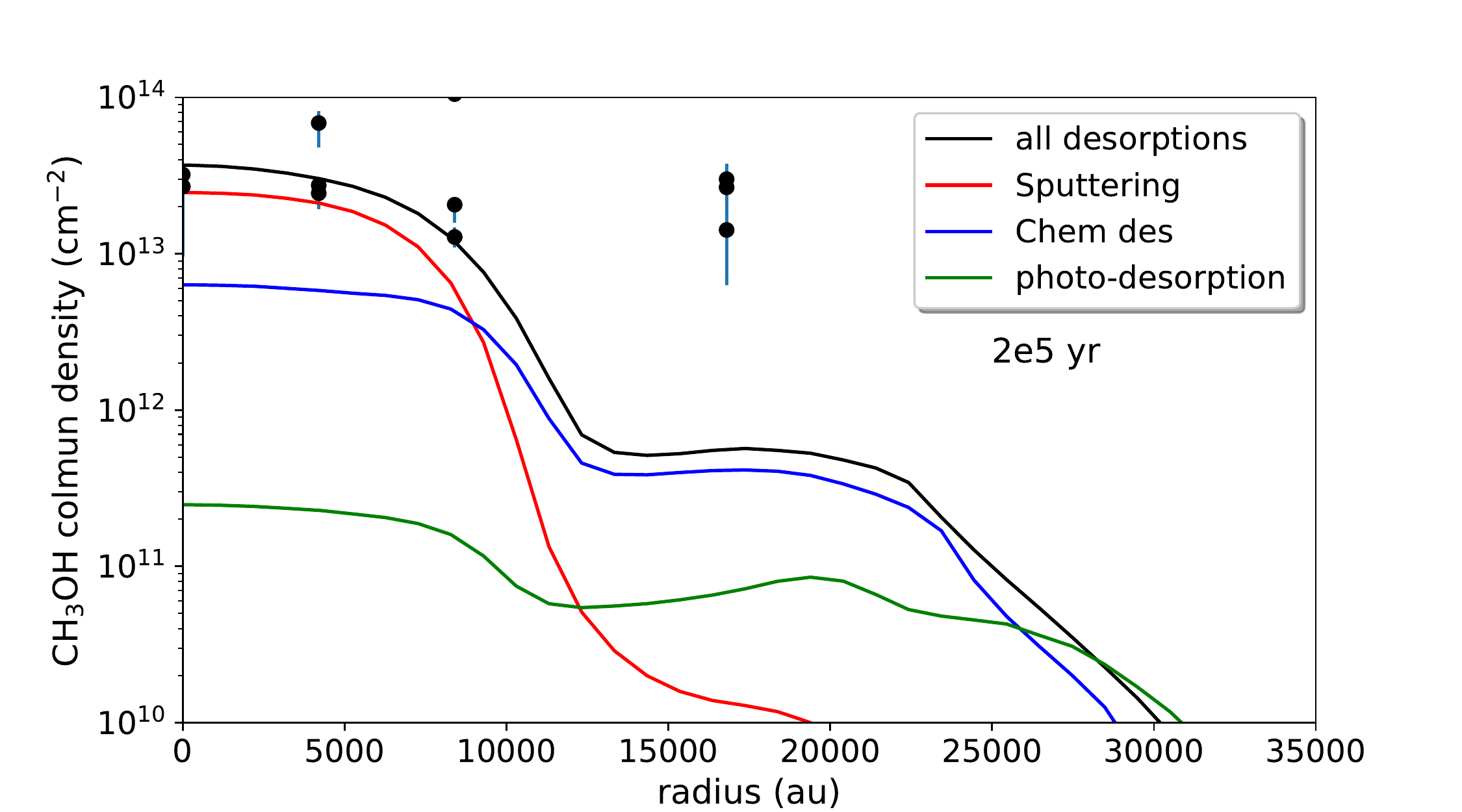}
\includegraphics[width=0.46\linewidth]{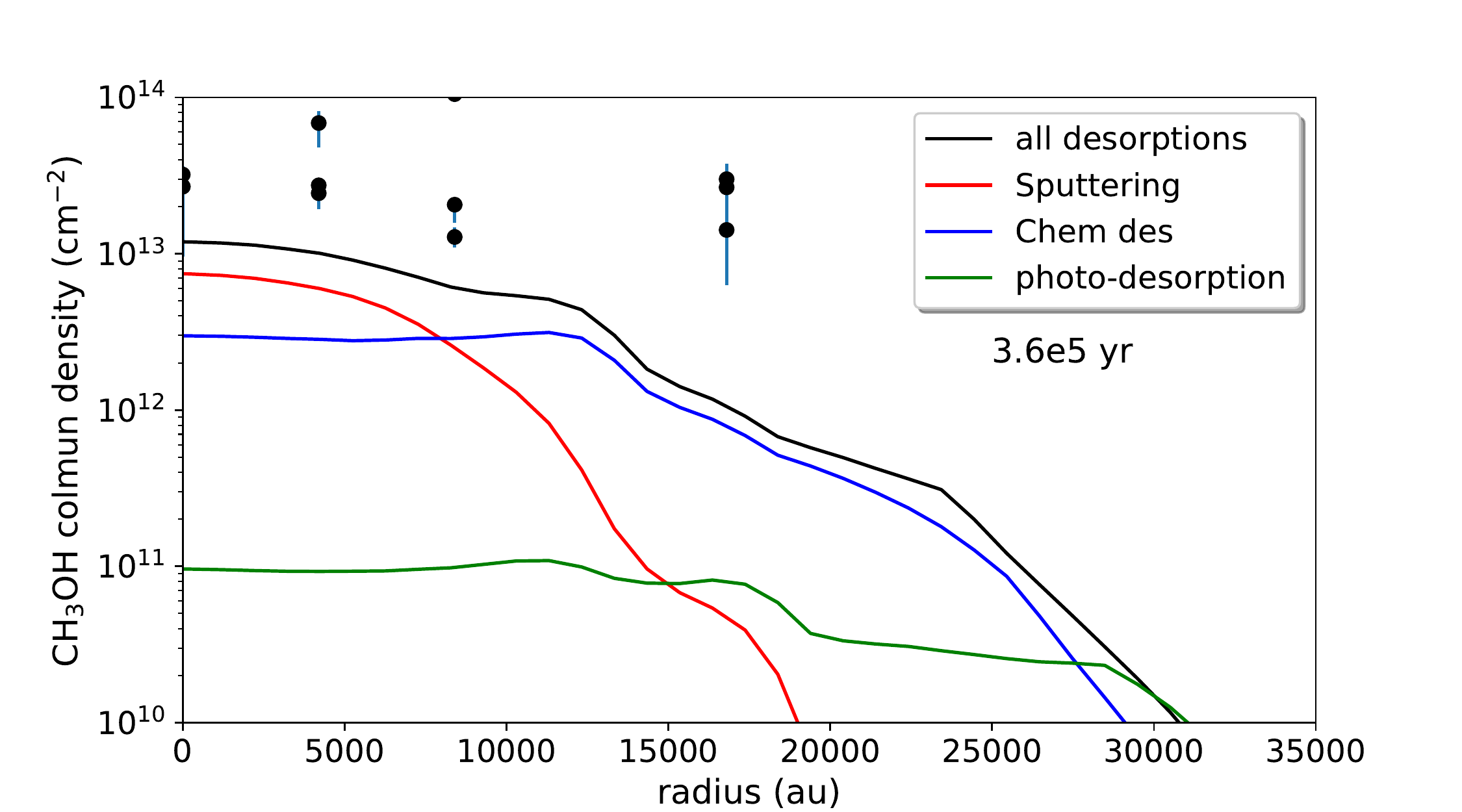}
\includegraphics[width=0.46\linewidth]{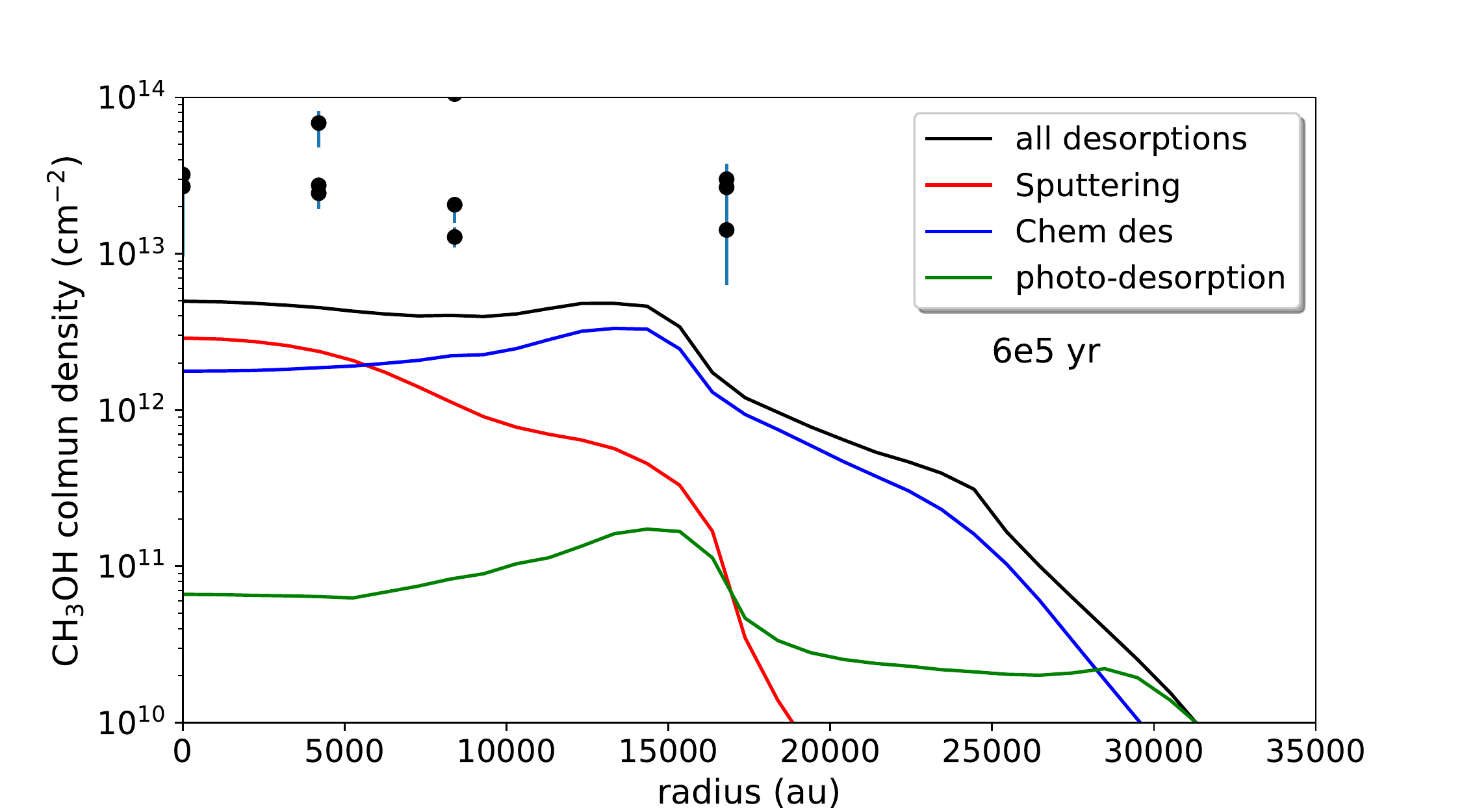}
\includegraphics[width=0.46\linewidth]{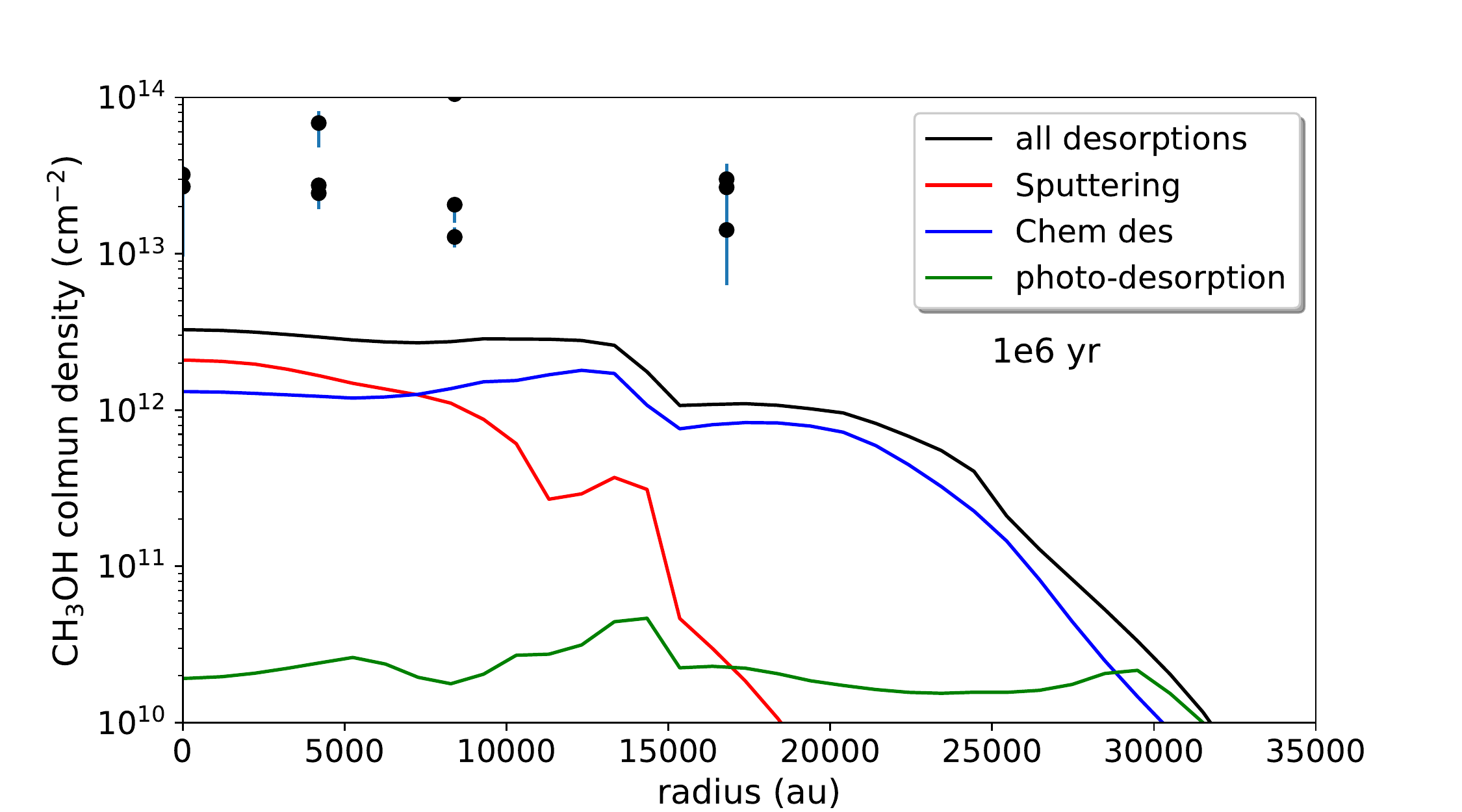}
\caption{Column densities of gas-phase methanol as a function of distance (in au) to the centre of the source. The different graphs represent different times. The lines in each box represent the models in which one of the non-thermal desorption processes has been added or all of them. The points are the observed column densities as described in the text and listed in Table~\ref{obs_CH3OH}.}
\label{coldens_CH3OH_allprocess}
\end{figure*}

In order to compare the model to the observations in a different way, we reconstructed the theoretical column densities predicted by our model. To do so, we assume a spherical geometry and integrated the column density along each line of sight \citep[see][for more details on the method]{Navarro2021}. We then obtained the column densities as a function of distance from the centre of the clouds shown in Fig~\ref{coldens_CH3OH_allprocess}. The observed radii have been computed for a source distance of 140 pc. We have four observed positions and three cores. 

The theoretical column densities decreases rapidly as we move away from the centre of the clouds. The conclusions on the most efficient processes as a function of radius (or density) discussed in the previous section stand here. Our model still underestimate the observed column densities at the larger radii (smaller densities).  At early times (1-2$\times 10^5$~yr), the inner column densities are better reproduced while the outer column densities (at 16800 au) are strongly underestimated. At later times, the inner predicted column density is smaller but the outer ones are larger. The predicted column density profile is then flatter. The overall observed column densities (which are rather constant across the clouds) are better reproduced for the times around $3.6 - 6\times 10^{5}$~yr. In that case, the predicted column densities are below the observed ones by less than a factor of 10.\\
From the observational point of view, the column densities are what is measured (with various assumptions for the radiative transfer analysis). The observed abundances used in the previous section were obtained by dividing these column densities with H$_2$ column densities computed from Herschel dust observations at the same positions and with a similar beam size. When computing the observed abundances with respect to H$_2$, the main hypothesis is that the methanol lines and the dust observations probe the same column of material. When we compared the observed and predicted abundances, we used the local densities measured at each position (and deduced from the molecular excitation). Since our model, at the positions of interests, has the 1D structure playing a minor role (the Av is high enough to prevent the direct UV photons to play a major role), the comparison does not depend on the assumed 1D structure, which may not represent the three sources altogether. When comparing the column densities, we are much more dependent on the assumed density profile.

 \section{Discussions}

\subsection{Effect of the initial atomic abundance}\label{initial_H}

Starting with no atomic initial abundance is a very common simplification of cold core chemical modeling. Molecular hydrogen is the first molecule to be formed in the interstellar medium. Its formation involves several micro-physical processes on top of interstellar grains, going from low energetic processes (sticking, diffusion and Langmuir-Hinshelwood mechanism) to more energetic ones (Eley-Rideal and "hot atom"). The efficiency of the H$_2$ formation depends strongly on the nature of the surfaces and their shape \citep[see][for a review of on the H$_2$ formation in the ISM]{2017MolAs...9....1W}. In astrochemical models of cold cores with large networks, namely, models that are not dedicated to the formation of H$_2$ itself \citep[such as][]{2001ApJ...553..595B,2010MNRAS.406L..11C,2011A&A...535A..27C,2014A&A...569A.100B}, only the Langmuir-Hinshelwood mechanism is considered for the formation of H$_2$, which may underestimate the formation efficiency of H$_2$ at moderate density. From an observational point of view, H$_2$ can only directly be observed in wavelengths (UV, near-IR, and mid-IR) and is not accessible in dense cold regions. In such regions, we can have some estimates of the H$_2$ abundance, and so the remaining atomic hydrogen, by measuring the residual atomic hydrogen fraction via HI Narrow Self- Absorption (HINSA) observations. Using such a method, \citet{2003ApJ...585..823L} and \citet{2005ApJ...622..938G} estimated the ratio between H and H$_2$ between $10^{-4}$ and a few $10^{-3}$. Using a H/H$_2$ ratio of $10^{-3}$ would lead to a H abundance of $5\times 10^{-4}$ with respect to the total proton density. We ran our models for various values of initial hydrogen ($5\times 10^{-4}$, $10^{-3}$, and $10^{-2}$). Initial atomic hydrogen abundances of $5\times 10^{-4}$ and $10^{-3}$ produce minor differences as compared to the results presented in the previous sections. Starting with a higher abundance of $10^{-2}$ (which may not be realistic) would strongly affect the methanol abundance (much less than the other species discussed in this paper) at early times and for densities larger than a few $10^3$~cm$^{-3}$. At a H density of $2\times 10^{4}$~cm$^{-3}$, the methanol gas-phase abundance could be increased by more than one order of magnitude with respect to the one computed by our "all desorption" model at $10^5$~yr. This difference fades with time and disappears at $10^6$~yr. For the present physical structure, an initial H/H$_2$ ratio different from one could help increase the modeled abundances for intermediate spatial points but only if the initial H abundance is high and only for early times (see Fig.~\ref{CH3OH_initialH}, to be compared to Fig.~\ref{CH3OH_allprocess}). 

 \subsection{Effect of the dust temperature}\label{sect_dust_t}

For the results presented here, we used the dust temperature obtained from Herschel observations \citep[following][]{2019A&A...624A.105F,2020A&A...637A..39N}. The obtained 1D dust profile is always above 10~K at all radii (see Fig.~\ref{1D_structure}) and above the gas temperature. The surface chemistry can be very dependent on this parameter. As such, our predicted abundance of ice CO is low while the CO$_2$ ice contains a large fraction of the oxygen (as well as water). This has consequences for the predicted ice and gas methanol. We redid all our simulations with a dust temperature equal to that of the gas, which mostly changes in the high-density regions ($\ge 2\times 10^4$~cm$^{-3}$). In Figs.~\ref{ice_fig_lowT} to \ref{COM_fig_lowT}, we show the abundance of the same species as in Section~\ref{model_results} (Figs.~\ref{ice_fig} to \ref{COM_fig}) computed with a dust temperature equal to the gas one at each radius of the 1D structure. There are a number of differences between the two sets of models at high density. As such, the following results are focused on these regions. Thus, we refer to the standard model as the one in which the dust temperature is determined by Herschel observations (and shown in the remainder of the paper). \\
The most obvious difference is the larger CO ice abundance, which propagates to the CH$_3$OH ice (and H$_2$CO ice not shown in the figures). As a consequence, the CO$_2$ ice is predicted much smaller going inward the cold core. The smaller grain temperature decreases the efficiency of the CO$_2$ ice formation (which possesses a barrier) and so more CO is available for H$_2$CO and CH$_3$OH. At $6\times 10^5$~yr, the inner CH$_3$OH ice abundance is two times larger with the smaller grain temperature (whatever the model presented on the figures as they all give the same abundance). Another consequence of the colder grains is a higher H$_2$O ice abundance by a factor of two, which is produced by a higher H$_2$ abundance on the grains (forming water by reacting with OH on the surface). The higher H$_2$ abundance in the ices is due to less efficient thermal desorption. Indeed, in our models, H$_2$ and He are both allowed to thermally desorb and considering their low binding energies, this process is efficient in our standard model, whereas it is negligible for H$_2$ in this lower dust temperature model. The CH$_4$ and NH$_3$ ices are not changed. \\
%CH$_4$ ice is also slightly more abundant than in the standard model because the higher H$_2$ ice abundance opens a new path to the formation of CH$_3$ on the surfaces through H$_2, ice$ + CH$_2, ice$ $\rightarrow$ H$_ice$ + CH$_3, ice$ and then CH$_3$ is hydrogenated. NH$_3, ice$ is not modified in the inner parts but rather decreased in this model for intermediate Av ($\sim$ 6). EXPLICATIONS?
The lower dust temperature produces some significant changes on the simple molecules commonly observed in the gas-phase (comparing Fig.~\ref{simplemol_fig} to Fig.~\ref{simplemol_fig_lowT}). If CO, CN, CS, and HC$_3$N do not show much difference, then SO, H$_2$O, HCN, and HCO$^+$ are strongly increased at high density in all models except for the CR heating model. The increase in SO, H$_2$O, and HCO$^+$ can be explained by a higher gas-phase abundance of OH and O$_2$. SO forms in the gas-phase by the reaction S + OH. H$_2$O in the gas is a product of the dissociative recombination of H$_3$O$^+$, itself formed by the reactions OH + H$^+$ $\rightarrow$ H + OH$^+$, H$_2$ + OH$^+$ $\rightarrow$ H + H$_2$O$^+$, and H$_2$ + H$_2$O$^+$ $\rightarrow$ H + H$_3$O$^+$. The higher OH gas-phase abundance is produced by more efficient production on the surface complex s-O...CO because of a higher grain coverage by CO. The reaction of  s-O...CO with s-H on the surface leads partly to desorbing OH, and interactions with UV photons leads to desorbing atomic oxygen \citep[see][]{2015MNRAS.447.4004R}. Overall, a higher grain coverage of CO produces a slight increase of the level of oxygen in the gas-phase, increasing the gas-phase abundances of SO, H$_2$O and HCO$^+$. The higher abundance of HCN, in the "photodesorption" and "no desorption models," is also due to a greater abundance of oxygen in the gas-phase but less directly. In our standard model, part of the HCN is destroyed by Si$^+$ (to form SiNC$^+$). In the colder dust model, the increase of OH abundance in the gas-phase reduces the Si$^+$ abundance by reacting with it to form SiO$^+$. To find these complex chemical effects, we have tested the model each time  by switching off and on the processes. \\
The methanol-increased abundance on the grains propagates to the gas-phase (comparing Fig.~\ref{COM_fig} to Fig.~\ref{COM_fig_lowT}). The other three COMs discussed in this paper (CH$_3$CHO, CH$_3$OCH$_3$, and HCOOCH$_3$) are not  significantly affected by the smaller grain temperature. The overall effects of the various non-thermal desorption processes previously discussed stand. The results presented in section~\ref{obs_methanol} are not significantly changed either. The methanol gas-phase abundance is just slightly more abundant at high density (see Fig.~\ref{CH3OH_allprocess_lowT}).\\

\section{Conclusions}

In this work, we used the full gas-grain model Nautilus and included a new non-thermal desorption process, which is the sputtering of ices by cosmic-ray particles, experimentally studied in the laboratory by \citet{2018A&A...618A.173D}. We tested its efficiency with respect to the non-thermal desorption mechanisms commonly included in astrochemical models: chemical desorption, grain heating by cosmic-rays, and photo-desorption. We also tested our model prediction with methanol observations in the TMC-1C, TMC-1(CP), and TMC-1(NH3) cold cores. We focused the discussions on the main ice components, simple molecules usually observed in cold cores (CO, CN, CS, SO, HCN, HC3N, and HCO$^+$), and complex organic molecules (COMs such as CH$_3$OH, CH$_3$CHO, CH$_3$OCH$_3$, and HCOOCH$_3$). Our conclusions are as follows:  
\begin{itemize}
\item [-] We found that all species are not sensitive in the same way to the non-thermal desorption mechanisms and the sensitivity also depends on the physical conditions. It is then mandatory to include all of them. Grains heating by cosmic rays appears to impact significantly some of the species formed in the gas-phase (such as CN, HCN, HC$_3$N, and HCO$^+$) through the desorption of CH$_4$ ices at high density. For molecules formed on the grains (such as H$_2$O, CH$_3$OH, CH$_3$OCH$_3$), the sputtering of the ices induced by cosmic-ray collision dominates the desorption at high density while photo-desorption dominates at low density (although resulting the gas-phase is much less high as the ice abundances are smaller). 
\item [-] A direct comparison of the gas-phase methanol in TMC-1C, TMC-1(CP), and TMC-1(NH3) cold cores with our model predictions shows that our models can reproduce the observations within a factor of 10 at $3.6\times 10^5$~yr for all densities. Chemical desorption seem essential to reproduce the observations for H densities smaller than $4\times 10^4$~cm$^{-3}$, while sputtering is essential above this density. 
\item [-] The models are, however, systematically below the observed abundances. Considering a more efficient chemical desorption, as measured on bare grains by \citet{2016A&A...585A..24M}, appears to head in the right direction for low densities but also seems a bit extreme. Using a higher efficiency for the sputtering, as measured in CO$_2$ ices, increases the methanol abundance in the gas at high density.
\item [-] Comparing CH$_3$OH observed and modeled column densities, rather than abundances, leads to similar conclusions except that the inner column densities are fully reproduced at early times (1-2$\times 10^5$~yr); however, in that case, the most external column densities are strongly underestimated by our models. The best agreement at all radii is obtained for times between 3.6$\times 10^5$ and 6$\times 10^5$~yr. In that case, the predicted column densities is flatter with the radius but below the observed column density by less than a factor of 10.
\item [-] Considering that a small fraction of the hydrogen is not in H$_2$ at the beginning of the chemical calculation only impacts the methanol abundance if this fraction is as high as H/H$_2$ = $2\times 10^{-2}$, which would be unrealistic. 
\item [-] In our simulations, we used the dust temperature observed with Herschel, which is a little above the gas temperature determined based on the molecular excitations. Setting the dust temperature equal to the gas one decreases the dust temperature in the inner regions of the cores by a few Kelvin. Such changes will affect the ice reservoir by producing less CO$_2$, increasing the CO and CH$_3$OH abundances. This higher CH$_3$OH ice abundance propagates in the gas-phase by increasing the gas-phase abundance by a factor of a few. The change in CO ice coverage slightly impacts  the oxygen quantity in the gas phase because, in our model, the atomic oxygen physisorbed on CO ice partly releases O and OH in the gas phase either by interaction with direct and secondary UV photons and hydrogenation.
\end{itemize}

In conclusion, the sputtering of ices by cosmic-ray collisions may be the most efficient desorption mechanism at high density (a few $10^4$~cm$^{-3}$ under the conditions studied here) in cold cores, while chemical desorption is still needed at smaller densities. Additional studies are needed on both chemical desorption and CR sputtering to assess their efficiency with respect to the main ice composition (especially in the presence of mixtures).

\section{Acknowledgements}

The authors acknowledge the CNRS program "Physique et Chimie du Milieu Interstellaire" (PCMI) co-funded by the Centre National d'Etudes Spatiales (CNES). DNA and AF thank the Spanish MICIU for funding support from AYA2016-75066-C2-2-P and PID2019-106235GB-I00.

%%%%%%%%%%%%%%%%%%%%%%%%%%%%%%%%%%%%%%%%%%%%%%%%%%



%\bibliographystyle{mnras}

\bibliographystyle{aa}
\bibliography{biblio}

%%%%%%%%%%%%%%%%%%%%%%%%%%%%%%%%%%%%%%%%%%%%%%%%%%

%%%%%%%%%%%%%%%%% APPENDICES %%%%%%%%%%%%%%%%%%%%%

\appendix

\section{CH$_3$OH and H$_2$ column densities observed at several positions of the cold cores TMC1-C, TMC1-CP, and TMC1-NH3}
\begin{table}
\caption{CH$_3$OH and H$_2$ column densities (in cm$^{-2}$) observed at several positions of the cold cores TMC1-C, TMC1-CP, and TMC1-NH3.  }
\begin{center}
\begin{tabular}{lllllll}
\hline
\hline
Offsets (") & \multicolumn{2}{c}{TMC1-C} &  \multicolumn{2}{c}{TMC1-CP} &  \multicolumn{2}{c}{TMC1-NH3}\\
& N$_{\rm CH3OH}$ & N$_{\rm H2}$ & N$_{\rm CH3OH}$ & N$_{\rm H2}$ & N$_{\rm CH3OH}$ & N$_{\rm H2}$ \\
\hline
 %& N$_{\rm CH3OH}$ (cm$^{-2}$) & N$_{\rm CH3OH}$ (cm$^{-2}$) & N$_{\rm CH3OH}$ (cm$^{-2}$) \\  
0 0 &   $(2.69^{+0.38}_{-0.30})\times 10^{13}$ & $19.8\times 10^{21}$ & $(3.21^{+0.7}_{-0.41})\times 10^{13}$ & $18.2\times 10^{21}$ & $(2.69^{+1.73}_{-0.53})\times 10^{13}$ & $16.9\times 10^{21}$ \\
+30 0 &  $(2.74^{+0.31}_{-0.27})\times 10^{13}$ & $18.5\times 10^{21}$ & $(2.44^{+0.51}_{-0.36})\times 10^{13}$ & $16.7\times 10^{21}$ & $(6.86^{+2.08}_{-1.35})\times 10^{13}$ & $15.6\times 10^{21}$\\
+60 0 &   $(2.06^{+0.48}_{-0.23})\times 10^{13}$ & $13.3\times 10^{21}$ & $(1.28^{+0.18}_{-0.19})\times 10^{13}$ &  $13.7\times 10^{21}$ & $(1.05^{+0.11}_{-0.15})\times 10^{14}$ &$12.9\times 10^{21}$ \\
+120 0 &  $(1.42^{+0.32}_{-0.48})\times 10^{13}$ & $4.8\times 10^{21}$  & $(2.66^{+2.03}_{-1.10})\times 10^{13}$ & $7.3\times 10^{21}$ & $(3.00^{+0.70}_{-0.43})\times 10^{13}$ & $10.0\times 10^{21}$  \\
\hline
\end{tabular}
The CH$_3$OH data will be published in Spezzano et al. (in prep) while the H$_2$ data are from \citet{2019A&A...624A.105F}. For some of the positions, Spezzano et al. found two velocity components. Here, we have summed the column densities of both components.
\end{center}
\label{obs_CH3OH}
\end{table}%

\section{Model results with an initial abundance of atomic hydrogen different from zero}

\begin{figure*}
\centering
\includegraphics[width=0.46\linewidth]{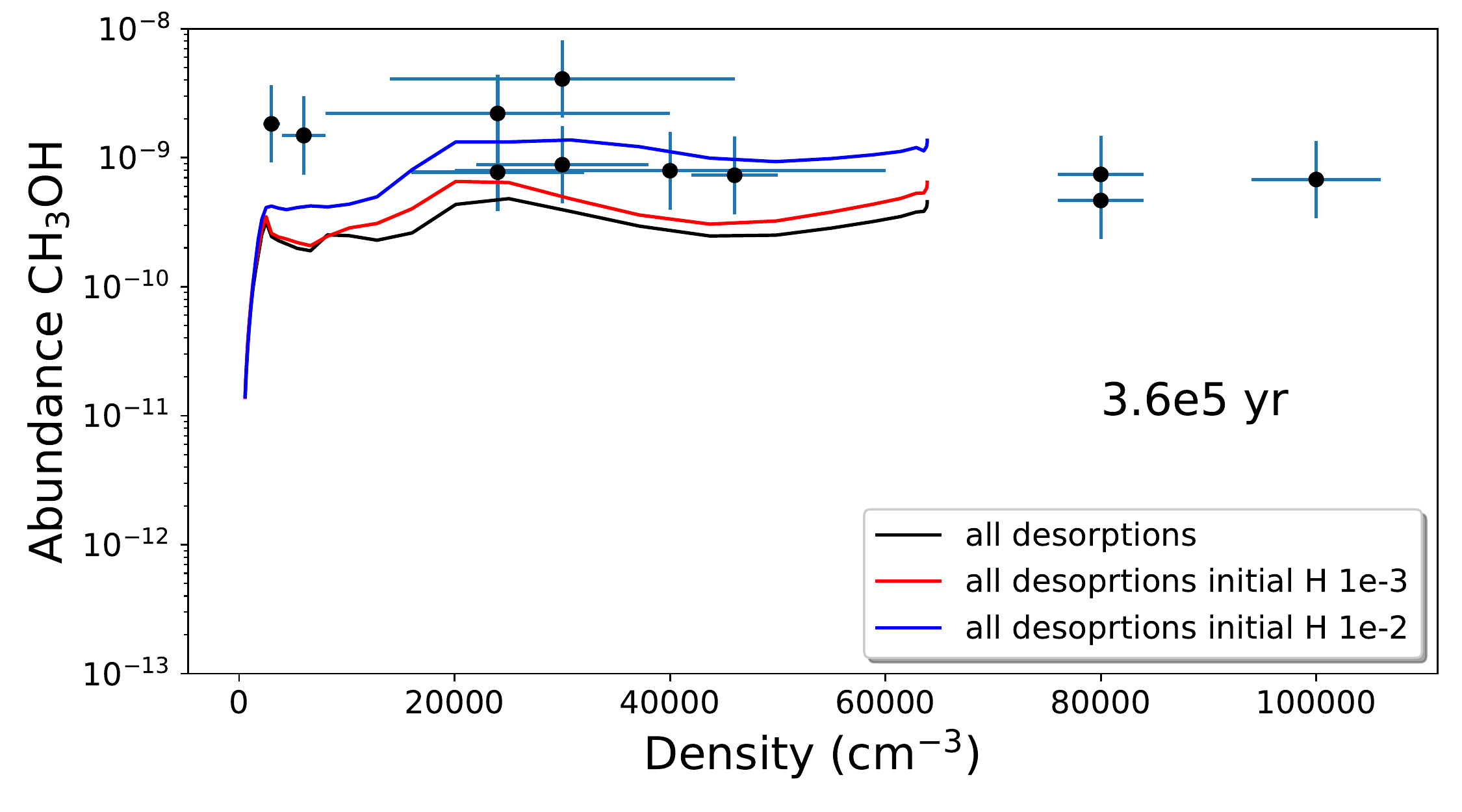}
\includegraphics[width=0.46\linewidth]{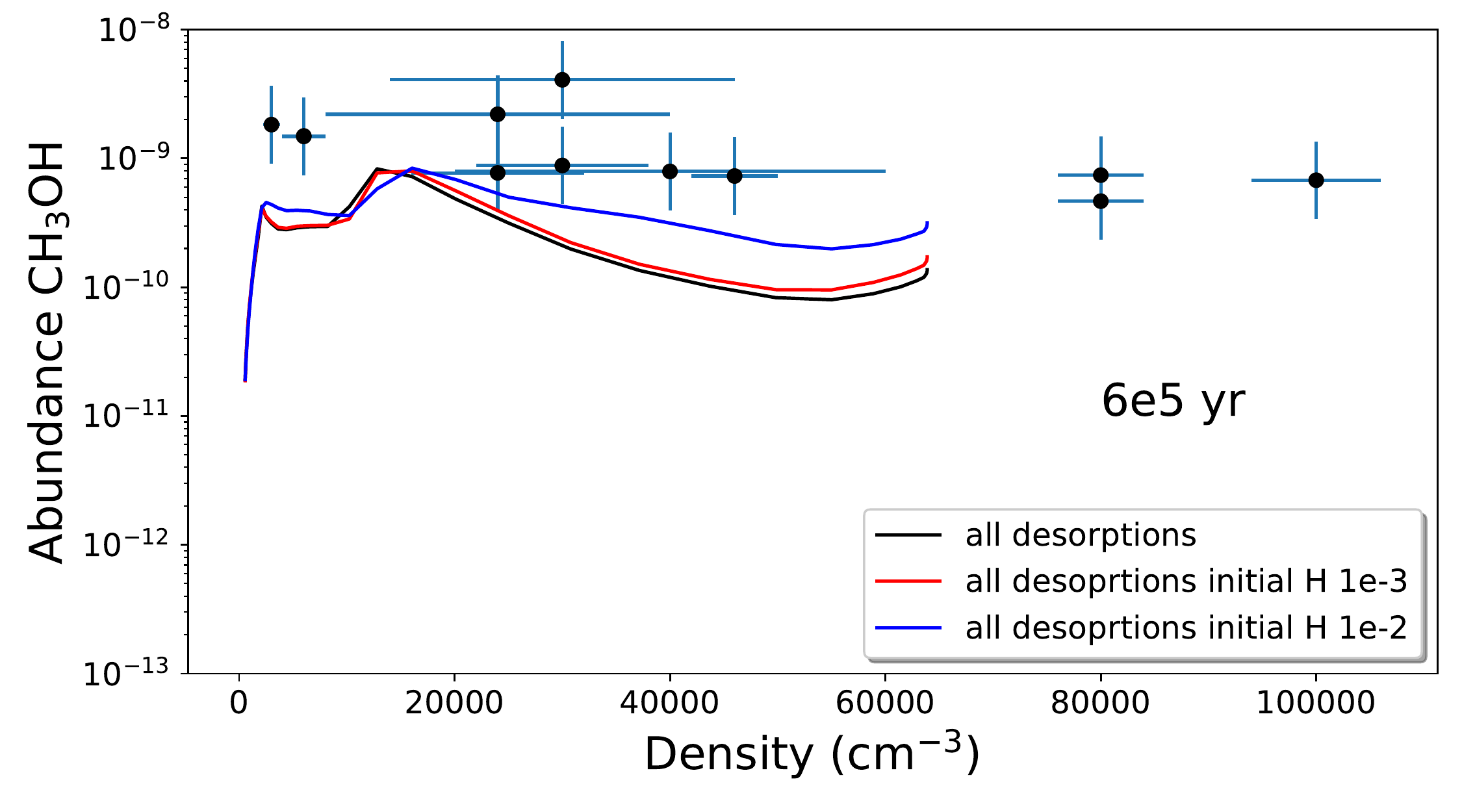}
\caption{Abundance of gas-phase methanol as a function of H density (in cm$^{-3}$) in the case of the model starting with initial abundances of zero, $10^{-3}$, and $10^{-2}$. Two times are shown: $3.6\times 10^{5}$ and $6\times 10^{5}$~yr. The points are the observed abundances as described in the text. In all cases, all the desorption mechanisms were included.}
\label{CH3OH_initialH}
\end{figure*}

\section{Model results when the dust temperature is set equal to the gas one}

\begin{figure*}
\centering
\includegraphics[width=0.46\linewidth]{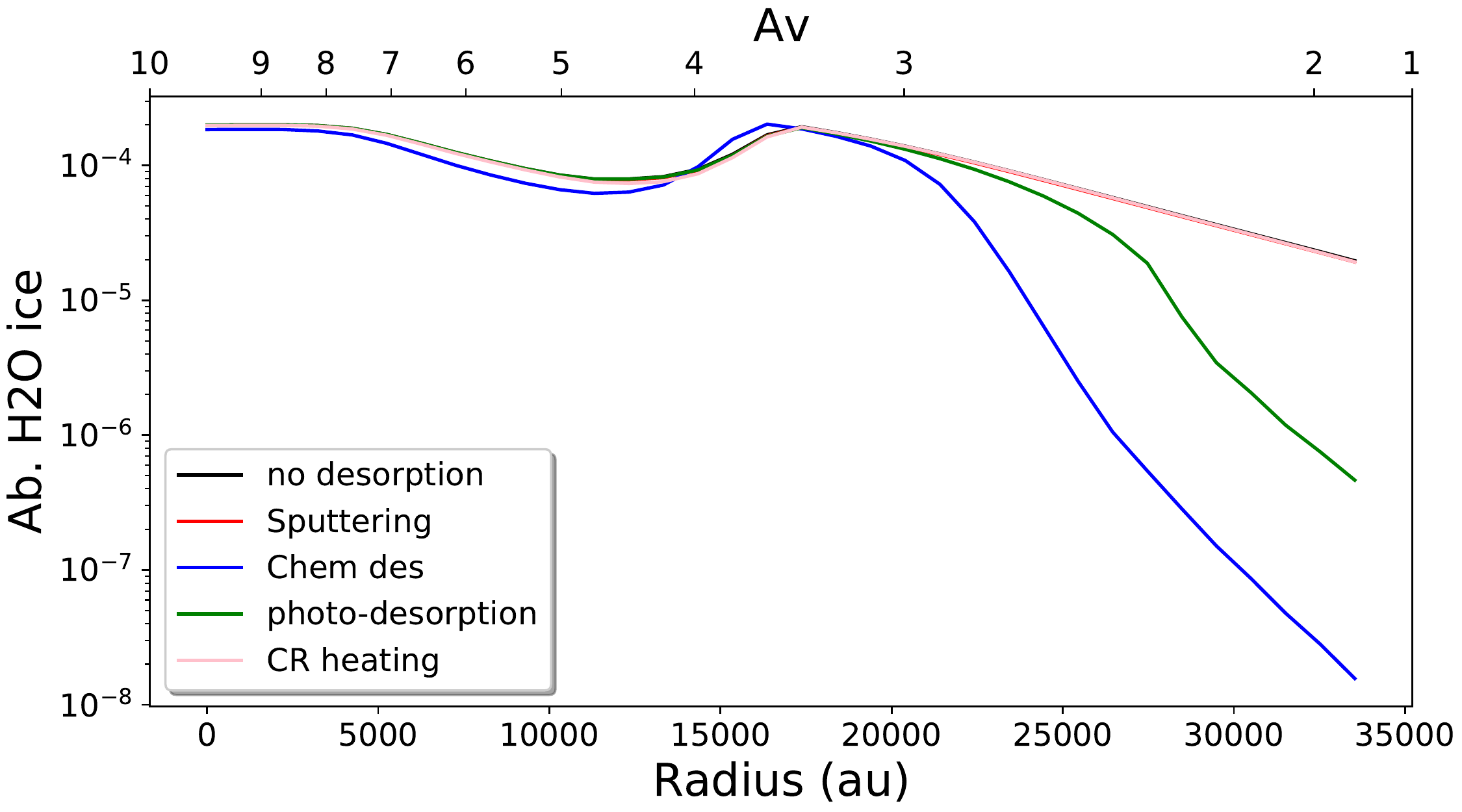}
\includegraphics[width=0.46\linewidth]{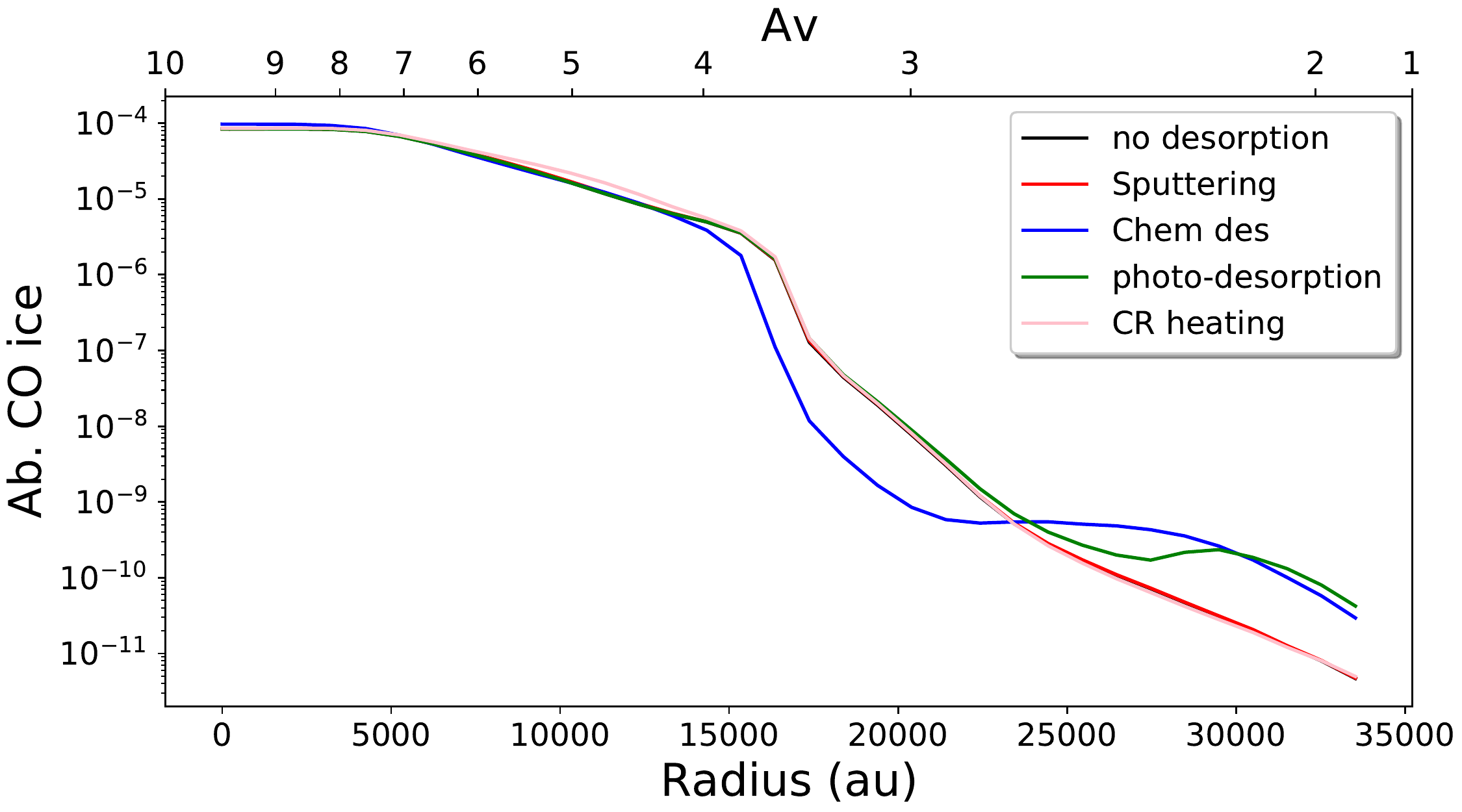}
\includegraphics[width=0.46\linewidth]{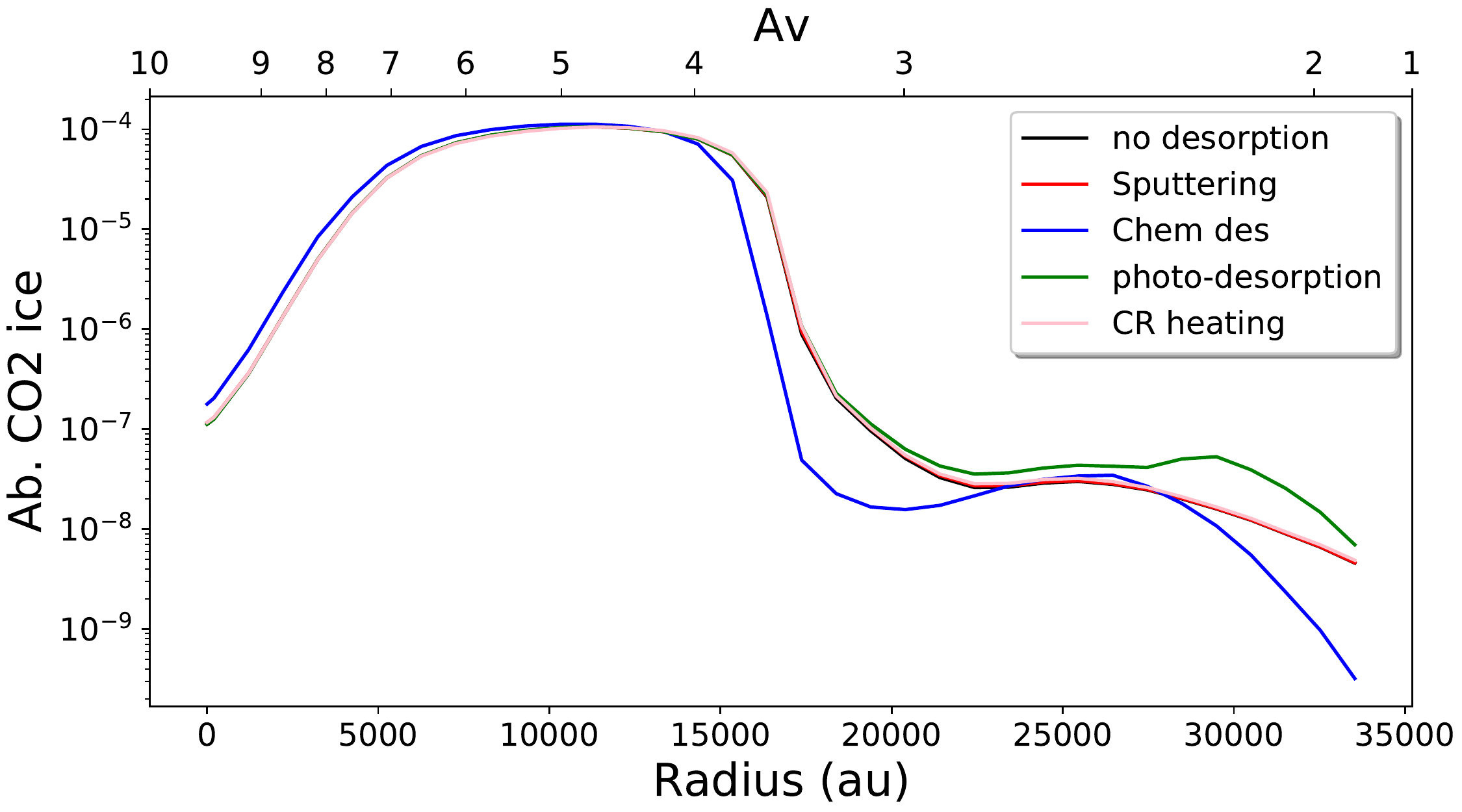}
\includegraphics[width=0.46\linewidth]{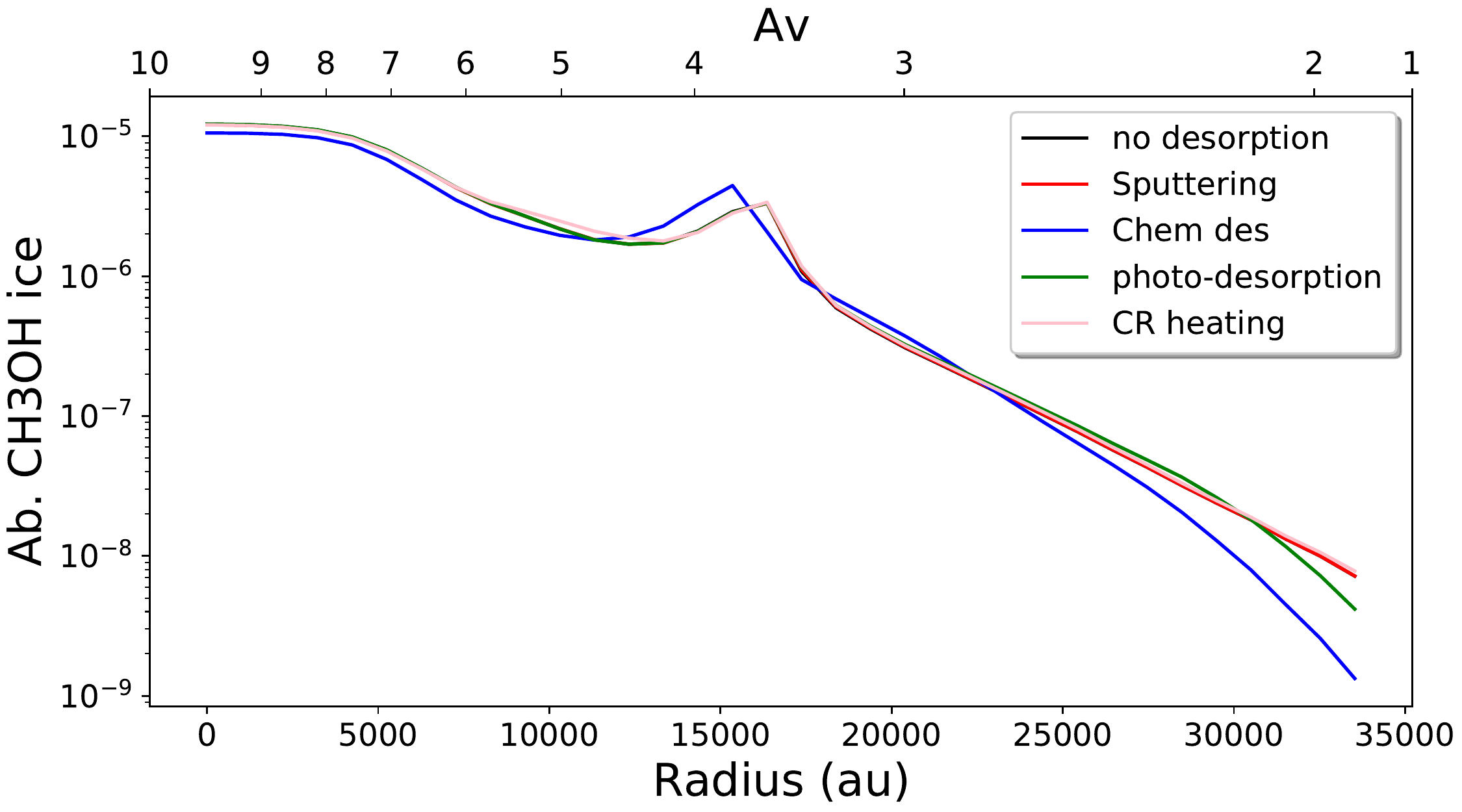}
\includegraphics[width=0.46\linewidth]{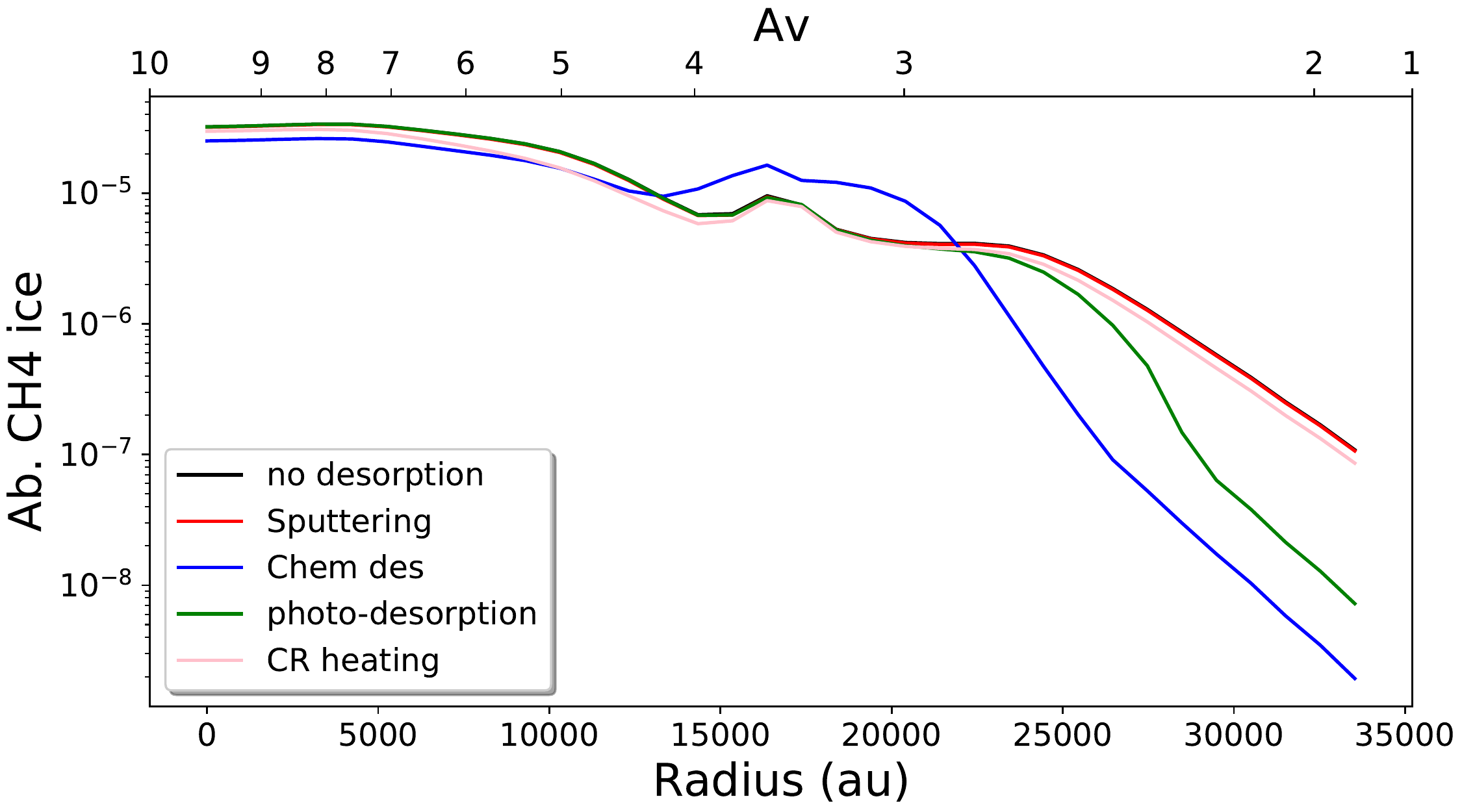}
\includegraphics[width=0.46\linewidth]{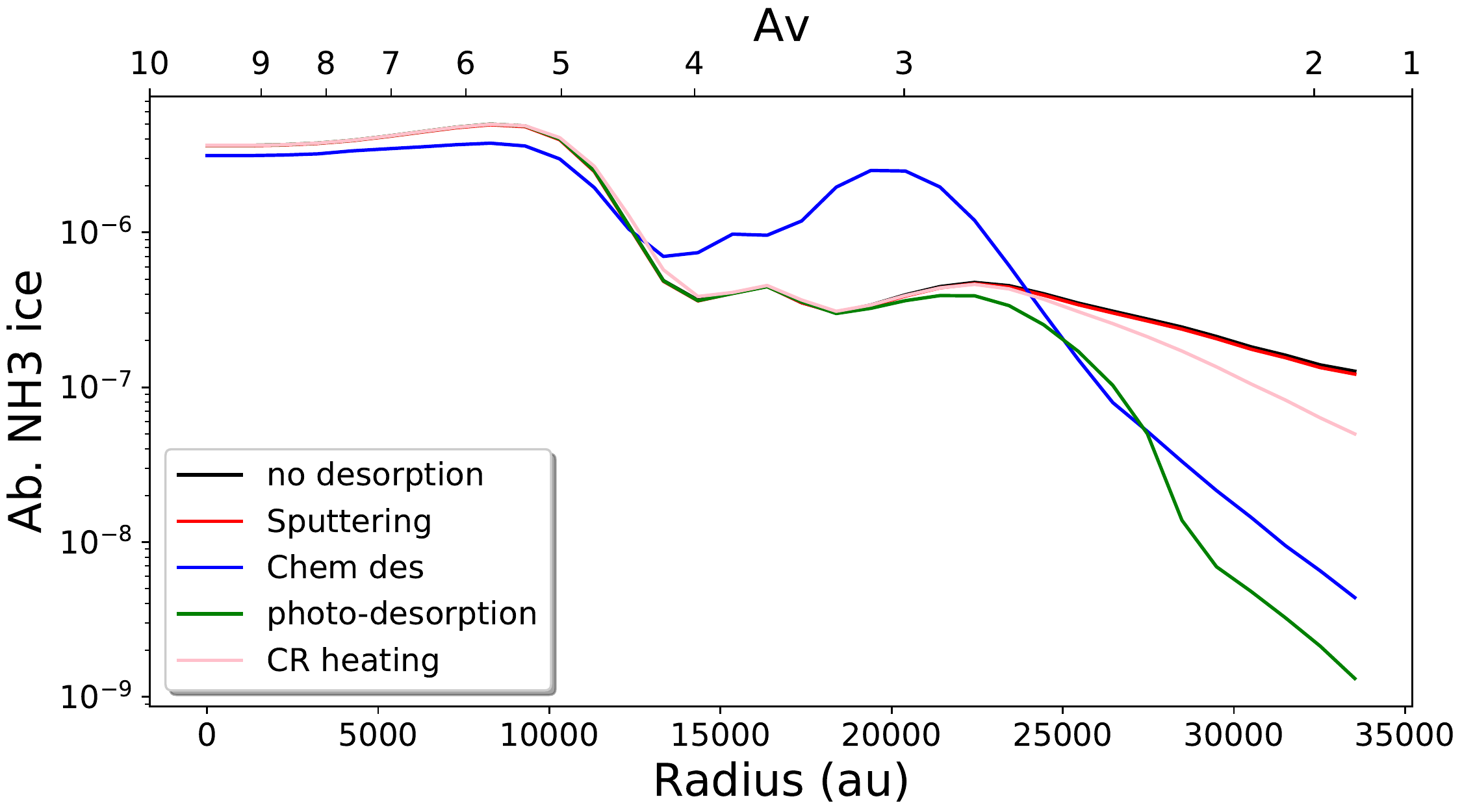}
\caption{Abundance of the main ice components as a function of radius (and visual extinction) for a time of $6\times 10^5$~yr, and the model in which the dust temperature is set equal to the gas one. The "no desorption" curve is almost the same as the "sputtering" one.}
\label{ice_fig_lowT}
\end{figure*}

\begin{figure*}
\centering
\includegraphics[width=0.46\linewidth]{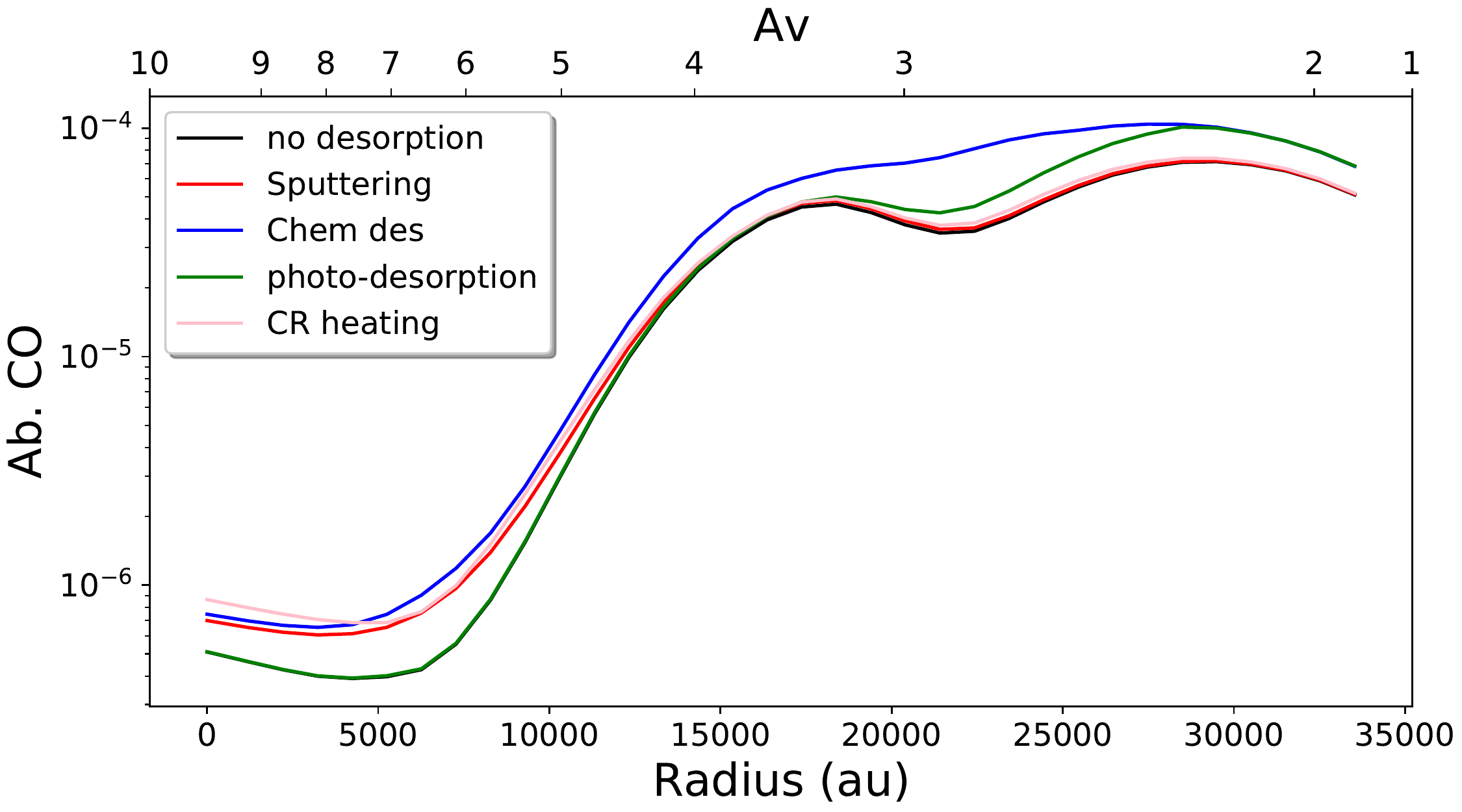}
\includegraphics[width=0.46\linewidth]{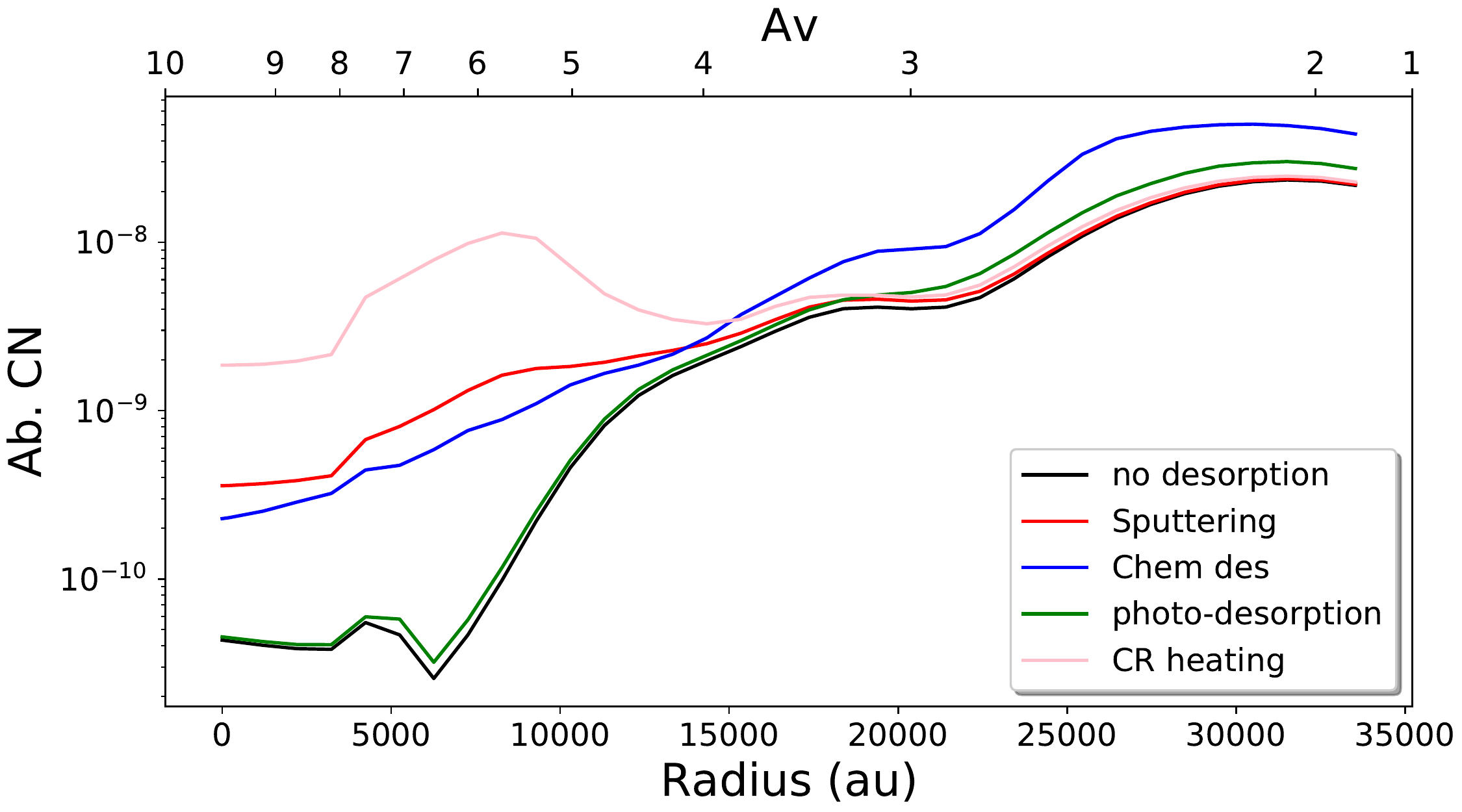}
\includegraphics[width=0.46\linewidth]{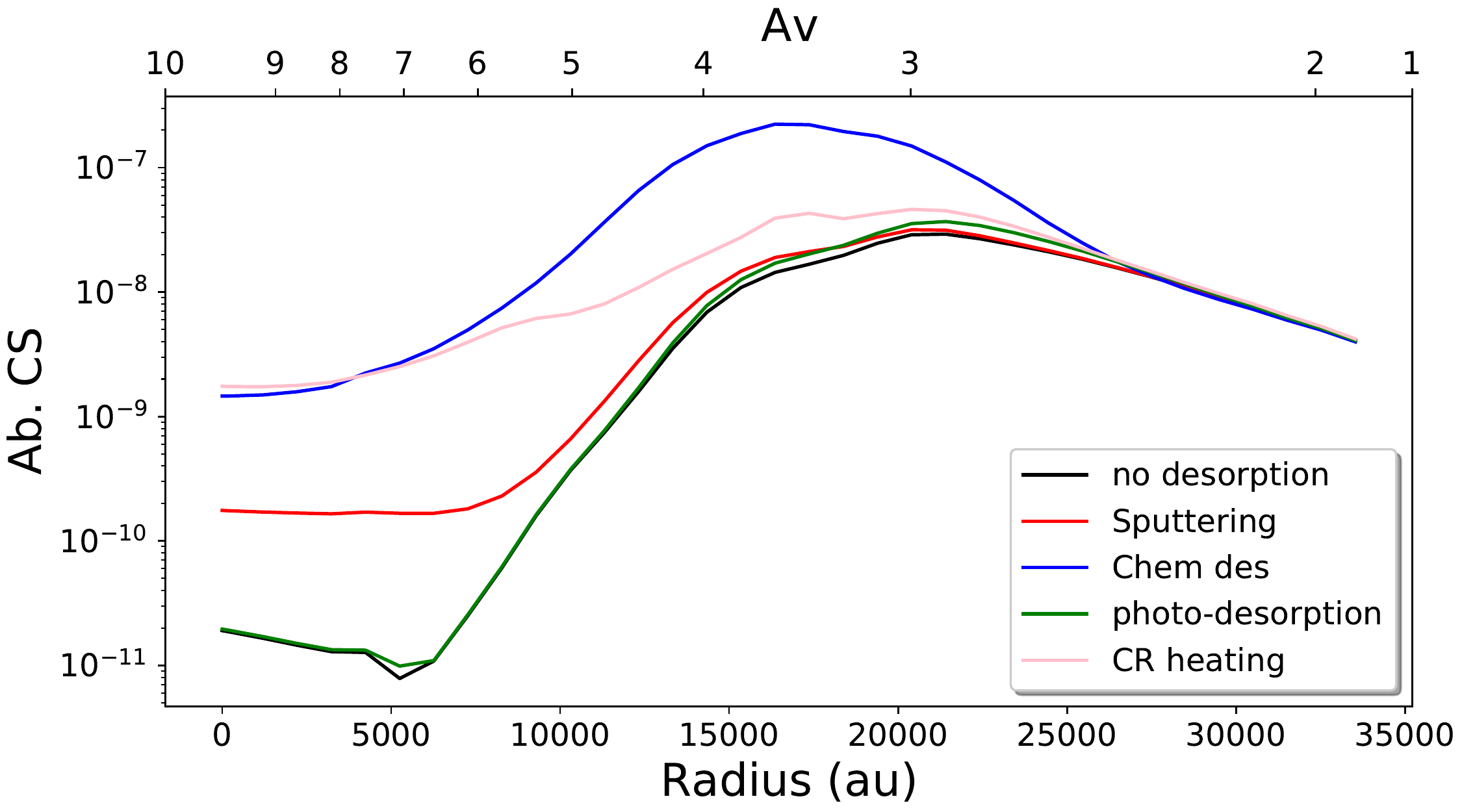}
\includegraphics[width=0.46\linewidth]{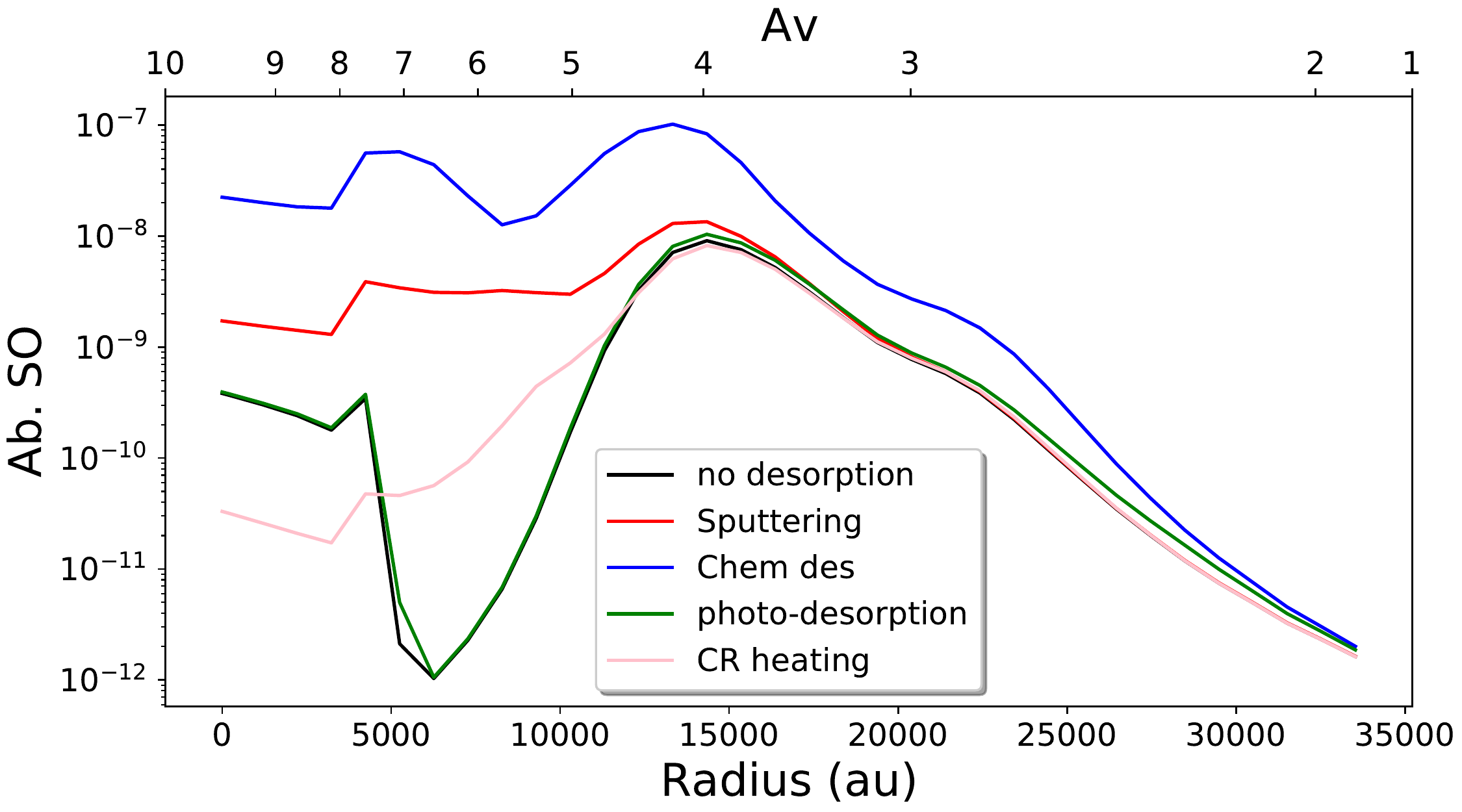}
\includegraphics[width=0.46\linewidth]{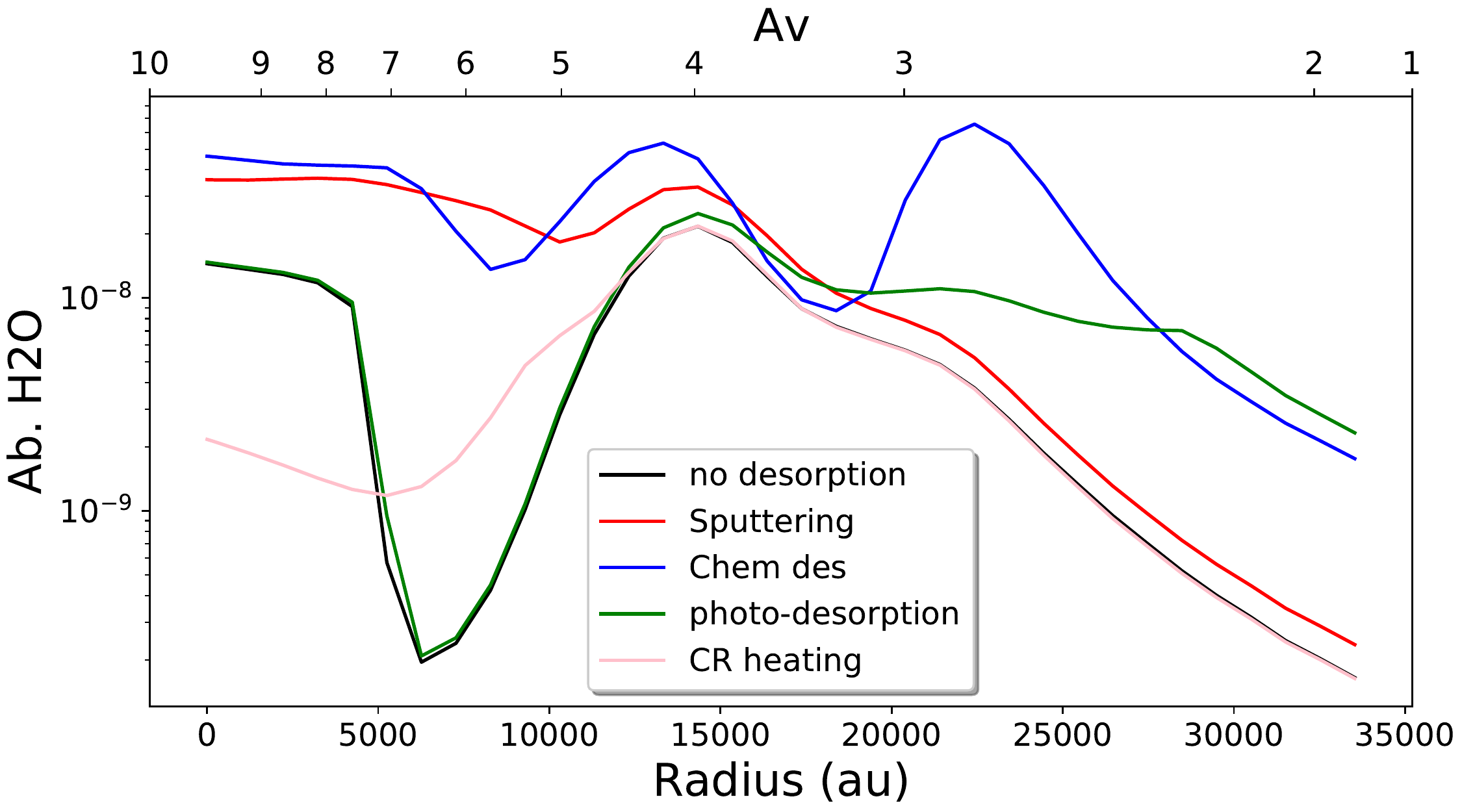}
\includegraphics[width=0.46\linewidth]{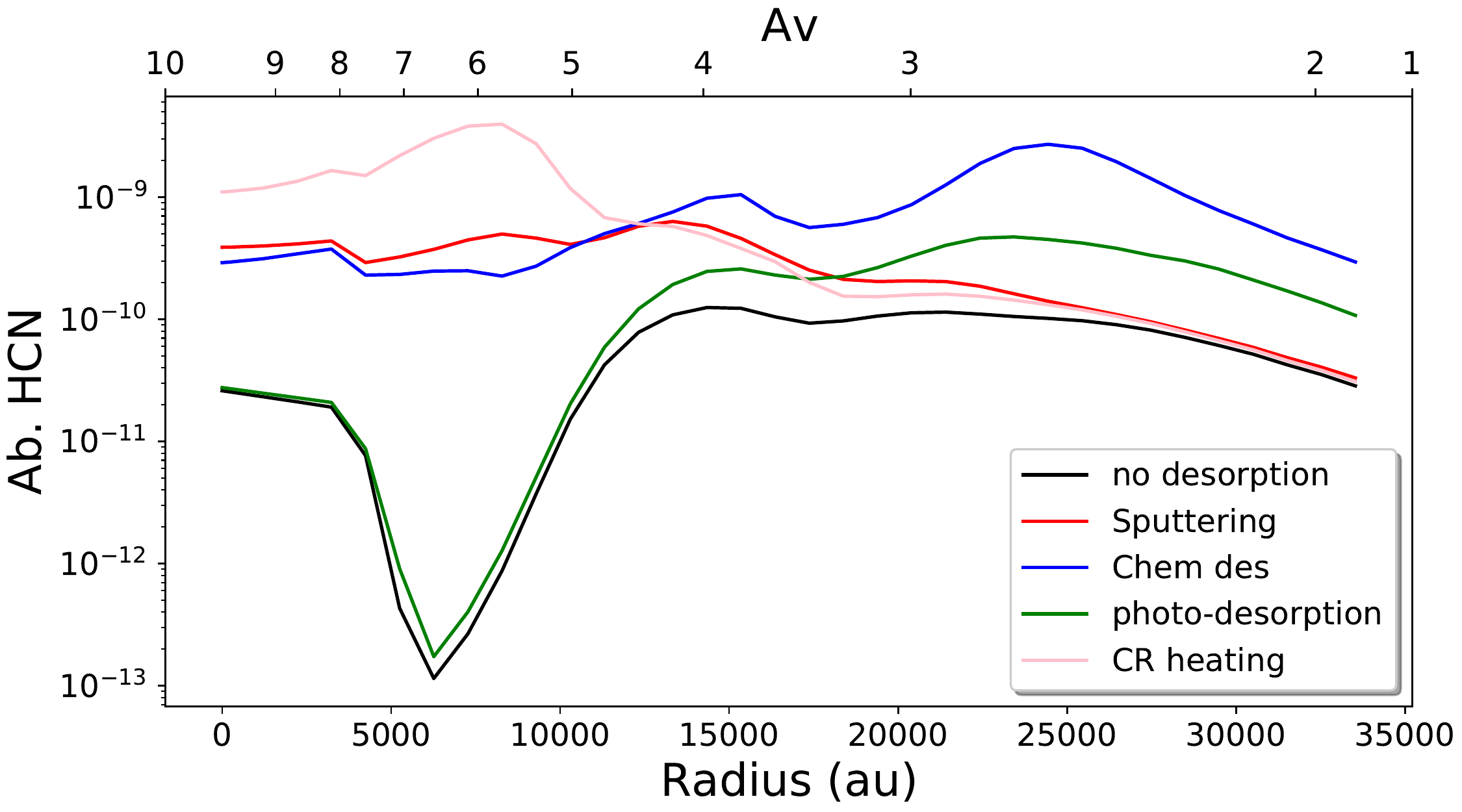}
\includegraphics[width=0.46\linewidth]{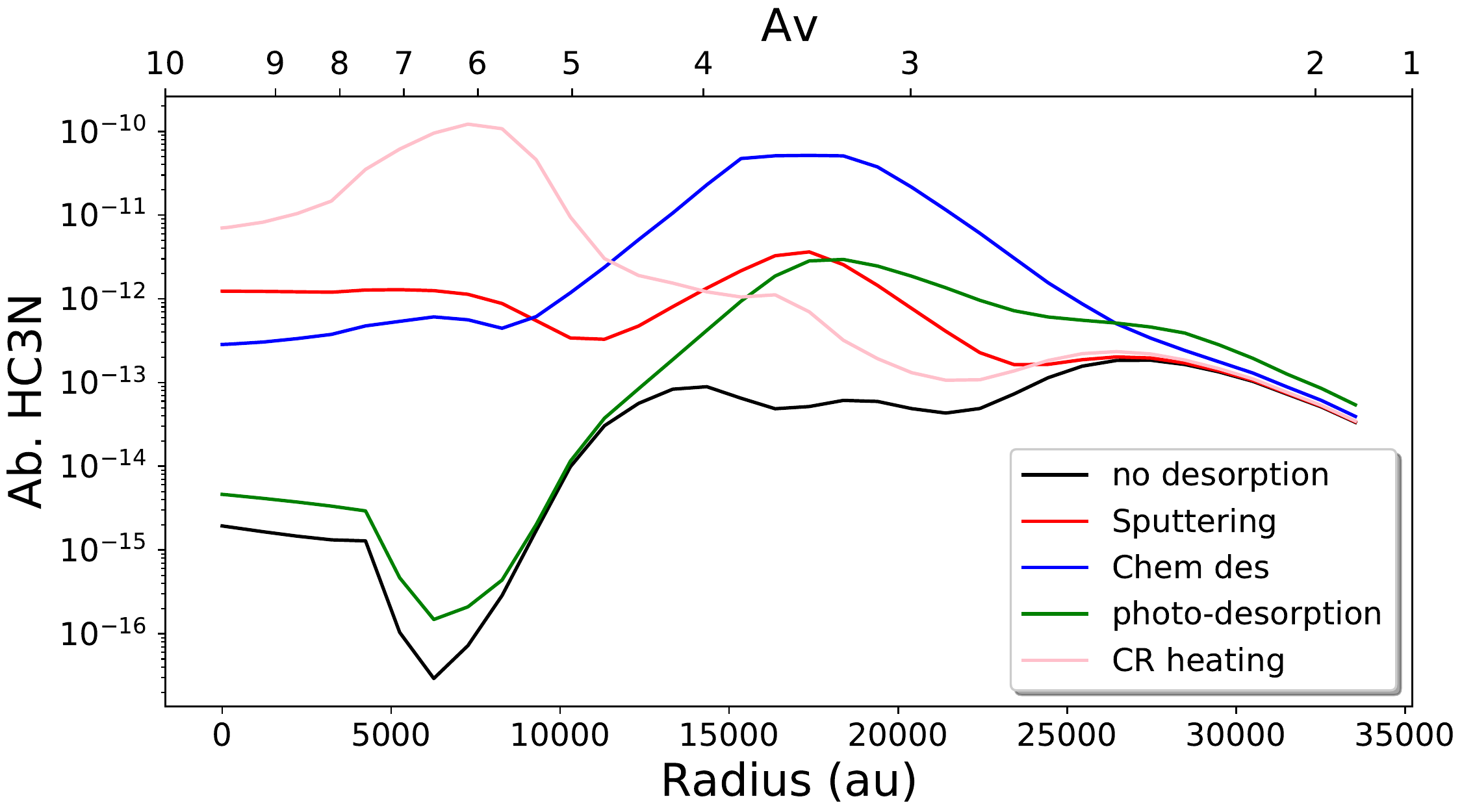}
\includegraphics[width=0.46\linewidth]{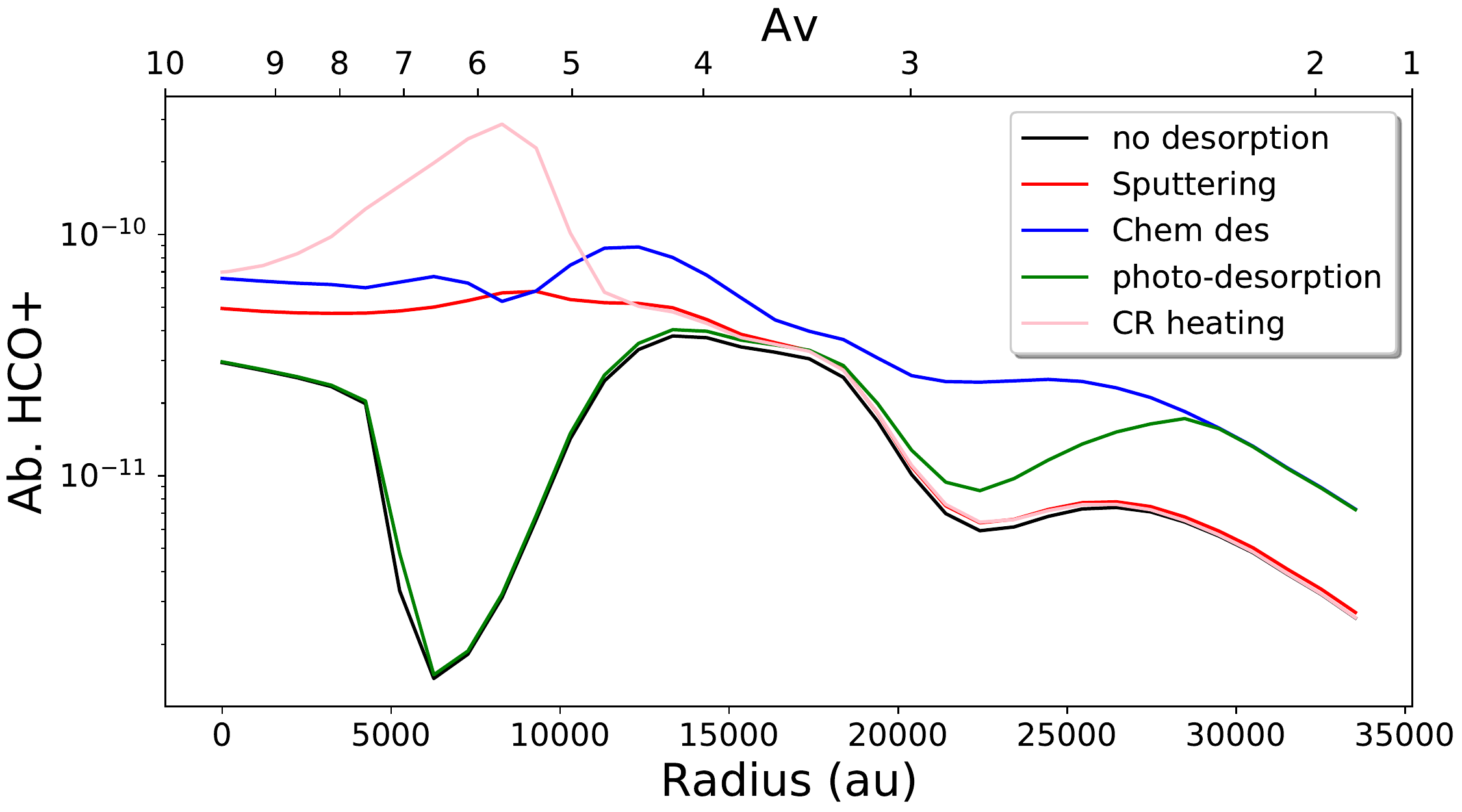}
\caption{Abundance of simple gas-phase molecules as a function of radius (and visual extinction) for a time of $6\times 10^5$~yr, and the model in which the dust temperature is set equal to the gas one.}
\label{simplemol_fig_lowT}
\end{figure*}

\begin{figure*}
\centering
\includegraphics[width=0.46\linewidth]{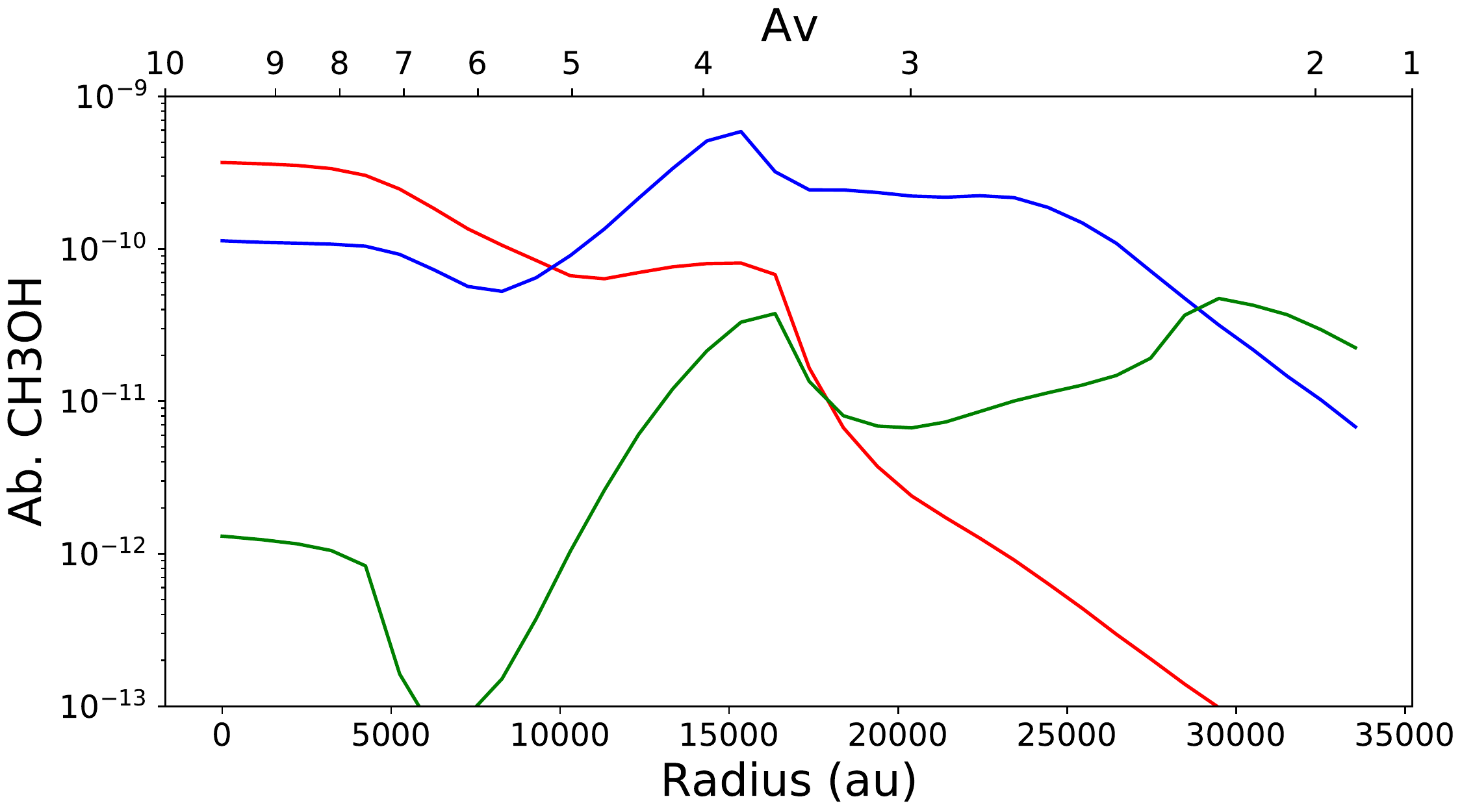}
\includegraphics[width=0.46\linewidth]{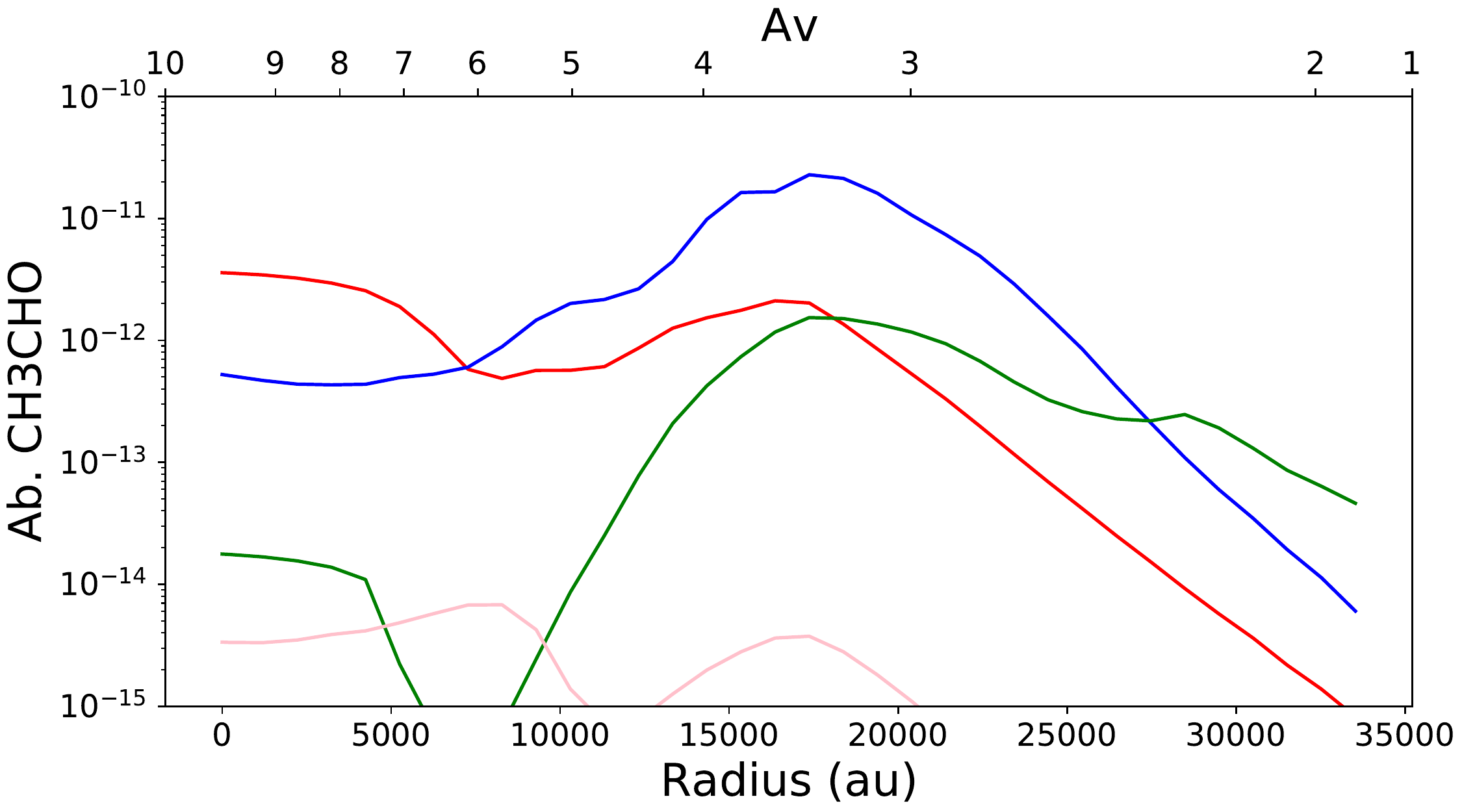}
\includegraphics[width=0.46\linewidth]{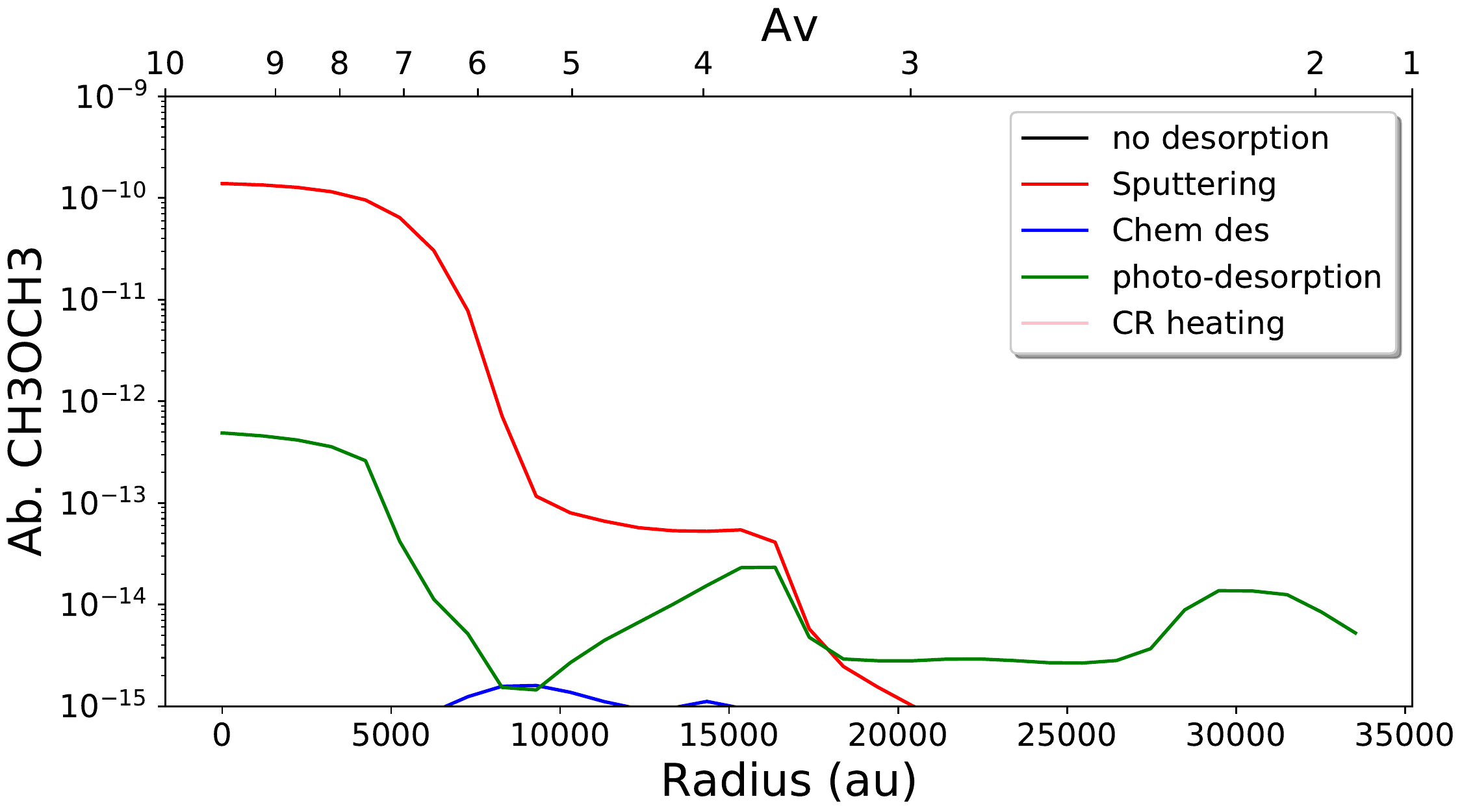}
\includegraphics[width=0.46\linewidth]{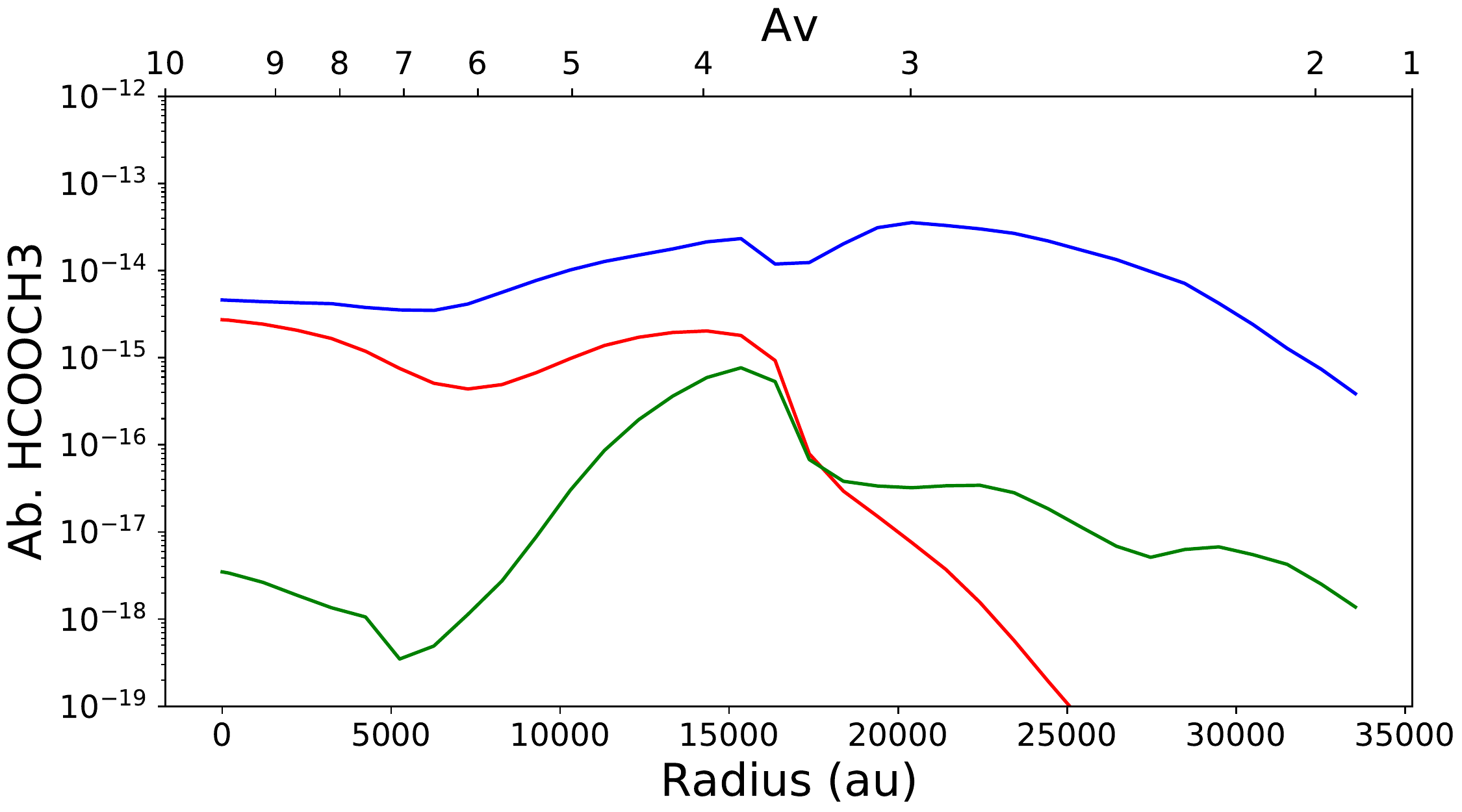}
\caption{Abundance of the complex organic molecules as a function of radius (and visual extinction) for a time of $6\times 10^5$~yr, and the model in which the dust temperature is set equal to the gas one.\label{COM_fig_lowT}}
\end{figure*}

\begin{figure*}
\centering
\includegraphics[width=0.46\linewidth]{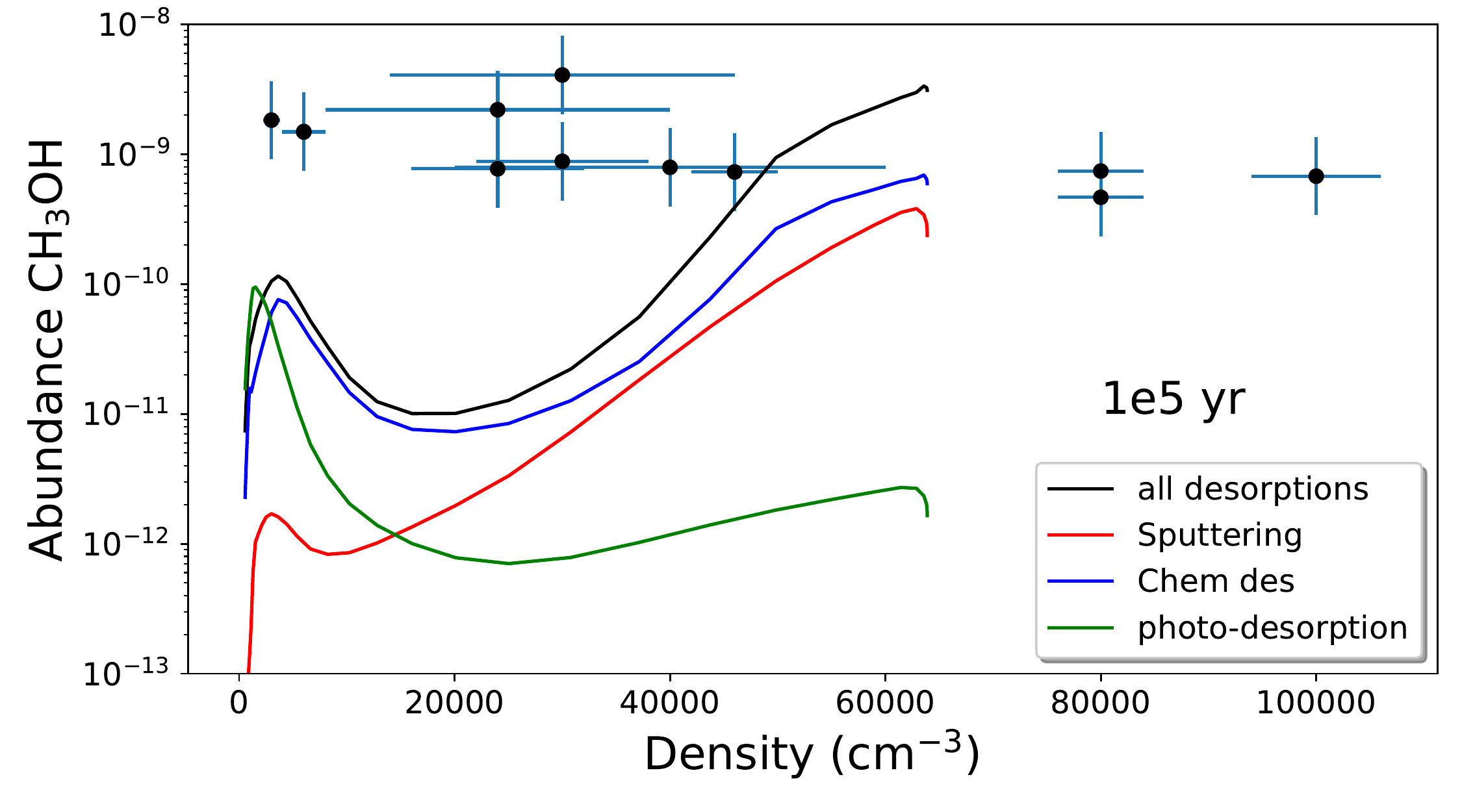}
\includegraphics[width=0.46\linewidth]{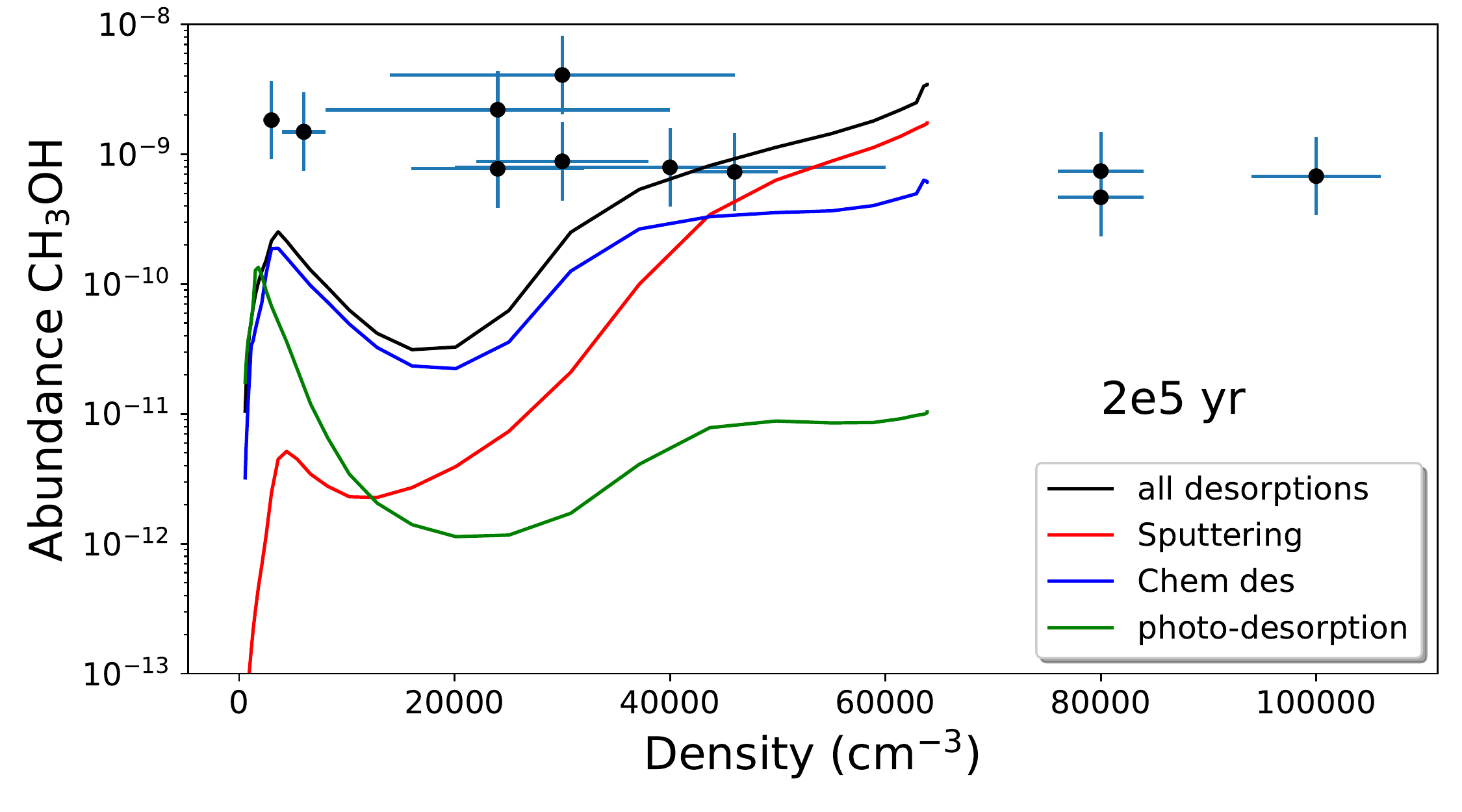}
\includegraphics[width=0.46\linewidth]{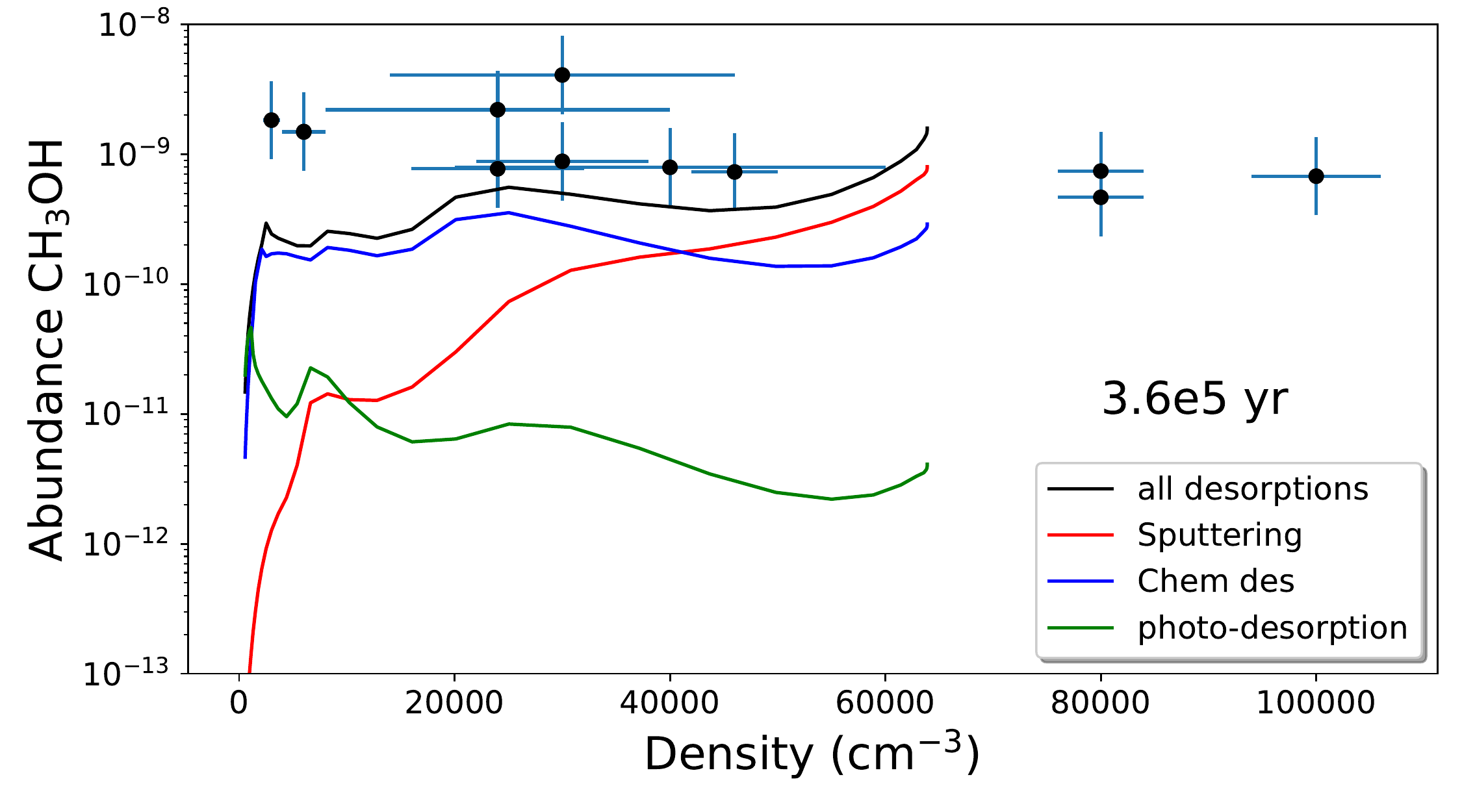}
\includegraphics[width=0.46\linewidth]{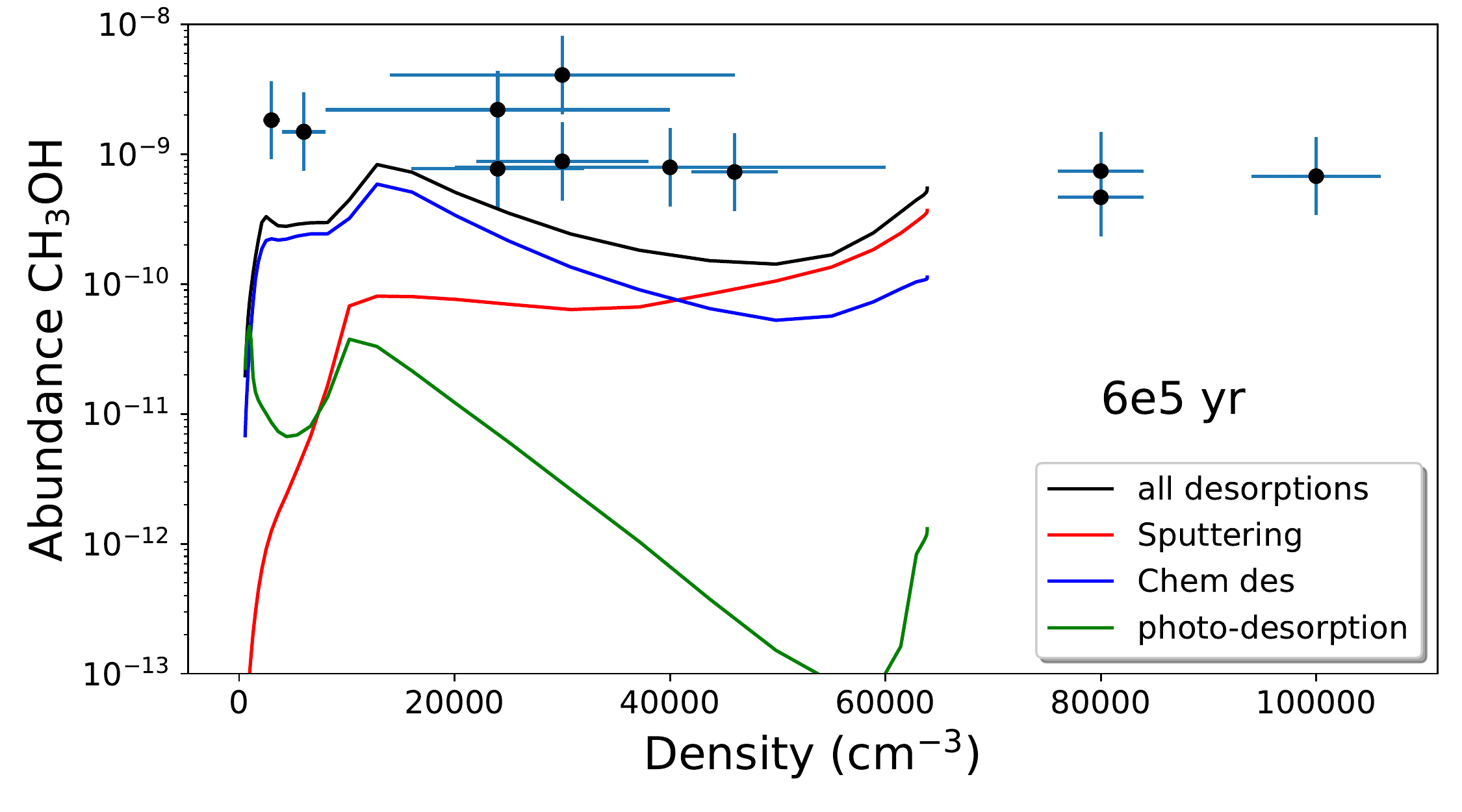}
\includegraphics[width=0.46\linewidth]{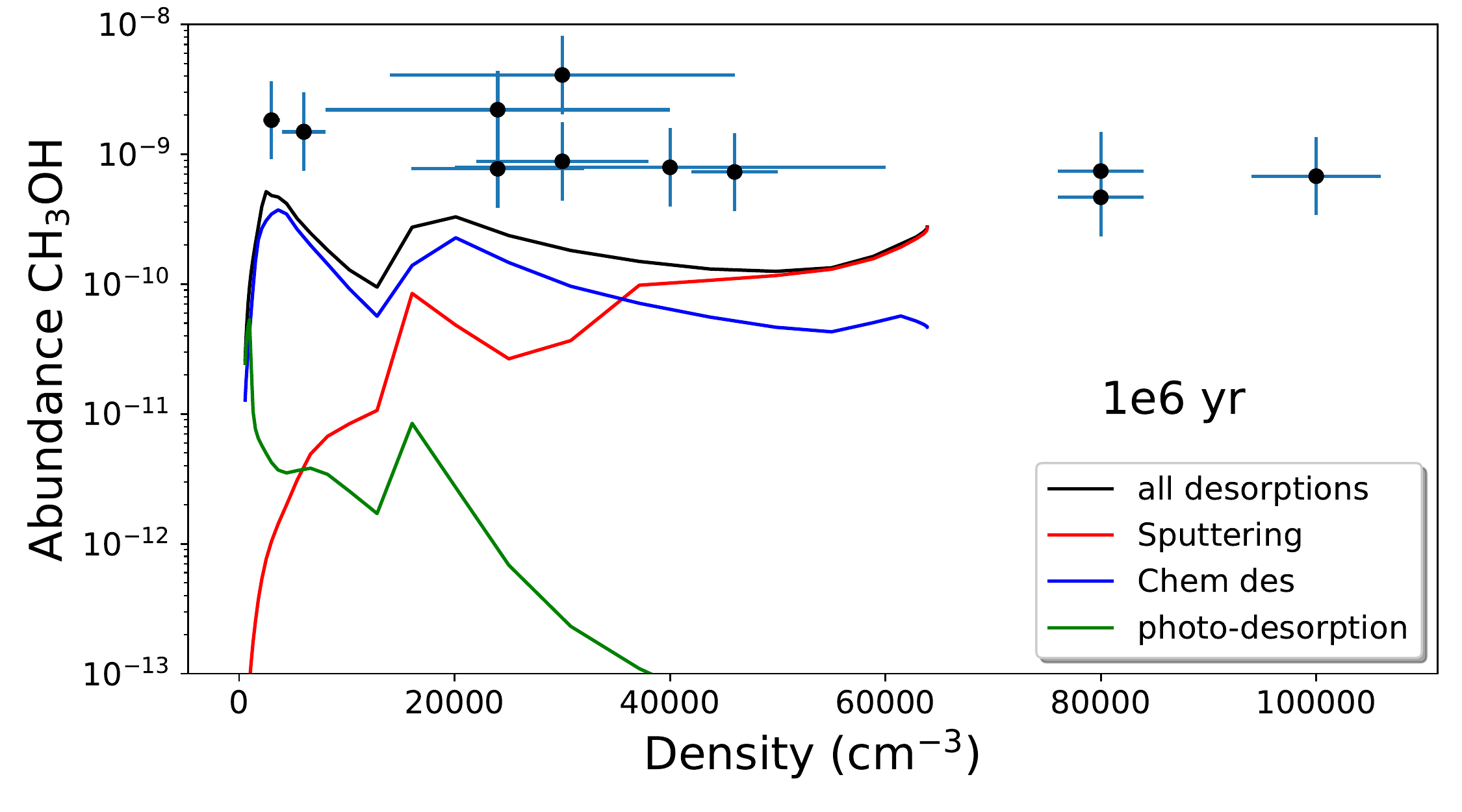}
\caption{Abundance of gas-phase methanol as a function of H density (in cm$^{-3}$) in the case of the model in which the dust temperature is set equal to the gas one. The different graphs represent different times. The lines in each box represent the models in which one of the non-thermal desorption process has been added or all of them. The points are the observed abundances as described in the text.}
\label{CH3OH_allprocess_lowT}
\end{figure*}

%%%%%%%%%%%%%%%%%%%%%%%%%%%%%%%%%%%%%%%%%%%%%%%%%%

% Don't change these lines
%\bsp   % typesetting comment
\label{lastpage}
\end{document}